\documentclass[%
 aip,
 amsmath,amssymb,
 reprint,%
groupedaddress
]{revtex4-1}

\usepackage{dcolumn} 
\usepackage{multirow}
\usepackage[T1]{fontenc}
\usepackage{dsfont}
\usepackage{verbatim}
\usepackage{bm}
\usepackage{hyperref} 
\usepackage{cleveref} 
\usepackage{floatrow}  
\usepackage{natbib}
\usepackage{graphicx}
\usepackage{ifpdf}
\usepackage{color}
\usepackage[normalem]{ulem}
\usepackage[usenames,dvipsnames]{xcolor}
\usepackage[caption=false]{subfig}
\begin{document}

\title{Active dipolar spheroids in shear flow and transverse field: Population splitting, cross-stream migration and orientational pinning}

\author{Mohammad Reza Shabanniya}
\thanks{shabannia@ipm.ir}
\affiliation{School of Physics, Institute for Research in Fundamental Sciences (IPM), P.O. Box 19395-5531, Tehran, Iran}
\author{Ali Naji}
\thanks{a.naji@ipm.ir (corresponding author)}
\affiliation{School of Physics, Institute for Research in Fundamental Sciences (IPM), P.O. Box 19395-5531, Tehran, Iran}

\begin{abstract} 
We study the steady-state behavior of active, dipolar, Brownian spheroids in a planar channel subjected to an imposed Couette flow and an external transverse field, applied in the `downward' normal-to-flow direction. The field-induced torque on active spheroids (swimmers) is taken to be of magnetic form by assuming that they have a permanent magnetic dipole moment, pointing along their self-propulsion (swim) direction. Using a continuum approach, we show that a host of  behaviors emerge over the parameter space spanned by the particle aspect ratio, self-propulsion and shear/field strengths, and the channel width. The cross-stream migration of the model swimmers is shown to involve a regime of linear response (quantified by a linear-response factor) in weak fields. For prolate swimmers, the weak-field behavior crosses over to a regime of full swimmer migration to the bottom half of the channel in strong fields. For oblate swimmers, a counterintuitive regime of reverse migration arises in intermediate fields, where a macroscopic fraction of swimmers reorient and swim to the top channel half at an acute `upward'  angle relative to the field axis. The diverse behaviors reported here are analyzed based on the shear-induced population splitting (bimodality) of the swim orientation, giving  two distinct, oppositely polarized, swimmer subpopulations (albeit very differently for prolate/oblate swimmers) in each channel half. In strong fields, swimmers of both types exhibit net upstream currents relative to  the laboratory frame. The onsets of full migration and net upstream current depend on the aspect ratio, enabling efficient particle separation strategies in microfluidic setups.
\end{abstract}


\maketitle

\section{Introduction}

Micro-/nanoscale active particles in fluid media include a wide range of artificial self-propellers and  microorganisms (or, for brevity, swimmers) that consume ambient free energy to enable their active motion \cite{Paxton2006_review,Lauga:RPP009,ramaswamyreview,Golestanian_review,Marchetti_review,Saintillan2013_ComptesRendus,Aranson2013_Active,Yeomans:EPJ2014,Saintillan2015,gompper_review,goldstein_review,cates_review,rusconi-stocker-review,bechinger_review,ZottlStark_review,Lauga:ANNREVF2016,Marchetti2016_ABPMinimalModel,Lowen:EPJST2016,KlumppReview2019}.  They have successfully been modeled in many cases as active Brownian particles self-propelling at constant speed (subject to thermal and athermal noises), a modeling approach that has also received   direct support from experiments  \cite{Howse2007_Self-Motile,Ebbens2010_Pursuit}. Swimmer motion in fluid channels has been of particular interest due to its relevance to nano-/microfluidic applications \cite{son_review}. Strong swimmer accumulation near the bounding walls is found to be a common motif in confined geometries; an effect known to arise from a number of different factors, including the persistent active motion of  swimmers, producing prolonged detention times near the walls, and hydrodynamic swimmer-wall couplings  \cite{Elgeti2016_Microswimmers, Elgeti2013_WallAccumulation, Hernandez-Ortiz1, li2011accumulation, li2011accumulation2,wallattraction,wallattraction2,wallattraction3, ardekani, Schaar2015_PRL, Mathijssen:2016b, Das_Nature_2015,Uspal_SM_2015,Chilukuri,Hill_Chirality_Upstream,ZottlPRL2012, ZottlPoiseuille2013,Mathijssen:2016a, Mathijssen:2016c, Uspal,KaturieaaoSciAdv, Costanzo,Kaya2009PRL,KAYA2012BJ,Nash2010,Ezhilan, Nili, Anand2019,Jiang2019,Potomkin_2017,Alonso-Matilla2019}.

The behavior of confined swimmers can change dramatically in an imposed shear flow  \cite{Mathijssen:2016a,Mathijssen:2016c,Uspal,Rusconi2014,Baker_2019,Nash2010,KAYA2012BJ,Morales_edgeSwimming,Chilukuri,Hill_Chirality_Upstream,Kaya2009PRL,MarcosPNAS,KaturieaaoSciAdv,Costanzo,ZottlPoiseuille2013,ZottlPRL2012,Jiang2019,Ezhilan, Nili, Anand2019,Jiang2019,Asheichyk2019,Alonso-Matilla2019,ZottlPRL2012, ZottlPoiseuille2013,Bretherton,sperm-rheotaxis,upstream2015prl,Potomkin_2017}. The celebrated upstream swim near no-slip walls emerges in a diverse class of sheared swimmers of elongated shape, such as sperm cells  \cite{Bretherton,sperm-rheotaxis,upstream2015prl}  and  {\em Escherichia coli} bacteria  \cite{Hill_Chirality_Upstream, KAYA2012BJ,Morales_edgeSwimming}. This effect has been attributed to a subtle balance between different factors contributing to the near-wall orientational dynamics of swimmers, giving a stable swim orientation in the upstream direction \cite{Mathijssen_weathervane,Hill_Chirality_Upstream, KAYA2012BJ,Mathijssen_weathervane,upstream2015prl,Lauga2013_PRL}.   The contributing factors include  the shear-induced Jeffery orbits  \cite{Jeffery,Bretherton1962}, the weathervane effect (caused by the fore-aft swimmer shape asymmetry), and intrinsic active torques (due, e.g.,  to near-wall hydrodynamic couplings) \cite{wallattraction2,upstream2015prl,Mathijssen_weathervane}.
Upstream swim has also been predicted to occur in the near-wall motion of spherical Janus particles, where the more intuitive mechanisms  applicable to elongated swimmers are absent \cite{Uspal,KaturieaaoSciAdv}. It emerges also within kinetic models of active Brownian particles in the absence of hydrodynamic interactions   (see Ref. \cite{Ezhilan} and references therein), with the upstream swim occurring here more generically near no-slip stationary walls due to a net,  shear-induced, reorientation of swimmers against the flow from their preferred, normal-to-wall, orientation in the absence of flow. 
In a Couette flow, the shear-induced bimodality of the swim orientation was addressed \cite{Nili} as a key to the understanding of the nonmonotonic behavior  \cite{Ezhilan} of the parallel-to-flow swimmer polarization. The  orientational bimodality was identified as a population splitting phenomenon \cite{Nili}, whereby a minority and a majority subpopulation of swimmers develop, swimming in opposite directions near the walls. 

Swimmer motion in fluid media can also be influenced by external stimuli and/or fields to which swimmers may respond actively and/or passively. Active response applies, e.g., to klinotaxis of microorganisms utilizing active reorientation mechanisms to align their effective path with the direction of external stimuli \cite{fraenkel_orientation_1961,Crenshaw1993}. This is applicable to certain types of chemotactic swimmers (e.g., sperm cells \cite{JulicherSpermChemo,JulicherPRL,Alvarez,JikeliSpermHelical,Crenshaw1993,Lushi}) and phototactic swimmers (e.g., the alga {\em Chlamydomonas reinhardtii} \cite{Bennett_Chlamy,goldstein_review,Photofocusing}), which  exhibit helical trajectories with an axis of motion that actively reorients in the direction of chemical gradients and light intensity, respectively. Passive response applies, e.g.,  to synthetic magnetic swimmers \cite{KlumppReview2019,Martin2014,Carlsen2014,Magdanz2013,Santon2017,Zhao2012,Schattling2017,Dreyfus2005, Tierno2009, Benkoski2011, Pak2011, Vach2015,Zhang2009_1, Zhang2009_2,Ogrin2008,Kokot_2017} and magnetotactic bacteria \cite{Schleifer1991, Frankel1997,  Bennet2014, Lefevre2014,Klummp2014BJ, FaivreReview2008,Klumpp2016, KlumppReview2019,Waisbord2016,Meng2018} that are aligned by external magnetic fields exerting a torque on their intrinsic magnetic dipole moment, and to gyrotactic swimmers \cite{Kessler1984, Kessler1985, Kessler1986b,Pedley1988,Pedley1990,PedleyReview1992,Pedley2015} that are  reoriented by the gravitational torque due to their  mismatching centers of gravity and buoyancy. 

It is this {\em passive} type of orientational response to an externally applied field, and its effects on the population splitting and transverse, cross-stream, migration of the orientational subpopulations of swimmers in a shear flow, that we aim to study in this paper. Being primarily interested in generic  (rather than swimmer specific) aspects of the problem, we adopt a minimal model \cite{gompper_review,bechinger_review,Marchetti_review,Marchetti2016_ABPMinimalModel,Saintillan2013_ComptesRendus} of active Brownian  spheroids in a planar channel subjected to a Couette flow in two dimensions. We include  the externally induced particle reorientation mechanism  through a cross-product torque, tending to align the assumed intrinsic (and fixed) dipole moment of the particles with an external field. As such, the present model can  kinematically be relevant to a wide range of problems, involving not only  swimmers possessing magnetic  \cite{Schleifer1991, Frankel1997, FaivreReview2008, Bennet2014, Lefevre2014,Klummp2014BJ, Klumpp2016, KlumppReview2019} or gravity-buoyancy force dipoles  \cite{Kessler1984, Kessler1985, Kessler1986b,Pedley1988,Pedley1990,PedleyReview1992,Pedley2015} but also those with electric dipoles in an applied electric field \cite{QuinckeRotor2019}. {\em Active} orientational response has also been modeled in some cases using  cross-product reorientation rates \cite{Taktikos,Lushi,Liebchen_2016},  producing preferential alignment of the particle symmetry (swim) axis with the direction of an external stimulus. For concreteness, and with no loss of generality, the dipole moment and the external field here are taken to be of magnetic type. The permanent dipole moment  is  fixed  along the swim direction \cite{FaivreReview2008, Klumpp2016, KlumppReview2019,Waisbord2016,Meng2018} and the external field  in the `downward' normal-to-flow direction (extensions of the current model will be discussed  elsewhere \cite{ZF,AA}). We study prolate as well as the seldom-studied oblate swimmers \cite{Michelin2017,Michelin2016}, and analyze and compare their probabilistic (and  certain aspects of their  deterministic) behaviors within a continuum approach in the dilute regime of noninteracting particles \cite{Ezhilan, Nili,Saintillan2013_ComptesRendus, Saintillan2015}.  

Our results reveal different regimes of field-induced, cross-stream, swimmer migration that are identified  and explained  based on the behaviors of the underlying orientational subpopulations of swimmers in the channel. They  include regimes of standard bimodality in weak fields, reverse bimodality (of prolate swimmers,  displaying reversed minority/majority populations in the top channel half) at intermediate field strengths, linear response (due to field-modified Jeffery oscillations of the swim orientation) in weak to intermediate fields, reverse migration (of a macroscopic fraction of oblate swimmers moving against the applied field due to a double-pinning or bistability of the swim orientation) in  intermediate to strong fields, full migration of swimmers to the bottom channel half and, eventually, a regime of unimodality on the bottom wall (due to single pinning of the swim orientation) as well as a regime of net upstream swimmer current in strong fields. Our results also suggest efficient routes to particle separation based on aspect ratio and motility strength that will be discussed in detail.

\begin{figure}[t!]
    \centering
    \vskip1mm
    \hskip5mm\includegraphics[width=8.5cm]{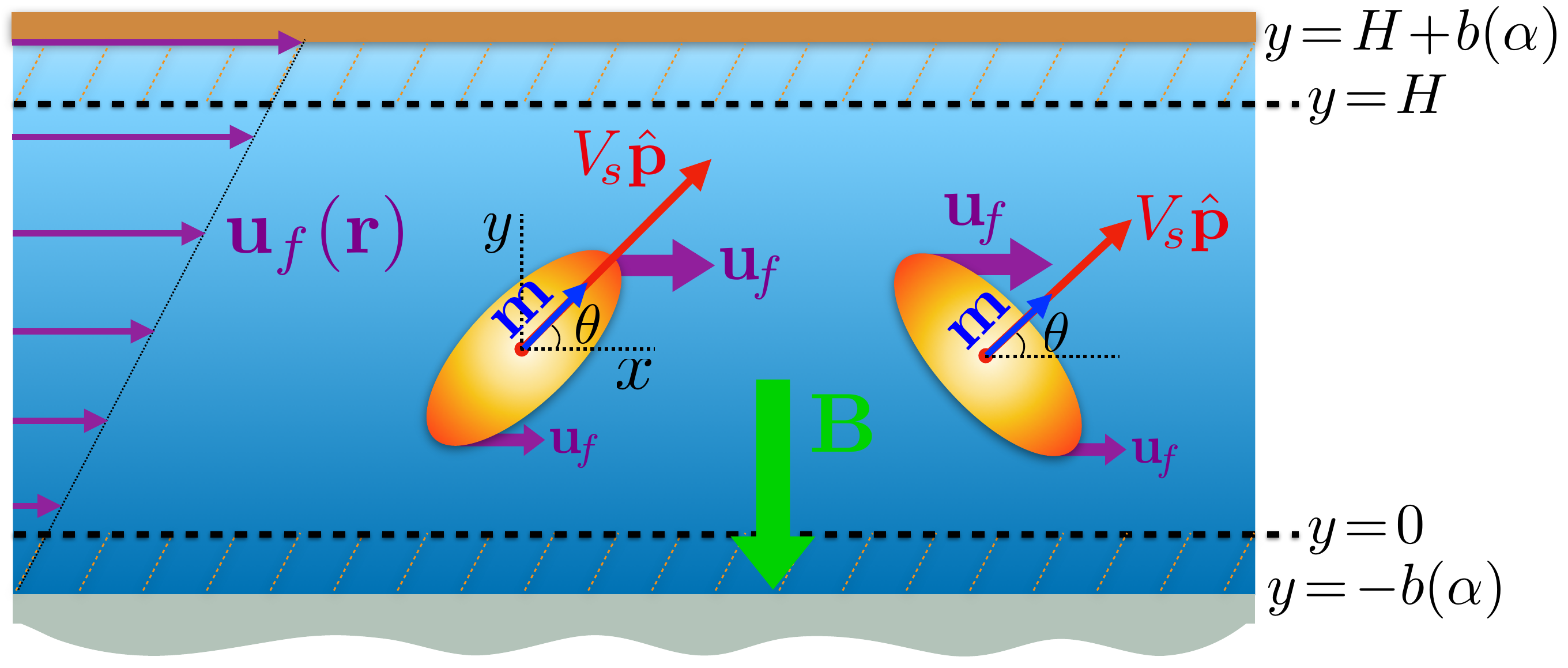}
    \caption{Schematic view of prolate/oblate  (left/right) swimmers with  self-propulsion velocity, $V_s{\hat{\mathbf p}}$, and  magnetic dipole moment, ${\mathbf m}$, in a Couette flow of linear velocity profile, $\mathbf{u}_f({\mathbf r})$ with ${\mathbf r}=(x,y)$,  and a `downward' magnetic field, $\mathbf{B}$.  The self-propulsion and magnetic axes coincide with the spheroidal symmetry axis. The no-flux channel walls (dashed lines) are placed at the  closest-approach distance (being equal to the semi-minor body-axis length of the spheroids), $b(\alpha)$, from the no-slip boundaries at $y=- b(\alpha)$ and  $H+ b(\alpha)$.}
    \label{fig:schematic}
\end{figure}

The paper is organized as follows: Our model and methods are introduced in Sections \ref{sec:model} and \ref{sec:Smoluchowski_eq}. The results for prolate (oblate) swimmers are given in Sections \ref{sec:prolates} (\ref{sec:oblates}), with the regimes of net upstream current and linear response discussed in Section  \ref{sec:flux} and Appendix \ref{sec:weak_field}, respectively. The results are summarized in Section \ref{sec:summary}, followed by further discussions on the applicability of the model to real systems in Section \ref{sec:discussion} and Appendix \ref{app:parameters}.

\section{Physical Specifications}
\label{sec:model}

We use a minimal  two-dimensional model of active Brownian particles (swimmers) that self-propel at fixed-magnitude  velocity $V_s{\hat{\mathbf p}}$ along the unit vector ${\hat{\mathbf p}}$, specifying their self-propulsion  (or swim) orientation; see Fig. \ref{fig:schematic}. The swimmers are modeled as spheroids of aspect ratio $\alpha$, whose  self-propulsion axis coincides with their axis of symmetry, being  the  {\em major} ({\em minor}) body axis in the case of {\em prolate} ({\em oblate}) swimmers with $\alpha>1$ ($\alpha<1$). They are further assumed to possess a permanent magnetic dipole moment along the same axis as ${\mathbf m}=m{\hat{\mathbf p}}$, with $m\geq 0$ (a more general case with the spheroidal magnetic axis deviating from the self-propulsion or symmetry axis  will be considered elsewhere \cite{ZF}). The in-plane rotational (Stokes) diffusivity of swimmers is denoted by $D_R(\alpha)$, and their translational diffusion tensor is expressed in terms of their parallel, $D_\parallel(\alpha)$, and perpendicular, $D_\perp(\alpha)$, (Stokes) diffusivities as \cite{gompper2006soft}
\begin{equation}
{\mathbb D}_T = D_\parallel(\alpha)\, {\hat{\mathbf p}}{\hat{\mathbf p}} + D_\perp(\alpha) \left(\mathbb{I} - {\hat{\mathbf p}}{\hat{\mathbf p}}\right).
\end{equation}

The swimmers are  confined within a long planar channel aligned with the $x$-axis (Fig. \ref{fig:schematic}). The channel has an {\em effective width} of $H$, representing the width that is sterically accessible to swimmers along the $y$-axis, as will be clarified  in Section \ref{sec:Smoluchowski_eq}. The channel is subjected to an incompressible stationary flow of {\em Couette}  type, with the fluid velocity, $\mathbf{u}_f({\mathbf r}) \!=\! u_f(y) \hat{\mathbf x}$, having a constant shear rate $\partial u_f(y)/\partial y =\! \dot\gamma\!\geq 0$. In Couette flow,  swimmers experience a position-independent torque due to the flow, presenting a scenario that enables unequivocal elucidation of  shear-induced phenomena \cite{Asheichyk2019,Nili,Chilukuri}. In the probabilistic approach to be adopted later, we shall require the ensuing shear-induced or Jeffery angular velocity, $\boldsymbol{\omega}_f$. For the no-slip spheroids assumed here, one has  \cite{Jeffery,Bretherton1962,kim_microhydrodynamics,gompper2006soft,doiedwards}
\begin{equation}
\boldsymbol{\omega}_f={\hat{\mathbf p}}\times \left[\big(\beta(\alpha)\, {\mathbb{E}}+{\mathbb{W}}\big)\cdot{\hat{\mathbf p}}\right],  
\label{eq:w_f0}
\end{equation}
where $\beta(\alpha)$  is the Bretherton number \cite{Bretherton1962}, 
\begin{equation}
\beta(\alpha)=\frac{\alpha^2-1}{\alpha^2+1}, 
\label{eq:breth_num}
\end{equation}
and ${\mathbb{E}}=\frac{1}{2}[({\nabla}\mathbf{u}_f)\!+\!({\nabla}\mathbf{u}_f)^T]$ and ${\mathbb{W}}=\frac{1}{2}[({\nabla}\mathbf{u}_f)\!-\!({\nabla}\mathbf{u}_f)^T]$ are the rate-of-strain and vorticity tensors, respectively. Using the parametrization ${\hat{\mathbf p}}=(\cos \theta, \sin \theta)$, where $\theta$ is the polar swim angle measured from the $x$-axis (Fig. \ref{fig:schematic}), one has  $ \boldsymbol{\omega}_f=  \omega_f(\theta) \,\hat{\mathbf z}$, where $\hat{\mathbf z}=\hat{\mathbf x}\times\hat{\mathbf y}$ is the unit vector normal  to the plane of the flow, and 
\begin{equation} 
{\omega}_f(\theta)=  \frac{\dot\gamma }{2} \bigg(\beta(\alpha) \cos2\theta-1 \bigg).
\label{eq:w_f}
\end{equation}  

The swimmers are also subjected to a uniform magnetic field, ${\mathbf B}$, applied transversally in the `downward' normal-to-flow direction, $\mathbf{B}=-B\hat{\mathbf y}$,  with $B\geq 0$ (the role of a parallel-to-flow field component will be discussed elsewhere \cite{ZF}). The field-induced angular velocity due to the corresponding torque that tends to align the permanent magnetic dipole moment of the particles with the external  field is  expressed  as 
\begin{equation}
\boldsymbol{\omega}_{\mathrm{ext}}= -\frac{D_R(\alpha)}{k_\mathrm{B} T} mB\left({\hat{\mathbf p}}\times \hat{\mathbf{y}}\right),
\label{eq:w_ext0}
\end{equation}
 where $D_R(\alpha)/k_{\mathrm{B}}T$ is the in-plane rotational mobility of the spheroids and $k_{\mathrm{B}}T$ is the ambient thermal energy scale. Explicitly, one has $\boldsymbol{\omega}_{\mathrm{ext}}=\omega_{\mathrm{ext}}(\theta) \,\hat{\mathbf z}$, with 
\begin{equation} 
{\omega}_{\mathrm{ext}}(\theta) = -\chi D_R(\alpha) \cos\theta, 
\label{eq:w_ext}
\end{equation} 
where $\chi$ is the dimensionless {\em field coupling strength} 
\begin{equation}
 \chi = \frac{m  B}{k_{\mathrm{B}}T}.
\end{equation} 
While $ \omega_f(\theta)$ is always negative, producing the standard, clockwise, Jeffery oscillations  \cite{Jeffery,Bretherton1962} of the spheroidal orientation in the absence of external fields and confining boundaries, $ \omega_{\mathrm{ext}}(\theta)$ can change sign in different $\theta$-quadrants. As such, cooperative/competitive interplays between shear- and field-induced torques, causing enhanced/suppressed oscillation rates, are  both realizable (see the supplementary material). 

Before proceeding further, we note that the magnetic dipole moment, $m$, will be assumed to remain fixed \cite{Waisbord2016,Meng2018,Zhou2017,GolestanianLensing} as the particle  aspect ratio, $\alpha$, is varied in our forthcoming analysis. Given the same magnetic material density, spheroids of differing  aspect ratio will thus have equal volumes \cite{Zhou2017,GolestanianLensing,Perrin,koenig},  ${\mathcal V}(\alpha)={\mathcal V}_0$, with the  radius of the {\em reference sphere} ($\alpha=1$), $R_{\mathrm{eff}}=(3{\mathcal V}_0/4\pi)^{1/3}$, to be used later for  rescaling the units of length. 

We also emphasize that, while, for concreteness, we have focused on magnetic response of active particles in a shear flow, the framework presented here is of a wider applicability, as cross-product angular velocities (reorientation rates) of the form \eqref{eq:w_ext0} can directly be used to model  orientational responses to other types of external fields/stimuli, as we discuss in Section \ref{sec:discussion}.

\section{Continuum Formulation}
\label{sec:Smoluchowski_eq}

Assuming that the swimmer suspension is sufficiently dilute, the steady-state properties of the system can be studied via a customarily used probabilistic approach based on a noninteracting Smoluchowski equation, governing the joint position-orientation probability distribution function (PDF) of swimmers, $\Psi({\mathbf{r}},{\hat{\mathbf p}})$, as \cite{Saintillan2013_ComptesRendus,Saintillan2015,gompper2006soft,doiedwards}
\begin{equation}
{\nabla}_{\mathbf{r}} \cdot \left( \mathbf{v}\Psi \right) + \hat{\mathcal R}_{\hat{\mathbf p}} \cdot \left(\boldsymbol{\omega}\Psi \right) = \nabla_{\mathbf{r}}\cdot{\mathbb D}_T\cdot \nabla_{\mathbf{r}} \Psi + D_R \hat{\mathcal R}_{\hat{\mathbf p}}^2\, \Psi.  
\label{eq:smoluchowski}
\end{equation}
Here, the rotation operator is defined as $\hat{\mathcal R}_{\hat{\mathbf p}}={\hat{\mathbf p}}\times \nabla_{\hat{\mathbf p}}$, where $\nabla_{\hat{\mathbf p}}$ represents unconstrained partial differentiation w.r.t. Cartesian components  of the unit vector ${\hat{\mathbf p}}$ (i.e., by treating the components as independent variables) \cite{gompper2006soft,doiedwards}. Also, $\mathbf{v}$ and $\boldsymbol{\omega}={\omega}(\theta)\hat{\mathbf z}$ are the net deterministic translational and angular  flux velocities, with the definitions   
\begin{align}
\mathbf{v} &= V_s {\hat{\mathbf p}} +u_f(y) \hat{\mathbf x}, \\
{\omega}(\theta) &={\omega}_f(\theta)+{\omega}_{\mathrm{ext}}(\theta),   
\label{eq:w_tot0}
\end{align}
where the signed magnitude of the net angular velocity ${\omega}(\theta)$ can explicitly be obtained using Eqs. \eqref{eq:w_f} and \eqref{eq:w_ext}. 

 Using  the spatial homogeneity of the system along the $x$-axis, the PDF can be expressed only as a function of $y$ and $\theta$ as $\Psi=\Psi(y, \theta)$.  Assuming also that the number of swimmers within the channel are fixed, we have  
\begin{equation}
\int_{-\pi/2}^{3\pi/2}\int_0^{H}{\mathrm{d}} y\,{\mathrm{d}}\theta\, \Psi(y, \theta)={\bar n}, 
\label{eq:norm_actual}
\end{equation}
in which the normalization constant ${\bar n}$ is an experimentally adjustable parameter (e.g.,  the number of swimmers per unit $x-z$ area of the channel in a three-dimensional realization of the system). Since the Smoluchowski equation is linear in $\Psi$, its solution can be rescaled with ${\bar n}$ and the numerical analysis of the problem can be done using the rescaled PDF, $\tilde \Psi(\tilde y, \theta)=  R_{\mathrm{eff}}\Psi(R_{\mathrm{eff}}\tilde y, \theta)/{\bar n}$, where  $\tilde y =  y/R_{\mathrm{eff}}$ is  the appropriately rescaled transverse coordinate. The timescales will be rescaled with the inverse of the rotational Stokes diffusivity of the reference sphere $D_{0R}=D_R(\alpha=1)=k_{\mathrm{B}}T /(8\pi\eta R_{\mathrm{eff}}^3)$, where $\eta$ is the fluid viscosity. The parameter space can thus be spanned by the dimensionless parameters $\{\alpha, \chi, \tilde H, Pe_s, Pe_f\}$, where
\begin{equation}
\tilde H = \frac{H}{R_{\mathrm{eff}}},\,\,\, Pe_s =  \frac{V_s/R_{\mathrm{eff}}}{D_{0R}},\,\,\,\,{\textrm{and}}\,\,\,\, Pe_f= \frac{\dot\gamma}{D_{0R}}, 
\end{equation}
are the rescaled channel width, the swim P\'eclet number, and the flow P\'eclet number, respectively. 

The dimensionless Smoluchowski equation \eqref{eq:smoluchowski} can then be expressed as 
\begin{align}
\label{eq:smoluchowski_rescaled}
&{Pe}_s \sin{\theta} \frac{\partial \tilde{\Psi}}{\partial \tilde y} + \frac{\partial}{\partial \theta} \bigg( \tilde \omega(\theta)\tilde{\Psi} \bigg)
\\
&\quad\,\, = \frac{4}{3} \bigg( \Delta_+(\alpha)- \Delta_-(\alpha)\cos{2\theta}\bigg)\frac{\partial^2 \tilde{\Psi}}{\partial \tilde y^2} + \Delta_R(\alpha) \frac{\partial^2 \tilde{\Psi}}{\partial \theta^2},  \nonumber
\end{align}
where the first term on the l.h.s. includes  particle self-propulsion and the second term incorporates  shear- and field-induced torques, with the rescaled net angular velocity,  $\tilde \omega(\theta)= \omega(\theta)/D_{0R}$, obtained from Eq. \eqref{eq:w_tot0} as  
\begin{equation}
\tilde \omega(\theta) = \frac{Pe_f}{2}\bigg(\beta(\alpha) \cos 2 \theta -1\bigg)- \chi \Delta_R(\alpha) \cos{\theta}. 
\label{eq:w_tot}
\end{equation}
The two terms on the r.h.s. of Eq. \eqref{eq:smoluchowski_rescaled} are due to the translational and rotational diffusions of swimmers,  where  $\Delta_\pm(\alpha) =  \left(\Delta_\parallel(\alpha)\pm \Delta_\perp(\alpha)\right)/2$  and the shape functions $\Delta_{\parallel,\perp}(\alpha)$ and $\Delta_R(\alpha)$, whose explicit forms are given in Appendix \ref{app:diffusivities}, are defined as the ratios of the spheroidal diffusivities and those of the reference sphere, i.e., 
\begin{equation}
\Delta_{\parallel,\perp}(\alpha) =  \frac{D_{\parallel,\perp}(\alpha)}{D_{0T}}\,\,\,\,{\textrm{and}}\,\,\,\,\Delta_R(\alpha) =  \frac{D_R(\alpha)}{D_{0R}}, 
\end{equation}
where $D_{0T}=k_{\mathrm{B}}T /(6\pi\eta R_{\mathrm{eff}})$ is the translational Stokes diffusivity of the reference sphere. 

\begin{figure*}[t!]
\begin{center}
\begin{minipage}[b]{0.55\textwidth}\begin{center}
	\begin{minipage}[t]{0.50\textwidth}\begin{center}
		\includegraphics[width=\textwidth]{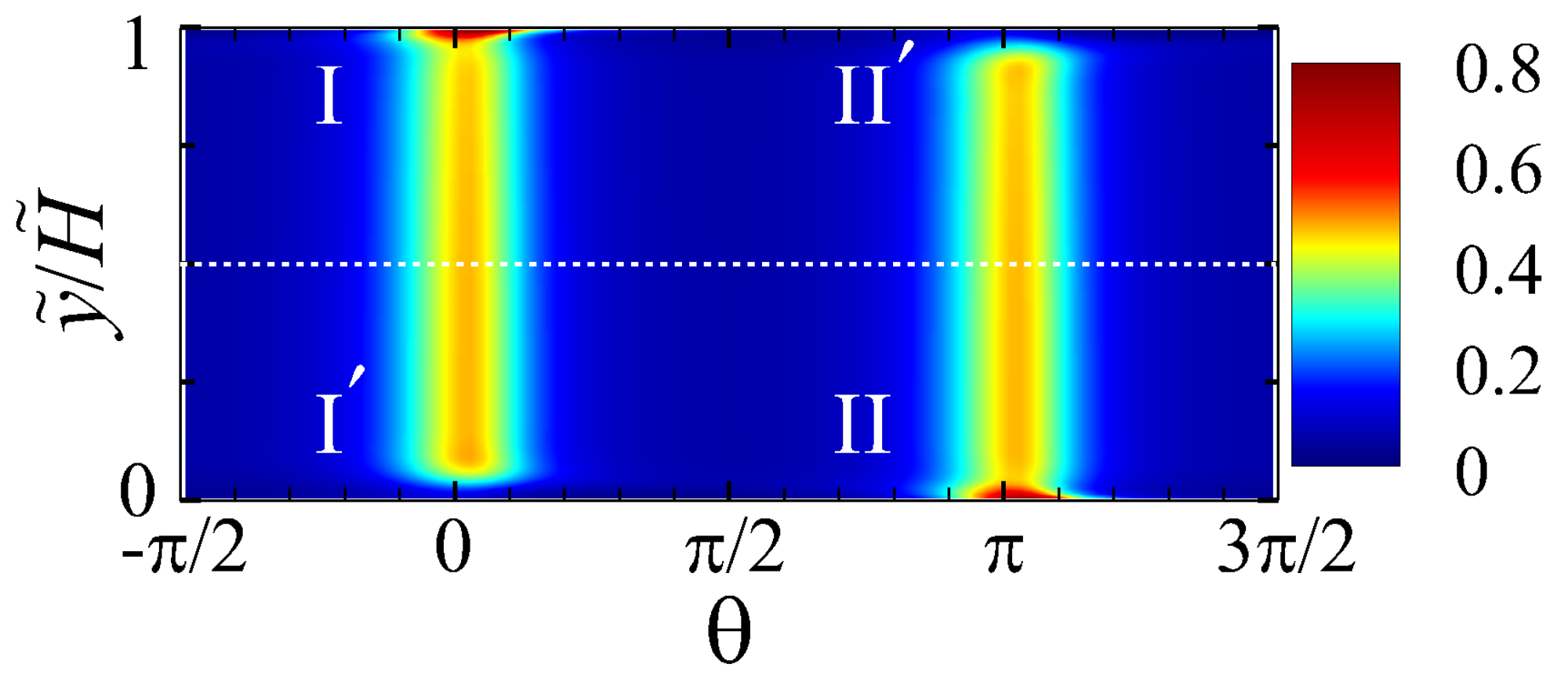} \vskip-1mm (a) $\chi=0$
	\end{center}\end{minipage}\hskip0mm	
	\begin{minipage}[t]{0.50\textwidth}\begin{center}
		\includegraphics[width=\textwidth]{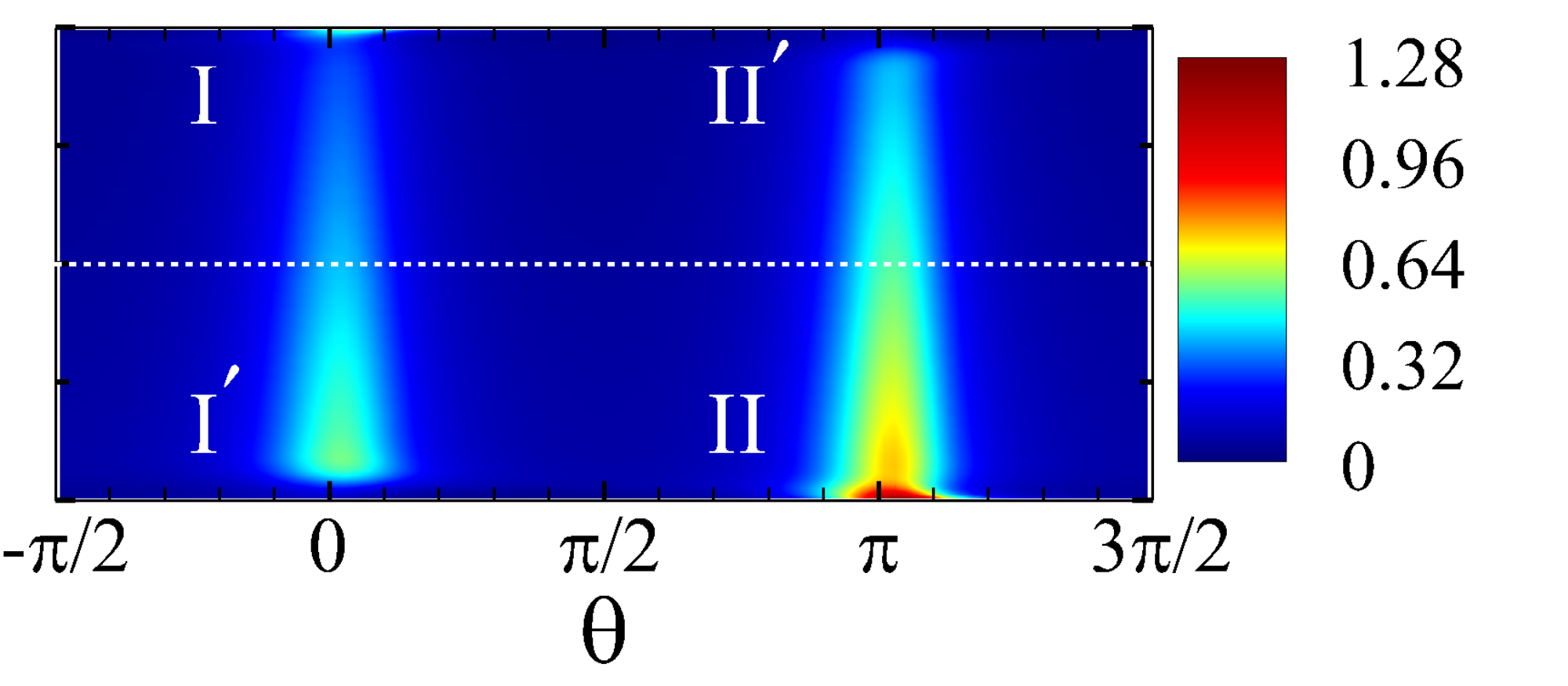} \vskip-1mm (b) $\chi=2$
	\end{center}\end{minipage}\vskip2mm
	\begin{minipage}[t]{0.50\textwidth}\begin{center}
		\includegraphics[width=\textwidth]{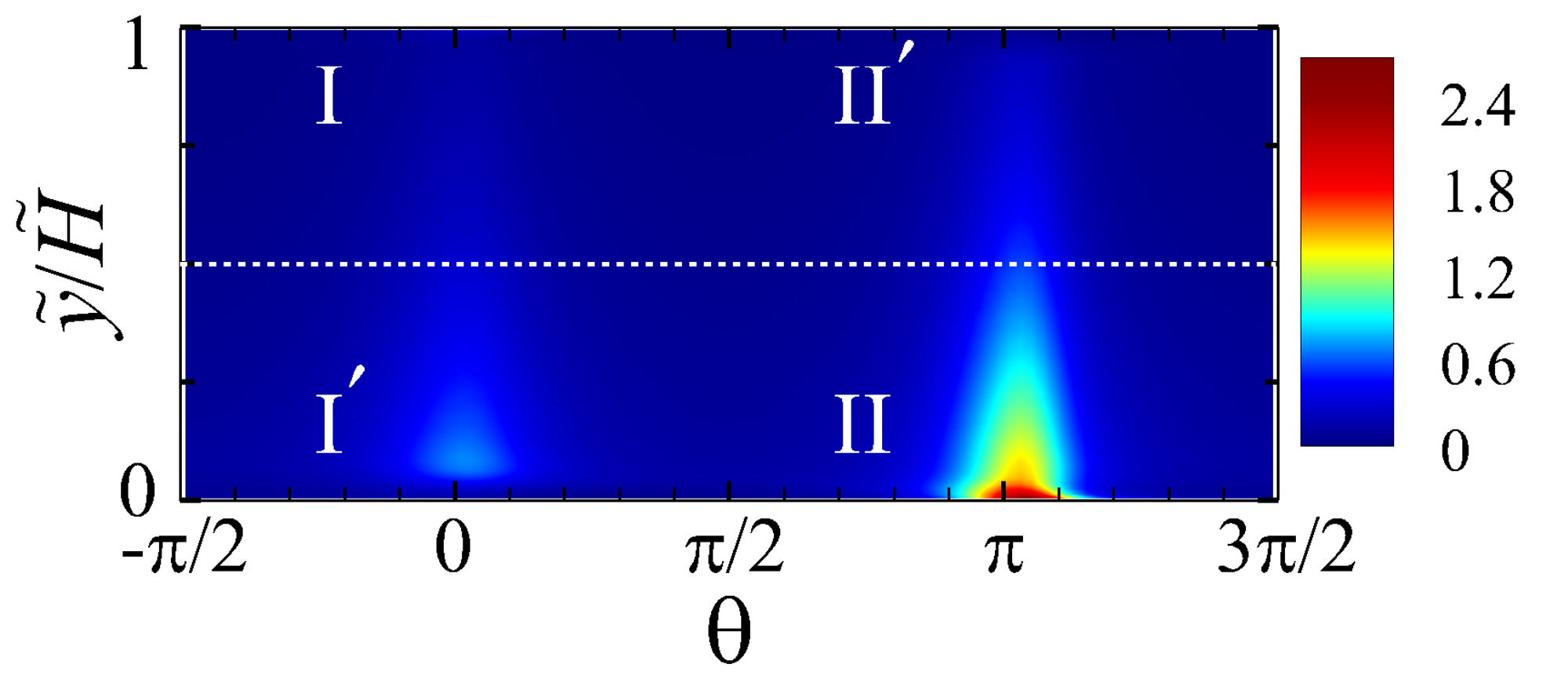} \vskip-1mm (c) $\chi=5$
	\end{center}\end{minipage}\hskip0mm	
	\begin{minipage}[t]{0.50\textwidth}\begin{center}
		\includegraphics[width=\textwidth]{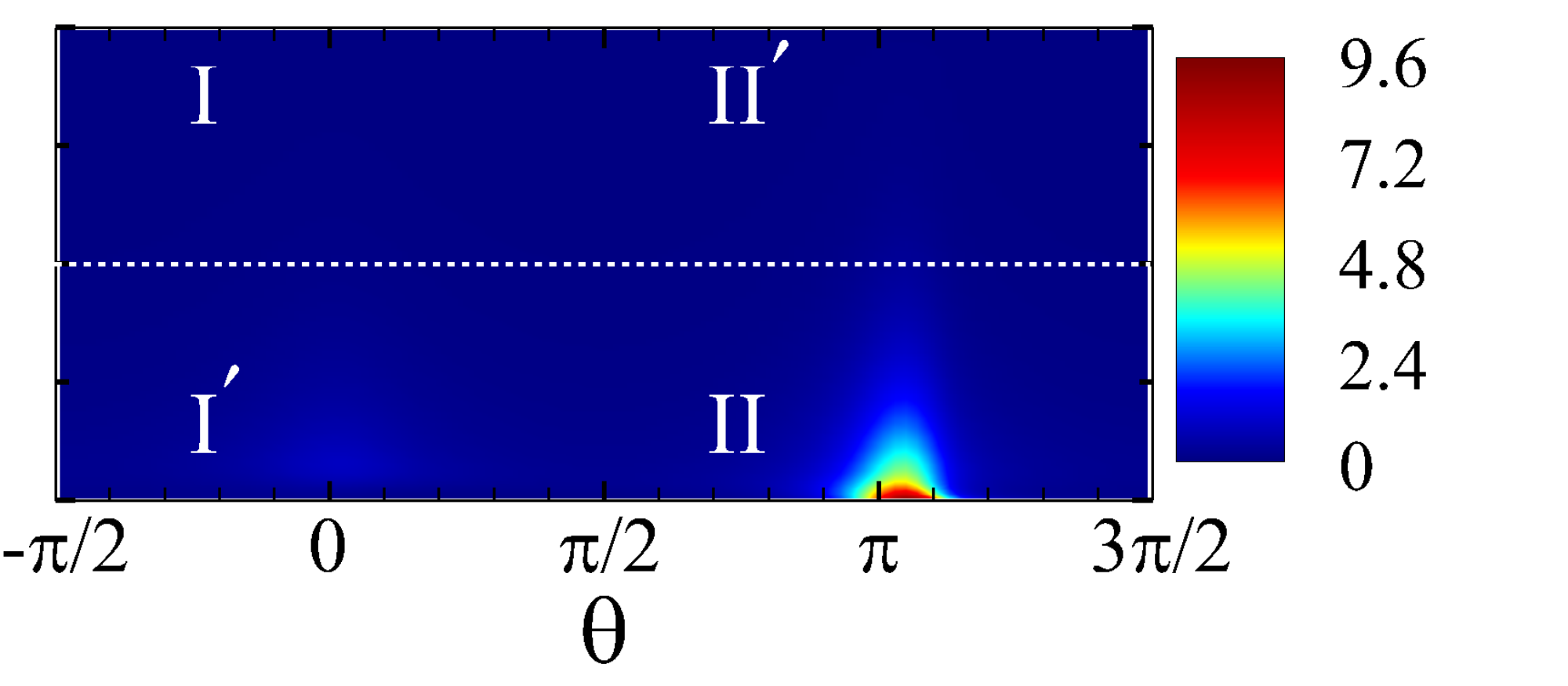} \vskip-1mm (d) $\chi=10$
	\end{center}\end{minipage}
\end{center}\end{minipage}\hskip0mm
\begin{minipage}[t]{0.365\textwidth}\begin{center}
		\includegraphics[width=\textwidth]{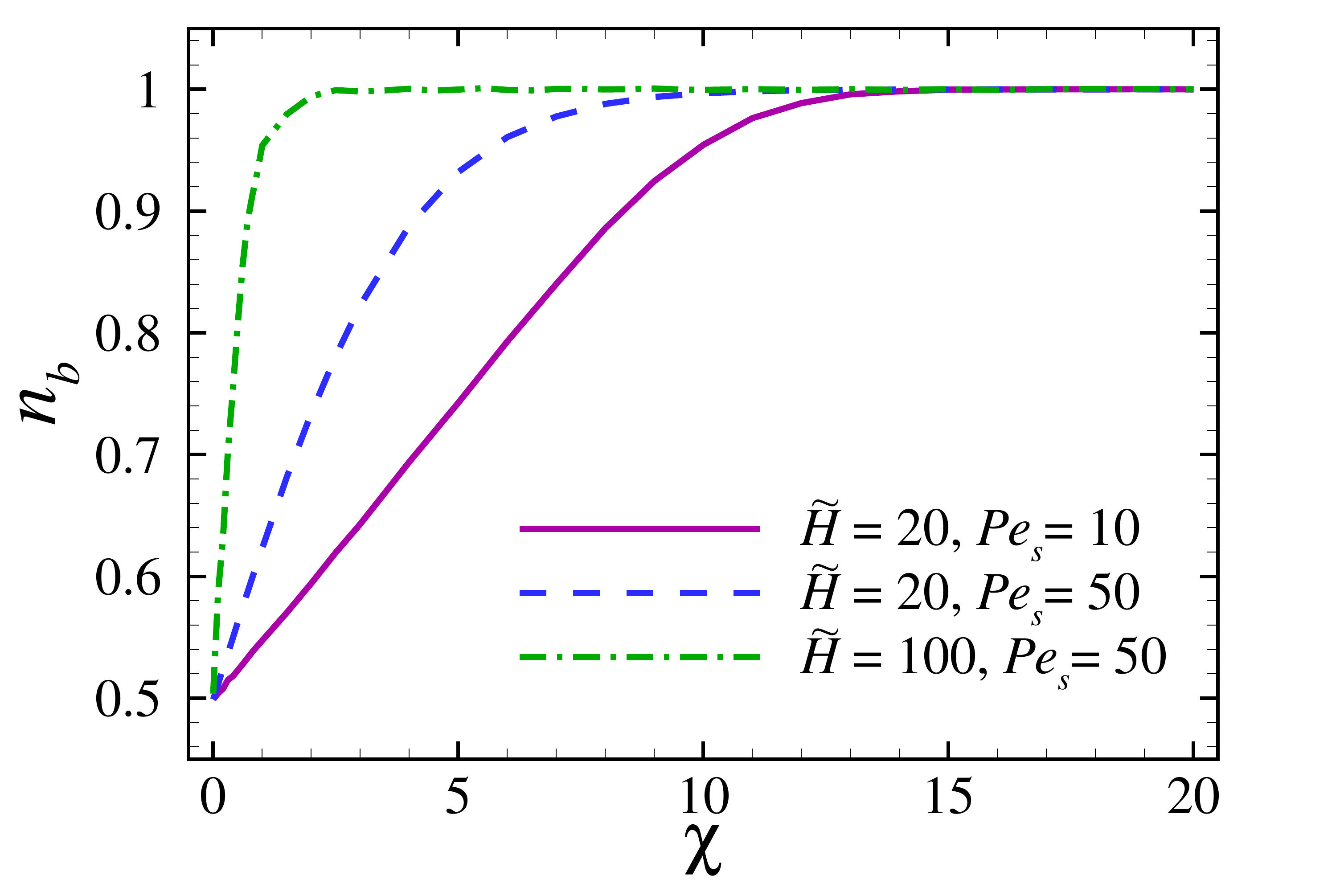}\vskip-1mm (e)
\end{center}\end{minipage}
	\vskip-4mm\caption{Density maps of the joint position-orientation PDF, $\tilde{\Psi}(\tilde y, \theta)$, for {\em prolate} swimmers of aspect ratio $\alpha=3$ within the computational coordinate domain for field coupling strengths (a) $\chi=0$, (b) $\chi=2$, (c) $\chi=5$ and  (d) $\chi=10$ at fixed $Pe_s=10$, $Pe_f=50$ and  $\tilde H=20$. Regions marked ${\textrm{I}}$ and ${\textrm{II}}$  (${\textrm{I}}'$ and ${\textrm{II}}'$) represent the majority (minority) subpopulations in the top and bottom halves of the channel, respectively. (e) Total fraction of swimmers in the bottom half of the channel, Eq. \eqref{eq:n_b_def}, as a function of $\chi$ for fixed $\alpha=3$ and $Pe_f=50$, and other parameter values shown on the graph.
}
\label{fig_contours_full_r3}
\end{center}
\end{figure*}

Equation \eqref{eq:smoluchowski_rescaled} is supplemented by the rescaled normalization constraint, $\int_{-\pi/2}^{3\pi/2}\int_0^{\tilde H}{\mathrm{d}}\tilde y\,{\mathrm{d}}\theta\, \tilde \Psi(\tilde y, \theta)=1$,  Eq. \eqref{eq:norm_actual}, and the {\em no-flux} boundary conditions at $\tilde y=0$ and $\tilde H$ as
\begin{equation}
\label{eq:no_flux}
\left[{Pe}_s \sin{\theta}\,\tilde{\Psi}\! -\! \frac{4}{3}\! \bigg( \!\Delta_+(\alpha)\!-\! \Delta_-(\alpha)\cos{2\theta}\!\bigg)\!\frac{\partial \tilde{\Psi}}{\partial \tilde y}\right]_{\tilde y=0,\tilde H}\!\!\!\!\!\!\!\!\!\!=0. 
\end{equation}
The no-flux  boundaries are taken at the closest-approach distance of spheroids from the actual no-slip  boundaries of the channel  (Fig. \ref{fig:schematic}); a geometric designation that can account for a finite exclusion layer at the hard boundaries. For simplicity and since, at sufficiently large shear rates that are of primary interest here, the major body axis of swimmers is typically aligned with the flow, we set the width of the exclusion layers equal to the semi-minor body-axis length, $b(\alpha)$, of the spheroids; or, in rescaled units $\tilde b(\alpha)=\alpha^{-1/3}$ (prolate) and $\alpha^{2/3}$ (oblate). The no-slip channel boundaries are thus implicitly assumed to be at $\tilde y=-\tilde b(\alpha)$ and $\tilde H+\tilde b(\alpha)$. These will be of relevance only in the calculation of the current density in the laboratory frame in Section \ref{sec:flux}. Having this in mind, we shall hereafter refer to the no-flux boundaries $\tilde y=0$ and $\tilde H$ as the {\em bottom} and {\em top `walls'} of the channel, respectively.  

Other choices of boundary conditions are possible  \cite{Chilukuri,li2011accumulation,li2011accumulation2,Nash2010,Ezhilan,AA}  (e.g., with nearly hard  swimmer-wall steric potentials \cite{Ezhilan,AA}), but their effects are expected to be relatively small, especially at elevated shear rates. Our later analysis of the field-induced migration of  sheared swimmers and its pertinent regimes of behavior is however done mainly by devising globally defined criteria and quantities (e.g., the integrated number, or fraction, of swimmers in entire channel halves), minimizing possible quantitative artifacts that may arise from our specific modeling of the near-wall regions. 

Our results are obtained by  solving Eq. \eqref{eq:smoluchowski_rescaled} numerically subject to the specified normalization and boundary conditions over the $\tilde y$ and $\theta$ domain, conveniently chosen as $\tilde y\in [0, \tilde H]$ and $\theta\in [-\pi/2, 3\pi/2)$ for prolate swimmers and $\theta\in [0, 2\pi)$ for oblate swimmers, with periodic boundary conditions imposed over $\theta$. Unless otherwise noted, the relative error in the reported data is $< 1\%$. The dimensionless parameters will be varied over a wide range of values consistent with realistic examples of swimmers; see Appendix \ref{app:parameters} for a few exemplary cases. The typical $\chi$-dependent behavior of prolate and oblate swimmers will however be detailed mainly by taking $\alpha=3$ and $1/3$, and using  the following two sets of parameters: $Pe_s=10\, (50)$, $Pe_f=50\, (100)$ and  $\tilde H=20\, (100)$.

\section{Magnetic prolate swimmers}
\label{sec:prolates}

\subsection{Shear-induced bimodality with no external field}
\label{subsec:no_field}

In the absence of an external field, the joint position-orientation PDF, $\tilde \Psi(\tilde y, \theta)$, of swimmers within the channel is  determined mainly by the interplay between swimmer self-propulsion  and the shear-induced torque on the swimmers, which is a constant across the channel width for the Couette flow (some of the features discussed below will be different in a Poiseuille flow that has a $y$-dependent shear rate; see, e.g., Refs. \cite{Rusconi2014,Ezhilan}).

The self-propulsion drives the swimmers to accumulate at the channel walls at their most-probable, normal-to-wall,  orientations with $\theta=\pm\pi/2$ for the top/bottom walls, respectively. It thus triggers strong midchannel depletion and an uneven spatial distribution of swimmers in $y$-direction, with localized peaks at the walls (not shown; see however Ref. \cite{Nili}). The shear-induced torque, on the other hand,  tends to drive the swimmers through Jeffery orbits along the flow \cite{Jeffery,Bretherton1962}, causing swimmer shear-trapping and an increasingly even spatial distribution in $y$-direction at sufficiently large shear rates. In this case, the PDF however becomes strongly polarized along the orientation coordinate, $\theta$, portraying two orientationally distinct, macroscopic populations for prolate swimmers. These populations appear as high-probability columnar regions over the computational coordinate domain shown in Fig. \ref{fig_contours_full_r3}a (the behavior turns out to be quite different for oblate swimmers as will be discussed in Section \ref{sec:oblates}).  This {\em bimodal} PDF incorporates a {\em majority} subpopulation of {\em upstream-swimming} particles ($\pi/2\leq \theta<3\pi/2$) together with a {\em minority} subpopulation of {\em downstream-swimming} particles ($-\pi/2\leq \theta<\pi/2$) in the bottom half of the channel, with the situation reversed in the top half, when the flow P\'eclet number, $Pe_f$, is large enough. The majority subpopulations can be discerned in Fig. \ref{fig_contours_full_r3}a as the high-probability, red-colored, spots centered at $\theta=0$ ($\pi$) on and near the top (bottom) wall. The shear-trapped extensions of the near-wall majorities (appearing as dark-yellow columns extending out from their base walls) form the majority subpopulations across the proximal half of the channel (regions marked by ${\textrm{I}}$ and ${\textrm{II}}$) and the minority subpopulations within the distal half (regions ${\textrm{I}}'$ and ${\textrm{II}}'$). This kind of shear-induced bimodality has previously been characterized as a {\em population splitting phenomenon} in the idealized case of needlelike particles in narrow channels \cite{Nili}. 

\subsection{Population splitting and field-induced migration}
\label{subsec:migration}

On applying a downward external field, the angular velocity component \eqref{eq:w_ext} triggers partial  migration of swimmers to the bottom half of the channel. This behavior can be seen by inspecting the PDFs in Figs. \ref{fig_contours_full_r3}a-d and it is quantifiable using the bottom-half swimmer fraction 
\begin{equation}
n_b= \int_{0}^{\tilde H/2}\!{\mathrm{d}}\tilde y\,\tilde{\phi}(\tilde y),   
\label{eq:n_b_def}
\end{equation} 
 where $\tilde{\phi}(\tilde y)= \int_{-\pi/2}^{3\pi/2}{\mathrm{d}}\theta\, \tilde \Psi(\tilde y, \theta)$ is the rescaled swimmer density,  integrating to one across the channel width.

Figure \ref{fig_contours_full_r3}e shows that $n_b$  increases linearly with the field coupling strength, $\chi$,  before it saturates at $n_b= 1$ due to full swimmer migration to the bottom half of the channel. 
The figure thus indicates an extended  {\em regime of linear response} to the external field (corresponding to the PDFs in Figs. \ref{fig_contours_full_r3}b and c). The linear increase in $n_b$ appears to occur at a faster rate as $Pe_s$ and/or $\tilde H$ are increased. This behavior can further be quantified  using a {\em linear migration response factor}, revealing intricate dependencies on system parameters that we detail in Appendix \ref{sec:weak_field}.      

The density maps of Figs. \ref{fig_contours_full_r3}b-d further reveal a subtle interplay between the minority and the majority subpopulations underlying the cross-stream  migration of swimmers, one that is not reflected by the orientationally averaged quantities such as $n_b$: For a moderately large field coupling (e.g., $\chi=2$), the probability density of both the majority and the minority subpopulations  in the top half of the channel is suppressed and the lateral spread of these subpopulations along the $\theta$-axis shrinks 
(Fig. \ref{fig_contours_full_r3}b, regions ${\textrm{I}}$ and ${\textrm{II}}'$). 
However, these effects occur more strongly for the {\em majority} subpopulation (${\textrm{I}}$). At higher field couplings  (e.g., $\chi=5$), the majority subpopulation of downstream swimming particles within the top half of the channel is almost completely diminished (Fig. \ref{fig_contours_full_r3}c, region ${\textrm{I}}$), even before the top-half minority subpopulation (${\textrm{II}}'$) is removed. It is interesting to note that some part of the diminishing top-half majority (${\textrm{I}}$) {\em persists} as a minority subpopulation in the {\em bottom} half of the channel (light-blue spot just above $\theta=0$ in Fig. \ref{fig_contours_full_r3}c, marked ${\textrm{I}}'$). Thus, in going from Fig. \ref{fig_contours_full_r3}a to d, the number of {\em downstream} swimming particles continuously decreases and eventually vanishes, as $\chi$ is increased, enforcing the swimmers to form a single population (${\textrm{II}}$) of only  {\em upstream} particles on the bottom wall (Fig. \ref{fig_contours_full_r3}d). 

The field-induced redistribution and cross-stream migration of prolate swimmers thus involve a well-defined sequence of eliminations in  the orientational subpopulations identified in the two channel halves. 
Such a migration pattern can be quantified by means of the fractions of downstream and  upstream swimmers in the top (bottom) half of the channel, $n_t^+$ and $n_t^-$ ($n_b^+$ and $n_b^-$),   
\begin{equation}
n_t^{\pm} = \int_{\tilde H/2}^{\tilde H}\!{\mathrm{d}}\tilde y\,\tilde{\phi}_{\pm}(\tilde y),\quad  n_b^{\pm} = \int_{0}^{\tilde H/2}\!{\mathrm{d}}\tilde y\,\tilde{\phi}_{\pm}(\tilde y), 
\label{eq:n_t_b_pm}
\end{equation}
where the number density of downstream, $\tilde{\phi}_{+}(\tilde y)$, and upstream  swimmers, $\tilde{\phi}_{-}(\tilde y)=\tilde{\phi}(\tilde y)-\tilde{\phi}_{+}(\tilde y)$,  are defined by integrating over the relevant angular intervals as 
\begin{equation}
\tilde{\phi}_{+}(\tilde y) \!= \!\!\int_{-\pi/2}^{\pi/2}\!{\mathrm{d}}\theta\, \tilde \Psi(\tilde y, \theta),\,\,\, \tilde{\phi}_{-}(\tilde y) \!= \!\!\int_{\pi/2}^{3\pi/2}\!{\mathrm{d}}\theta\, \tilde \Psi(\tilde y, \theta). 
\label{eq:phi_t_b_pm}
\end{equation}
In the shear-dominant regime considered here, the primary contribution to these (orientationally averaged) fractions comes from the respective densely populated regions, ${\textrm{I}}$, ${\textrm{I}}'$, ${\textrm{II}}$ and/or ${\textrm{II}}'$, where the most-probable swim orientations are approximately $\theta= 0$ and $\pi$.

\subsubsection{Reverse bimodality (top half)}
\label{subsubsec:rev_bm}

\begin{figure}[t!]
\sidesubfloat[]{%
\hskip-1mm%
\includegraphics[width=0.75\textwidth]{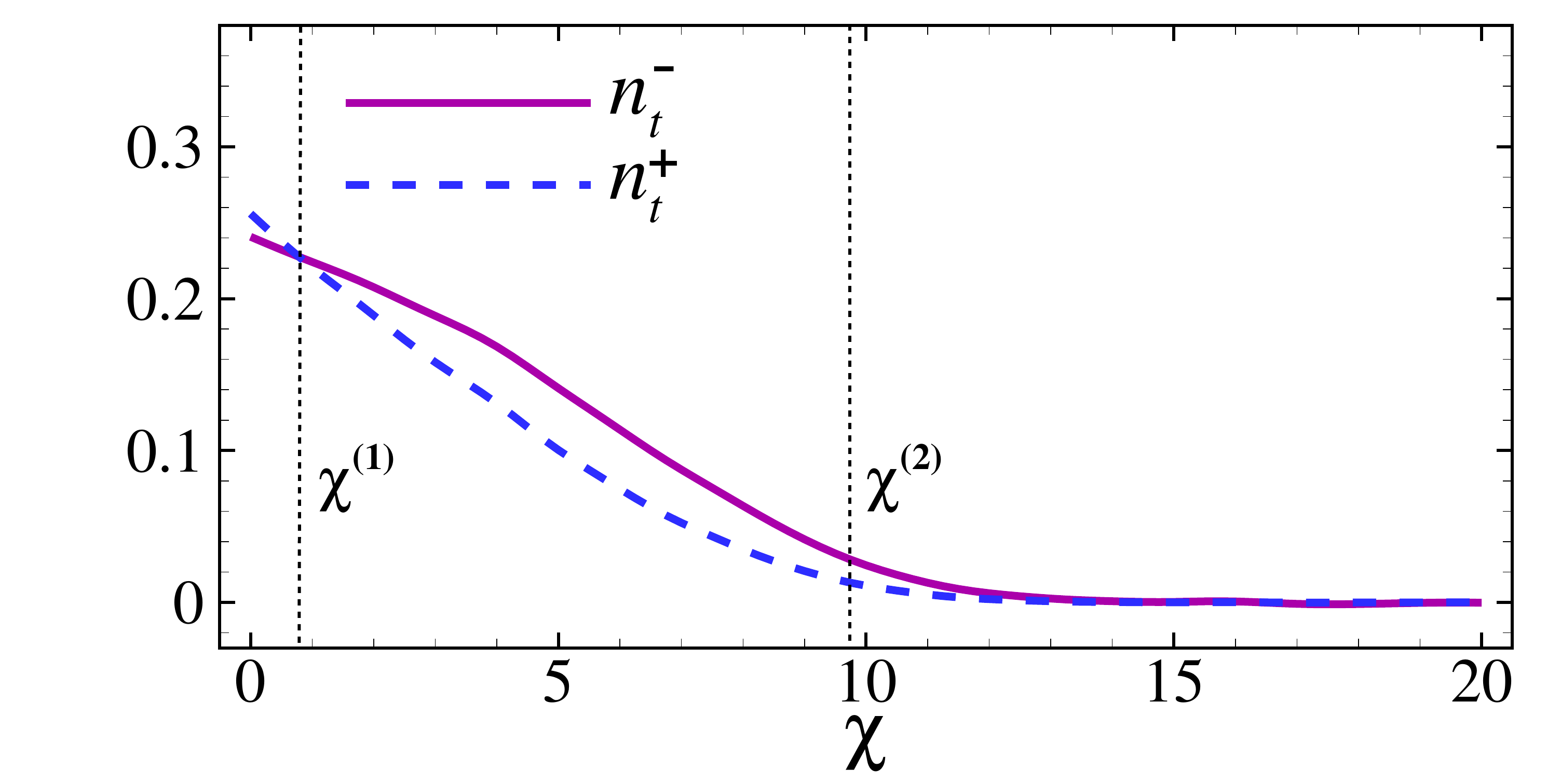}%
}\vskip1mm%
\hskip-2mm%
\sidesubfloat[]{%
\hskip8mm%
\includegraphics[width=0.63\textwidth]{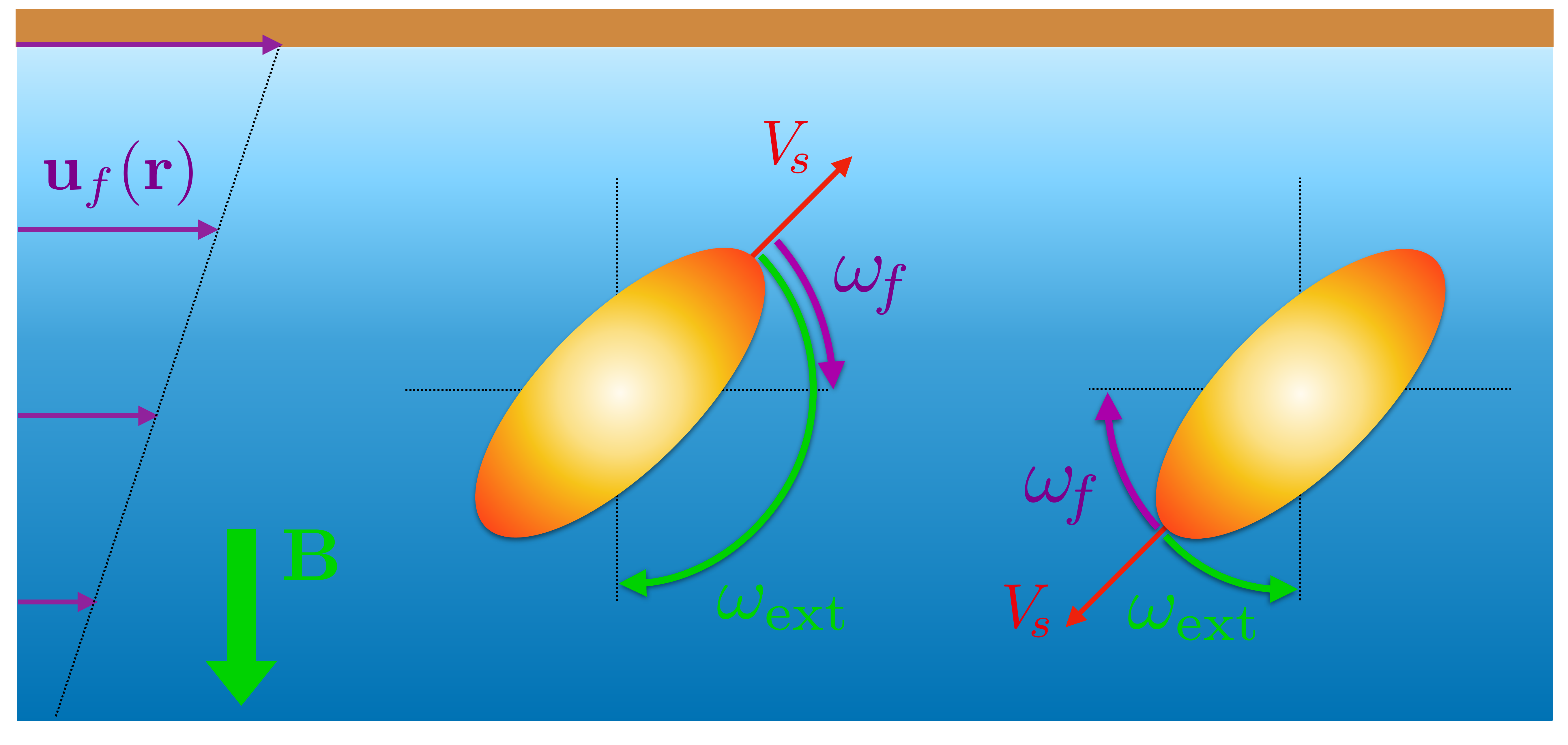}%
}\vskip3mm%
\sidesubfloat[]{%
\hskip-1mm%
\includegraphics[width=0.75\textwidth]{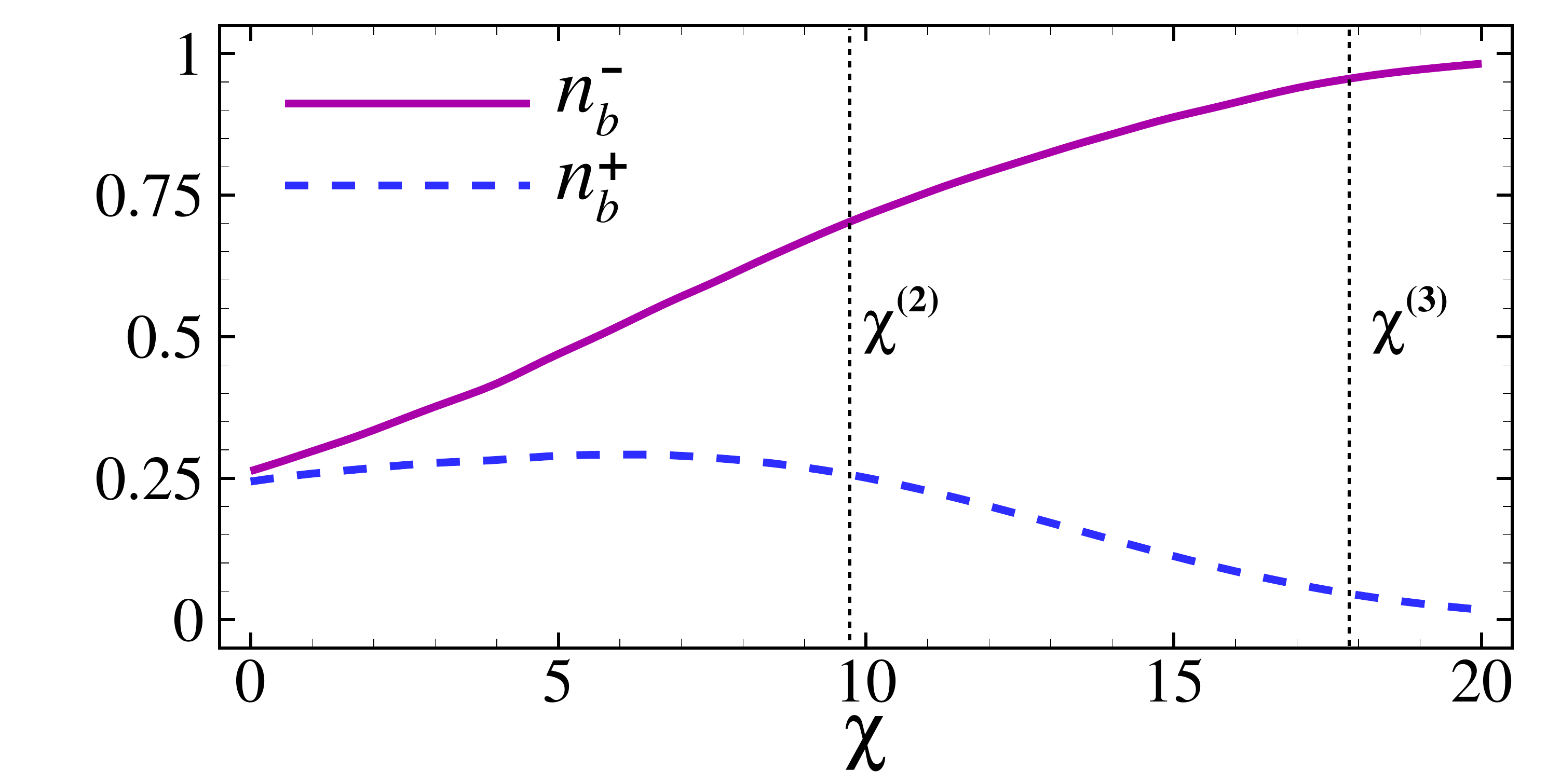}%
}
\vskip-2mm\caption{(a) Downstream ($n_t^+$) and upstream ($n_t^-$) fractions of prolate swimmers in the top half of the channel (Eq. \eqref{eq:n_t_b_pm}) as functions of the field coupling strength, $\chi$, for fixed $Pe_s=10$, $Pe_f=50$ and  $\tilde H=20$. 
(b) Cooperative and competitive nature of the shear- and field-induced angular velocities for the top-half majority (left) and minority (right) subpopulations, respectively,  results in the phenomenon of reverse bimodality. (c) Same as (a) but here we show the bottom-half downstream ($n_b^+$) and upstream ($n_b^-$) fractions of swimmers. See the text for the definitions of $\chi^{(1)}$, $\chi^{(2)}$ and $\chi^{(3)}$ shown on the graph. 
}
\label{fig_r3_submigration}
\end{figure}

The field-induced variations of the up-/downstream swimmer fractions in the {\em top half} of the channel, $n_t^{\pm}$, are shown in Fig. \ref{fig_r3_submigration}a, revealing two different regimes of orientational bimodality: For sufficiently small field strengths, $\chi<\chi^{(1)}$, where $\chi^{(1)}\simeq 0.8$ for the parameters in the figure (indicated by a dotted vertical line), the fraction of downstream swimmers in the top half is larger than that of the upstream swimmers, $n_t^+>n_t^-$. This is in accordance  with the {\em standard bimodality} (sBm) scenario discussed in the field-free case in Section \ref{subsec:no_field}. The sBm regime extends (shrinks), i.e., $\chi^{(1)}$ becomes larger (smaller), as $Pe_f$ is decreased (increased). For larger field strengths, $\chi>\chi^{(1)}$, we find an extended regime of {\em reverse bimodality} (rBm), where the upstream swimmers in the top half outnumber the downstream ones, $n_t^+<n_t^-$. 

The rBm regime crosses over to a regime of {\em full migration} (FM), with swimmers almost fully migrated  to the bottom half of the channel. We conventionally define this to occur, when $n_t = n_t^++n_t^-< 0.05$, or when the field coupling is stronger than another crossover value, $\chi>\chi^{(2)}$, where $\chi^{(2)}\simeq  9.8$  (indicated by a second dotted vertical line) in Fig. \ref{fig_r3_submigration}a. In the FM regime, the notion of bimodality becomes obsolete. The preceding criterion also means $n_b = 1- n_t > 0.95$; i.e., $\chi^{(2)}$ represents the value at which the linearly increasing $n_b$ curves shown in Fig. \ref{fig_contours_full_r3}e nearly level off and converge to $n_b=1$.  The density maps of  Fig. \ref{fig_contours_full_r3} exemplify the regimes of standard bimodality (panel a), reverse bimodality (panels b and c) and full migration (panel d).

The origin of the reverse bimodality can be understood by examining the interplay between the shear- ($ \omega_f$) and field-induced ($ \omega_{\mathrm{ext}}$) angular velocities, or torques, on individual swimmers in the top half of the channel. The typical swimmer orientations are shown in Fig. \ref{fig_r3_submigration}b, where the left and right configurations belong to regions ${\textrm{I}}$ (downstream majority) and  ${\textrm{II}}'$  (upstream minority), respectively, with the self-propulsion-induced tilt toward the wall also depicted in each case. The purple and green arrows indicate $ \omega_f$ and $ \omega_{\mathrm{ext}}$, respectively, with the arrow endpoints indicating the orientation angles, where the corresponding   torque magnitudes are zero or minimized. As seen, the torques are {\em cooperative} and {\em competitive} in the case of majority and minority subpopulations, respectively, clearly indicating why the majority subpopulation near the top wall (${\textrm{I}}$) is eliminated more rapidly on increasing the field strength.

\subsubsection{Minority persistence and beyond (bottom half)}
\label{subsubsec:min_persis}

The quantitative behaviors of the up- and downstream fractions of swimmers in the {\em bottom half} of the channel, i.e., $n_b^-$ (solid curve) and $n_b^+$ (dashed curve), respectively, reveal further features (see Fig. \ref{fig_r3_submigration}c): As $\chi$ is increased up to $\chi^{(2)}$ and concurrently the top-half majority and minority  subpopulations  (${\textrm{I}}$ and ${\textrm{II}}'$)  transfer to the bottom half, $n_b^+$  remains nearly a constant or, quite weakly increases slightly above $n_b^+\simeq 0.25$. This establishes the previously noted observation in Fig. \ref{fig_contours_full_r3}c that the bottom-half downstream minority of swimmers (${\textrm{I}}'$) at typical orientation $\theta\simeq 0$ persists despite the maximally exerted clockwise torque due to the external field at this orientation. During this process, which occurs concurrently with the reverse bimodality ($\chi<\chi^{(2)}$),  the bottom-half upstream majority, $n_b^-$,  continuously increases (Fig. \ref{fig_r3_submigration}c). Thus, on migration to the bottom half of the channel, the swimmers coming from the top half predominantly convert to the upstream majority (${\textrm{II}}$) in the bottom half.
At higher couplings, $\chi>\chi^{(2)}$, $n_b^+$  decreases as the system goes through the full migration (FM) regime. In this case, not only the swimmers mostly migrate to the bottom half, but they are also accumulated mainly within a thin layer of thickness $R_{\mathrm{eff}}$ next to the bottom wall.

Our results in Fig. \ref{fig_r3_submigration}c enables another criterion to identify a subsequent regime of behavior: When $n_b^+<0.05$ (or $n_b^->0.95$), which occurs at a crossover value of $\chi^{(3)}$, one ends up with a {\em unimodal regime}  (Um) of upstream swimmers in the bottom half (in the figure, $\chi^{(3)}\simeq 18$). At the onset of the Um regime, the most-probable swim orientation angle is close to $\pi$ (with PDF density maps resembling Fig. \ref{fig_contours_full_r3}d). As $\chi$ is increased to infinity, the swim angle increases gradually within the third $\theta$-quadrant toward $3\pi/2$ (or $-\pi/2$), which is the anticipated limiting (normal-to-wall) swim orientation. 

\begin{figure}[t!]
\begin{center}
\includegraphics[width=0.8\textwidth]{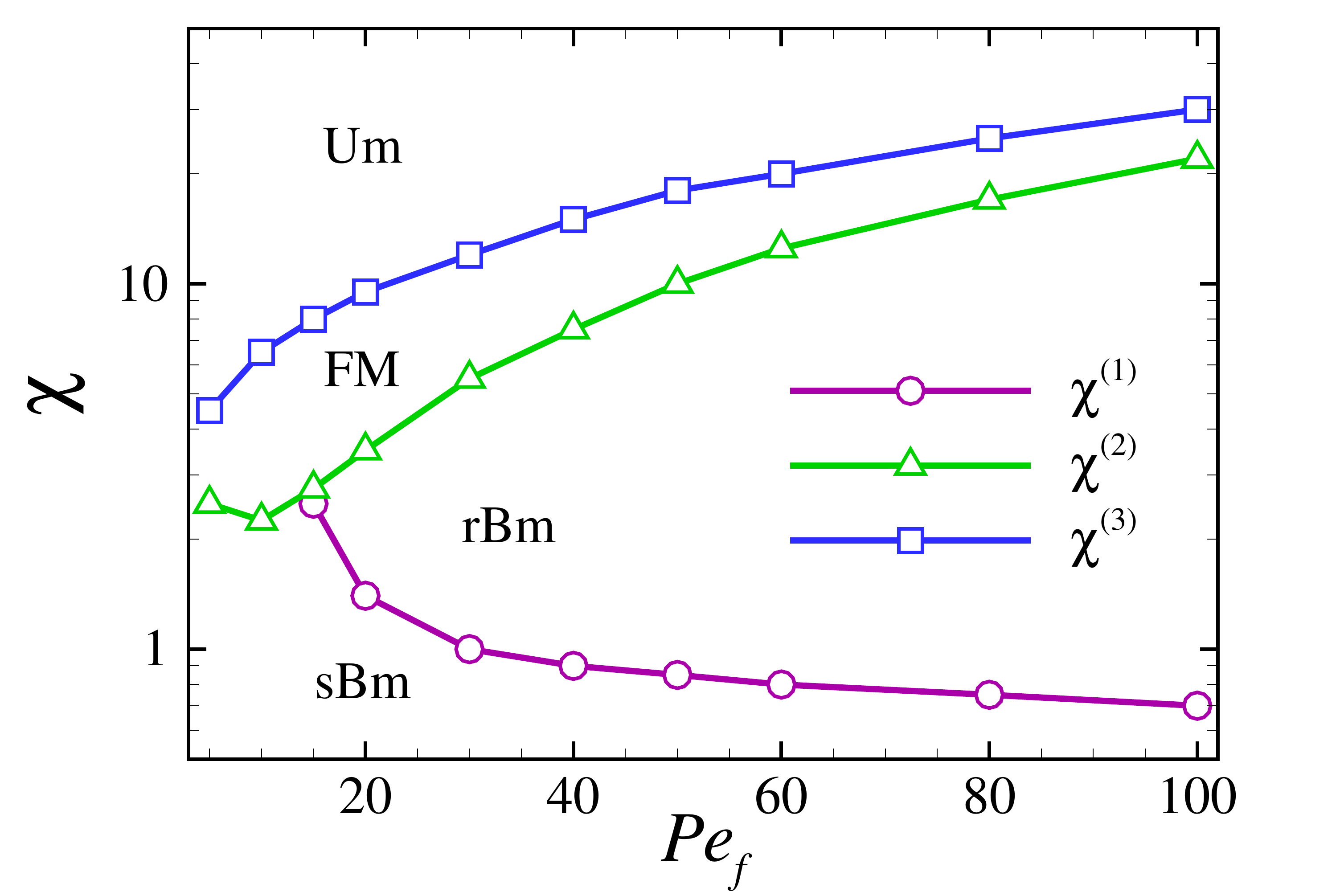}
\vskip-4mm
\caption{Different regimes of behavior for the field-induced migration of prolate swimmers across the channel width in terms of the field coupling strength, $\chi$, and the flow P\'eclet number, $Pe_f$, at fixed $\alpha=3$, $Pe_s=10$, $\tilde H=20$. Symbols show computed data for the crossover couplings $\chi^{(1)}$, $\chi^{(2)}$ and $\chi^{(3)}$, and solid curves are drawn to guide the eye. See the text for definitions of the different regimes  shown on the graph. 
}
\label{fig_r3_submigration_PD}
\end{center}
\end{figure}

\subsubsection{Migration pattern of prolate swimmers: A summary}
\label{subsubsec:summary_prolates}

The cross-stream migration of magnetic prolate swimmers to the bottom half of the channel at sufficiently large $Pe_f$ and upon increasing $\chi$ can thus be considered to set in effectively only above $\chi^{(1)}$, followed sequentially by specific changes in the top/bottom minority and majority subpopulations, expressible symbolically as 
\begin{align}
\left\{\begin{array}{ll}
\triangleright\,\,\textrm{Standard bimodality (sBm):}&\\
\,\,\,\,\,\{{\textrm{I}}>{\textrm{II}}'\}\,\,\&\,\,\{{\textrm{I}}'<{\textrm{II}}\}&\hskip0mm\quad\chi<\chi^{(1)},\\ 
\\
\triangleright\,\,\textrm{Reverse bimodality (rBm)/}&\\
\,\,\,\,\,\textrm{\& Minority persistence:}&\\
\,\,\,\,\,\{{\textrm{I}}<{\textrm{II}}'\}\,\,\&\,\,\{{\textrm{I}}'<{\textrm{II}}\}
&\hskip-0mm\quad\chi^{(1)}<\chi<\chi^{(2)},\\
\\
\triangleright\,\,\textrm{Full migration (FM):}&\\
\,\,\,\,\,\{{\textrm{I}}= {\textrm{II}}'\simeq \emptyset\}\,\,\&\,\,\{{\textrm{I}}'< {\textrm{II}}\}&\hskip-0mm\quad\chi^{(2)}<\chi<\chi^{(3)},\\
\\
\triangleright\,\,\textrm{Unimodality (Um):}&\\
\,\,\,\,\,\{{\textrm{I}}={\textrm{I}}'={\textrm{II}}'\simeq  \emptyset\}\,\,\&\,\,\{{\textrm{II}}\}&\hskip-0mm\quad\chi>\chi^{(3)}.
\end{array}\right.
\end{align}

The crossover values $\{\chi^{(1)}, \chi^{(2)}, \chi^{(3)}\}$ of the field coupling strength  can generally vary depending on the  system parameters. Figure \ref{fig_r3_submigration_PD} gives a broader view of the  migration regimes of prolate swimmers across the whole range of $Pe_f$ at fixed $\alpha=3$, $Pe_s=10$ and $\tilde H=20$. As seen, at sufficiently low $Pe_f\lesssim 15$, the FM regime can be reached continuously from the sBm regime upon increasing $\chi$ and without passing through the rBm regime. Also, the rBm regime expands, while the other regimes shown in the figure shrink, as $Pe_f$ is increased. 

The regime of linear response (see Fig. \ref{fig_contours_full_r3}e and Appendix \ref{sec:weak_field}) falls within the sBm and rBm regimes. The crossover from the linear to the nonlinear response occurs continuously for prolate swimmers, showing no particular fingerprints in the quantities analyzed in the preceding sections. The situation turns out to be very different for oblate swimmers as we shall discuss in Section \ref{sec:oblates}.

\begin{figure*}[t!]
\begin{center}
\begin{minipage}[b]{0.55\textwidth}\begin{center}
	\begin{minipage}[t]{0.50\textwidth}\begin{center}
		\includegraphics[width=\textwidth]{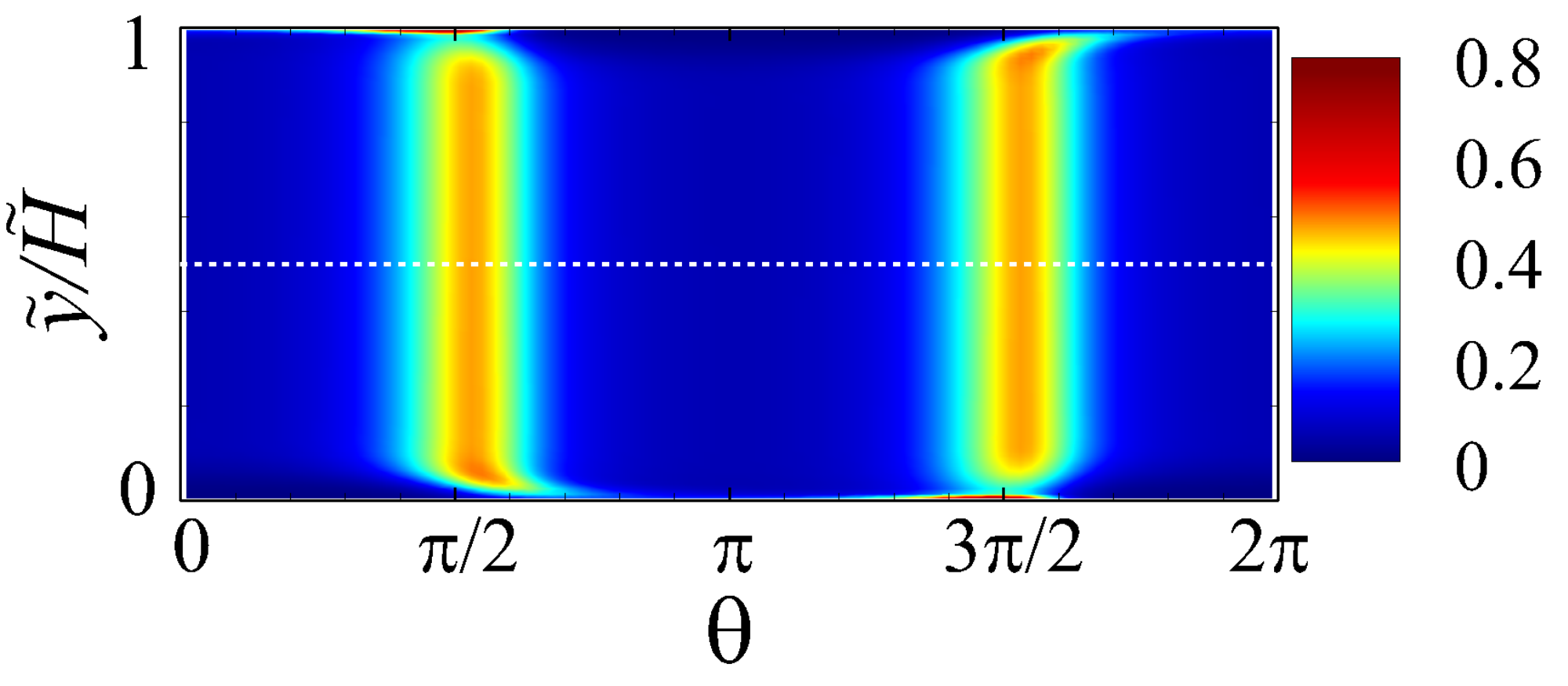} \vskip-2mm (a) $\chi=0$
	\end{center}\end{minipage}\hskip0mm	
	\begin{minipage}[t]{0.50\textwidth}\begin{center}
		\includegraphics[width=\textwidth]{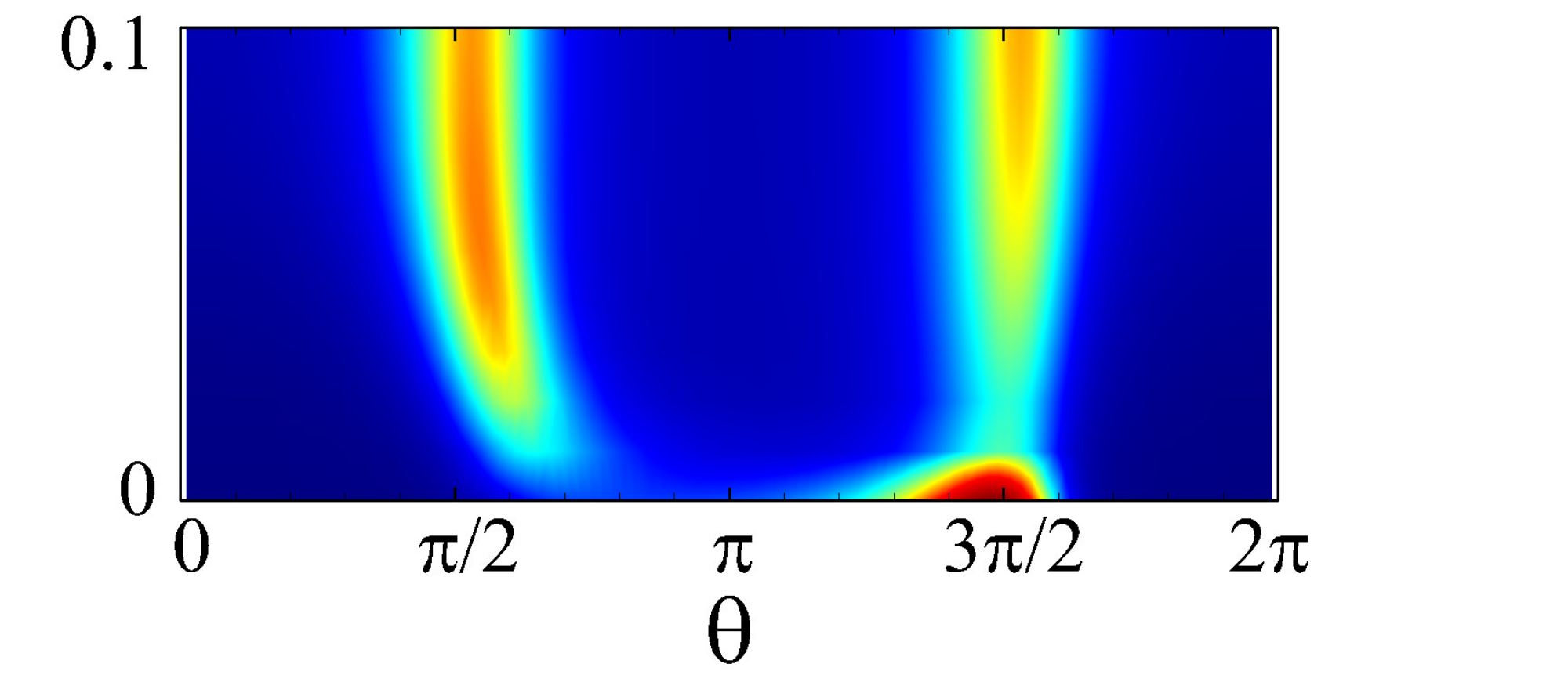} \vskip-2mm (b) $\chi=0$
	\end{center}\end{minipage}\vskip2mm
	\begin{minipage}[t]{0.50\textwidth}\begin{center}
		\includegraphics[width=\textwidth]{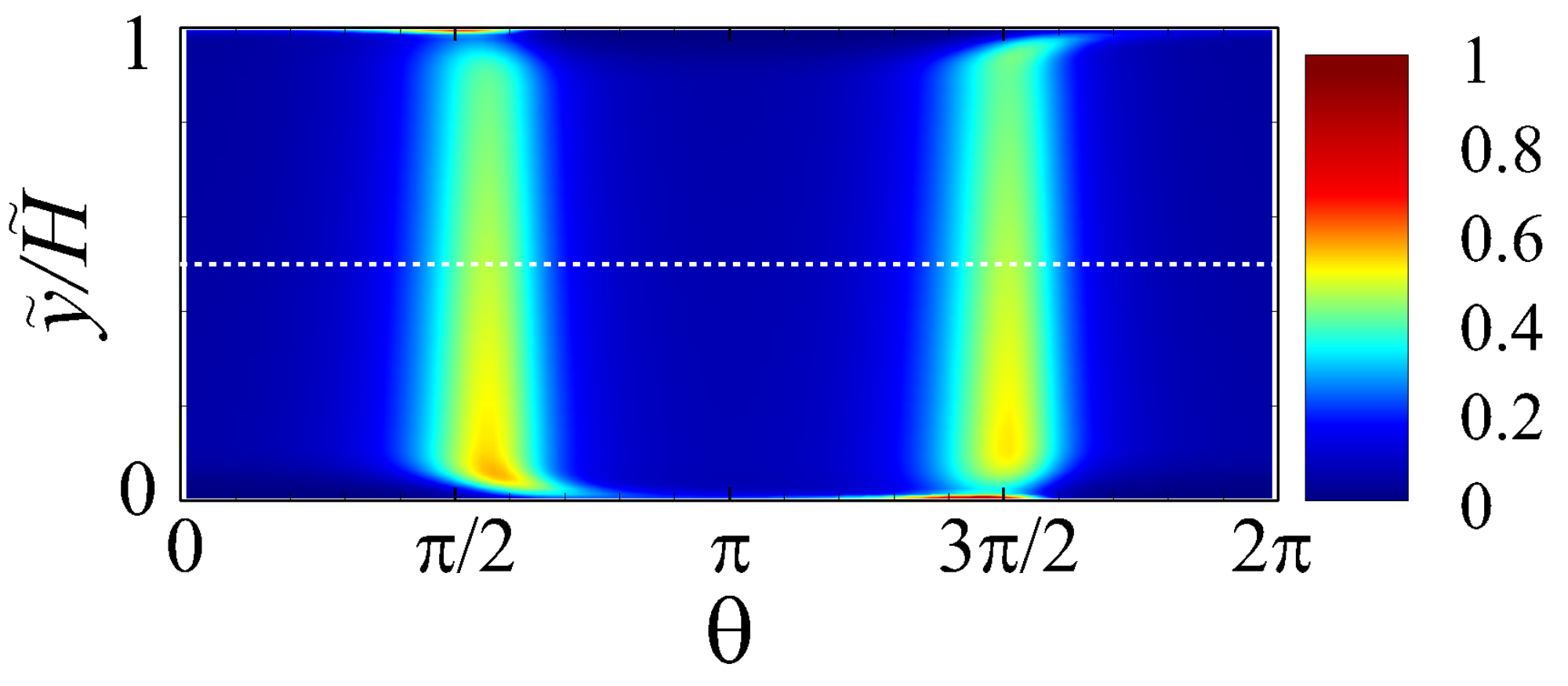} \vskip-2mm (c) $\chi=10$
	\end{center}\end{minipage}\hskip0mm	
	\begin{minipage}[t]{0.50\textwidth}\begin{center}
		\includegraphics[width=\textwidth]{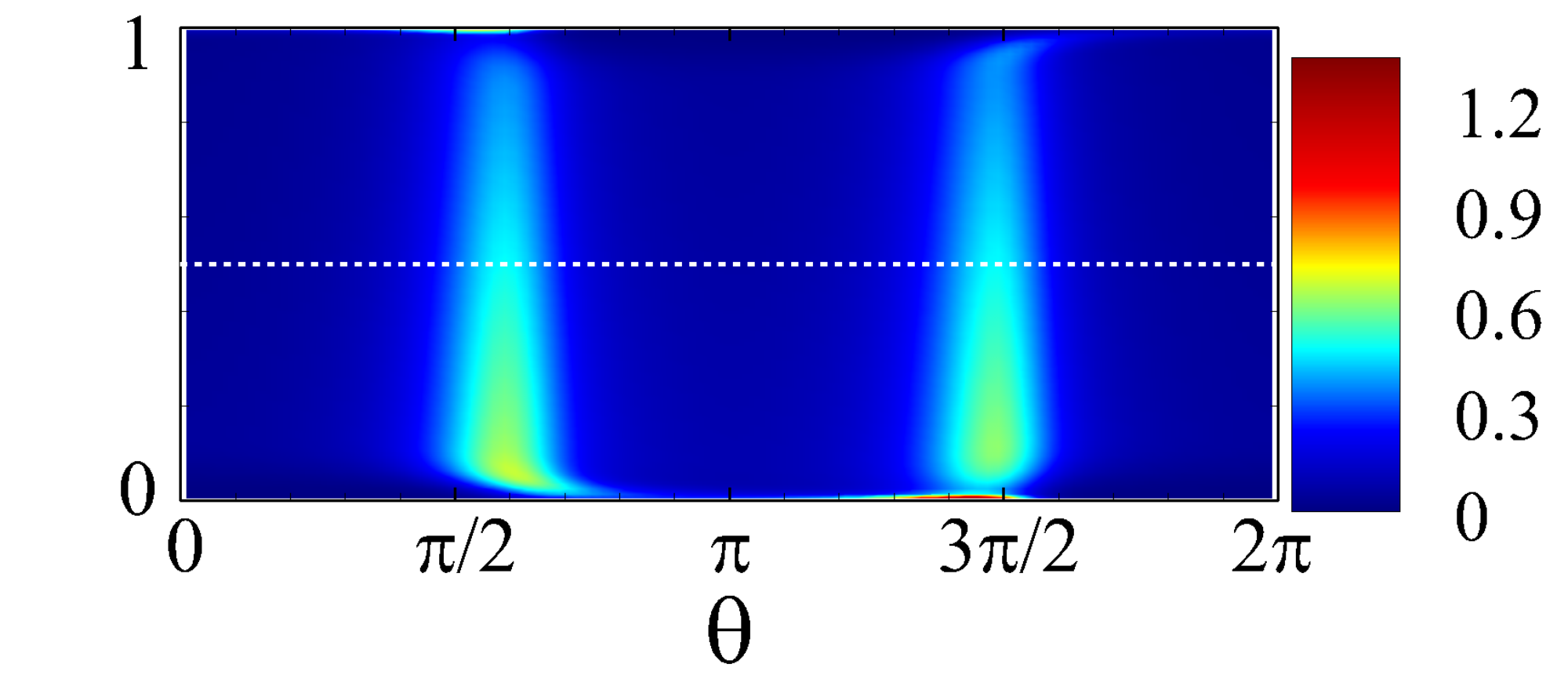} \vskip-2mm (d) $\chi=20$
	\end{center}\end{minipage}
\end{center}\end{minipage}\hskip2mm
\begin{minipage}[t]{0.365\textwidth}\begin{center}
		\includegraphics[width=\textwidth]{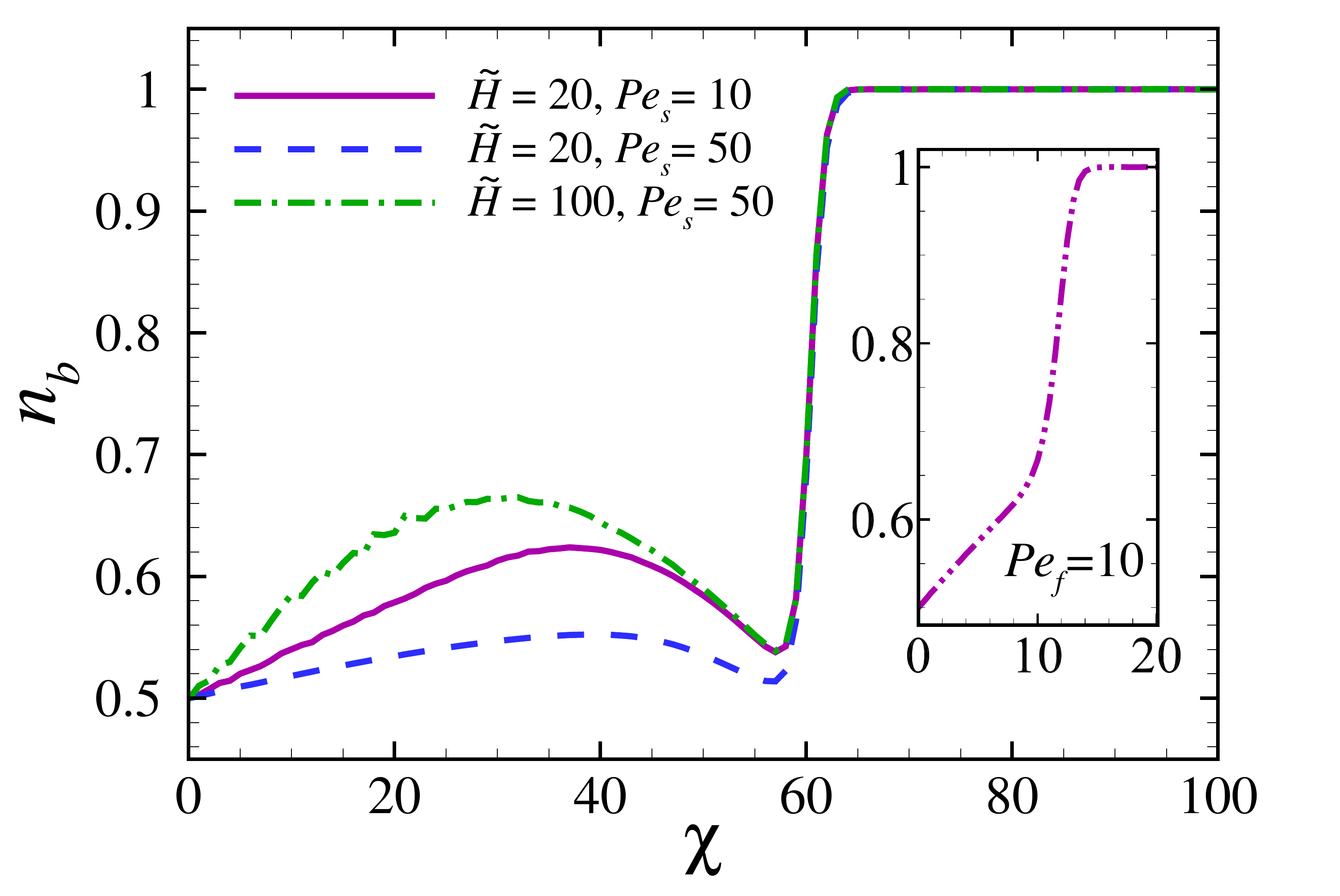}\vskip-1mm (e)
\end{center}\end{minipage}
\vskip-4mm
	\caption{
	Density maps of the PDF, $\tilde{\Psi}(\tilde y, \theta)$, for {\em oblate} swimmers of aspect ratio $\alpha=1/3$ for (a) $\chi=0$ (global view), (b) $\chi=0$ (closeup of the near-wall region close to $\tilde y=0$), (c) $\chi=10$ and  (d) $\chi=20$ at fixed $Pe_s=10$, $Pe_f=50$ and  $\tilde H=20$. (e) Total fraction of swimmers in the bottom half of the channel, Eq. \eqref{eq:n_b_def}, as a function of $\chi$ for fixed $\alpha=1/3$ and $Pe_f=50$, and other parameter values shown on the graph. 
	}
\label{fig_contours_full_r033}
\end{center}
\end{figure*}

\subsection{Orientational dynamics and pinning}
\label{subsec:pinning}

Some of the key aspects of the cross-stream migration of prolate swimmers can be elucidated further using the {\em deterministic} dynamics of the swim orientation  (noise effects are indeed subdominant in regimes of interest here). This noise-free dynamics is governed by the overdamped equation $\dot \theta(t)= \omega(\theta(t))$, where the net angular velocity, $ \omega(\theta)=\omega_f(\theta)+\omega_{\mathrm{ext}}(\theta)$, Eq. \eqref{eq:w_tot0}, is the sum of the shear-/field-induced contributions,  respectively. With more details provided in the supplementary material, here we discuss only a few of those aspects. 

For prolate particles, $| \omega_f|$ is minimized at the nematic orientations $\theta =0$ and $\pi$, while $| \omega_{\mathrm{ext}}|$ is maximized at these orientations. Indeed, the field-induced torque tends to stabilize the swim orientation at the downward angle $\theta= -\pi/2$, where it vanishes and changes its sign from negative (within the first/fourth $\theta$-quadrants) to positive (within the second/third $\theta$-quadrants). Evidently, the magnitude of the net angular velocity, $ \omega(\theta)$, is also increased (decreased)  in the first/fourth (second/third)  $\theta$-quadrants, giving a decreased (increased) rotational residence time  in the said angular quadrants relative to their corresponding  base, zero-field,  values. This can be interpreted as an increased (decreased) {\em flip-down} ({\em flip-up}) rate of the swim orientation, as it performs clockwise, field-modified Jeffery oscillations (see Section II  of the supplementary material). The partial migration of swimmers from the top to the bottom half of the channel can thus be associated with the increased flip-down rate of the swim orientation as compared with its flip-up rate, giving the swimmers larger likelihoods in the bottom half, as corroborated by the increase in $n_b$ in Fig.  \ref{fig_contours_full_r3}e. The  flip-up/-down rate difference  can be shown to vary linearly with $\chi$ to the leading order, explaining the origin of the linear-response regime (see Appendix \ref{sec:weak_field} and Section III of the supplementary material). The above line of argument  can also be used to justify the emergence of the reverse bimodality/minority persistence regimes, as it captures the basic mechanism depicted in Fig. \ref{fig_r3_submigration}b. Given our numerical analyses of the preceding sections, the full migration behavior can also be inferred to result  from strong modifications in Jeffery oscillations of the swim orientation. Above a certain threshold, $\chi_\ast$, the situation becomes wholly different as a saddle-node bifurcation \cite{Strogatz2000} takes place, leading to a  {\em stable fixed point}, $ \omega(\theta_\ast)=0$,  at an angular orientation, 
$\theta_\ast$, in the third $\theta$-quadrant (see Section IV of the supplementary material). The swim orientation is thus {\em pinned},  Jeffery oscillations are eliminated  and the system is brought into the Um regime. 

Although useful in understanding certain aspects of the problem at hand, the above reasoning is based only on the  rotational dynamics of swimmers and ignores their self-propulsion and, as such, largely underestimates 
the onset of the Um regime from its numerically computed  value, $\chi^{(3)}$ (the predicted pinning angle, $\theta_\ast$, however approximates the computed most-probable orientation within a relative margin of only a few percent). For the parameters in Fig. \ref{fig_r3_submigration_PD} and  $Pe_f=100$, the predicted onset, $\chi_\ast$, underestimates $\chi^{(3)}$ by 30\%, with larger deviations occurring at smaller $Pe_f$. 
Such deviations between the crossover values of $\chi$ as predicted by the deterministic  orientational dynamics and those obtained numerically are more significant in the case of prolate swimmers as opposed to the case of oblate swimmers (Section \ref{sec:oblates}). This is because the self-propulsion mechanism tends to bring the major body axis of prolate swimmers in the normal-to-wall orientation, which is strongly opposed by the flow. 

\section{Magnetic oblate swimmers}
\label{sec:oblates}

\subsection{Shear-induced bimodality with no external field}
\label{sec:no_field_ob}

Sheared, field-free, oblate swimmers also exhibit bimodal PDFs, $\tilde{\Psi}(\tilde y, \theta)$,
across the channel width at elevated shear rates, albeit with notable qualitative differences from the prolate swimmers   (compare Fig. \ref{fig_contours_full_r033}a with Fig. \ref{fig_contours_full_r3}a). As in the case of prolate particles, the shear-induced angular velocity, $ \omega_f$, Eq. \eqref{eq:w_f}, creates clockwise Jeffery oscillations for the oblate particle orientation between the most-probable angles $\theta =\pi/2$ and $3\pi/2$, where $| \omega_f|$ is minimized. In these up-/downward orientations, the major body axis of swimmers is nematically aligned with the flow, while the swim orientation (being along the minor body axis; Fig. \ref{fig:schematic})  will be pointed normal to the channel walls, making the said orientations also strongly favored by the self-propulsion mechanism near the walls.

Thus, rather than the up-/downstream populations found in the case of prolate swimmers, the population splitting of sheared oblate swimmers occurs as two predominantly upward- and downward-pointing populations develop across the central regions of the channel (dark-yellow columns formed around $\theta= \pi/2, 3\pi/2$ in Fig. \ref{fig_contours_full_r033}a). On closer inspection, however, each of these midchannel populations is found to originate as a continuous extension of a single population of wall-accumulated, oppositely pointing, population; e.g., the upward-pointing midchannel population (dark-yellow column at $\theta= \pi/2$) stems from a majority population (red-colored spot centered around $\theta=3\pi/2$) near the bottom wall. This is seen more clearly in the closeup view of the near-wall region in Fig. \ref{fig_contours_full_r033}b. This behavior is in stark contrast with the case of prolate swimmers shown in Fig. \ref{fig_contours_full_r3}a, indicating that, as one moves away from the bottom wall into the channel, the most-probable orientation of oblate swimmers {\em flips} in clockwise direction from downward to upward orientation. Hence, a  narrow probability stretch is formed through the second and  third $\theta$-quadrants  (Fig. \ref{fig_contours_full_r033}b), with the probability density fading away around $\theta=\pi$, where the magnitude of the shear-induced angular velocity, $ \omega_f$, is maximized. For the parameters in Figs. \ref{fig_contours_full_r033}a or b, the flipping of the swim orientation occurs  within the  distances $|\delta \tilde y|\simeq 0.4$ from the walls, or $|\delta  y|\simeq b$ in actual unit,
 measured from the no-flux walls (exclusive of the closet-approach distance, $b$, from the no-slip boundaries; Section \ref{sec:Smoluchowski_eq}). Also, as in the case of prolate swimmers (Fig. \ref{fig_contours_full_r3}a), the PDF remains unimodal in the close proximity of the  walls (Figs.  \ref{fig_contours_full_r033}b).

\subsection{Population splitting and nonmonotonic  response}
\label{subsec:migration_ob}

The shear-induced polarization of the bimodal swim orientation at $\theta=\pi/2$ and $3\pi/2$ suggests that the effects of an applied field should be relatively small in the case of oblate swimmers  (Figs. \ref{fig_contours_full_r033}a-d) as compared  with the case of prolate swimmers (Figs. \ref{fig_contours_full_r3}a-d). This is because the angular velocity, $ \omega_{\mathrm{ext}}$, due to the {\em downward-pointing} field used in our analysis, Eq. \eqref{eq:w_ext}, vanishes at those same  orientations. 
The fraction of swimmers in the bottom half of the channel, $n_b$ (Eq. \eqref{eq:n_b_def}), being  plotted as a function of $\chi$ in Fig. \ref{fig_contours_full_r033}e, shows that, not only the full migration of oblate swimmers to the bottom half of the channel ($n_b\simeq 1$) occurs at substantially larger field coupling strengths, but it is also preceded by a {\em nonmonotonic} regime, distinctly different from the monotonic one found for prolate swimmers (compare Figs. \ref{fig_contours_full_r033}e and  \ref{fig_contours_full_r3}e). 

Figure \ref{fig_contours_full_r033}e  shows a regime of {\em linear response} for $n_b$ (exemplified by the PDFs in Figs. \ref{fig_contours_full_r033}a-d) in sufficiently weak fields, followed by a local peak, marking the onset of its nonlinear behavior: By increasing $\chi$ beyond the peak location, $n_b$  decreases down to a local dip, entailing a counterintuitive regime of {\em reverse migration}  (RM) from the bottom to the top channel half. The inset of Fig. \ref{fig_contours_full_r033}e shows that the RM regime can disappear at smaller values of $Pe_f$ (we have used $Pe_f=50$ for the main set and $Pe_f=10$  for the inset of Fig. \ref{fig_contours_full_r033}e). We shall return to these points in the following sections and only note here that the reverse migration takes place only partially as $n_b$ remains above its base value of $n_b(\chi=0)=1/2$.

\begin{figure}[t!]
\begin{center}
\sidesubfloat[]{%
\hskip-1mm%
\includegraphics[width=0.75\textwidth]{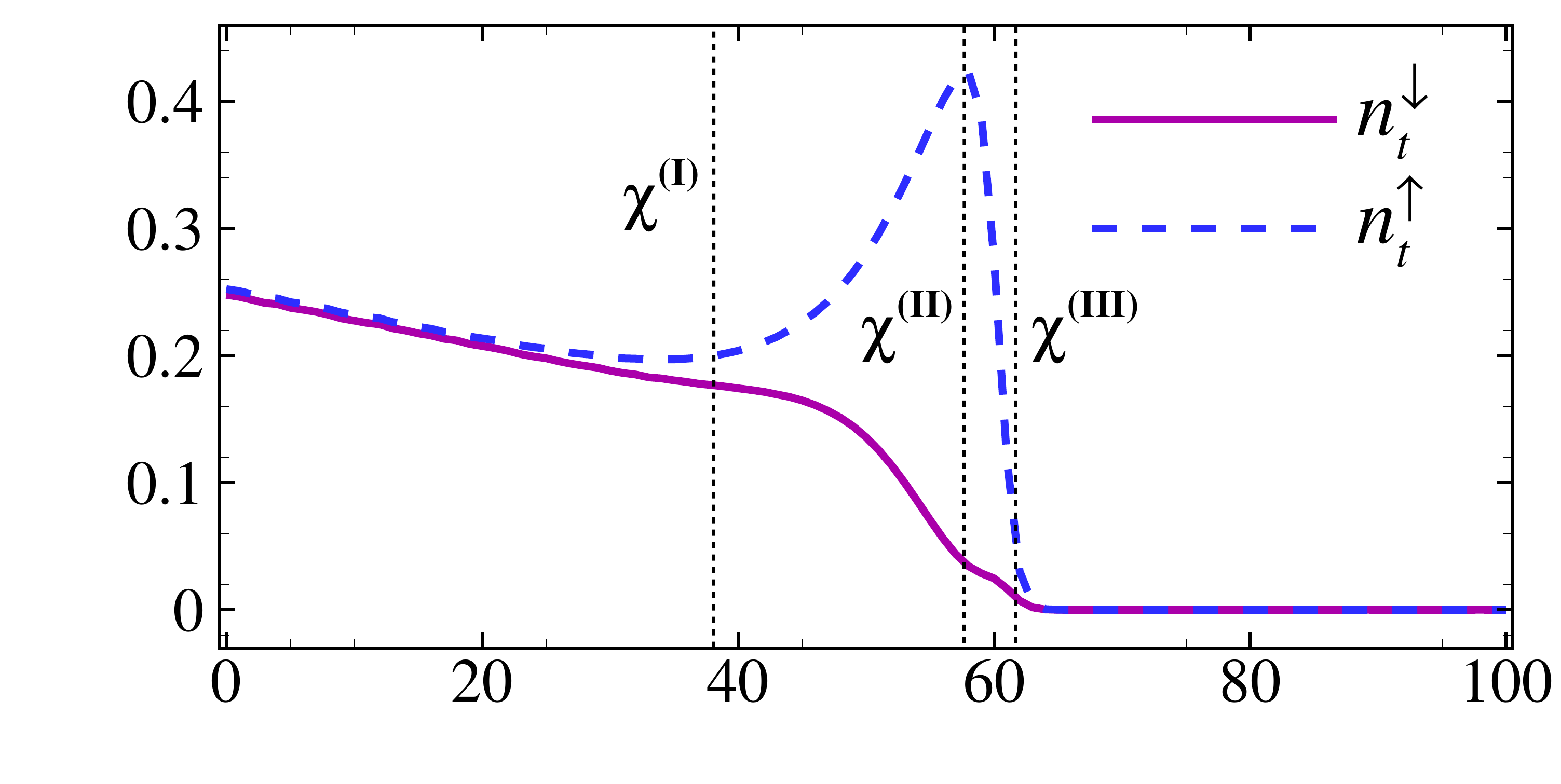}%
}%
\vskip-2mm%
\sidesubfloat[]{%
\hskip-1mm%
\includegraphics[width=0.75\textwidth]{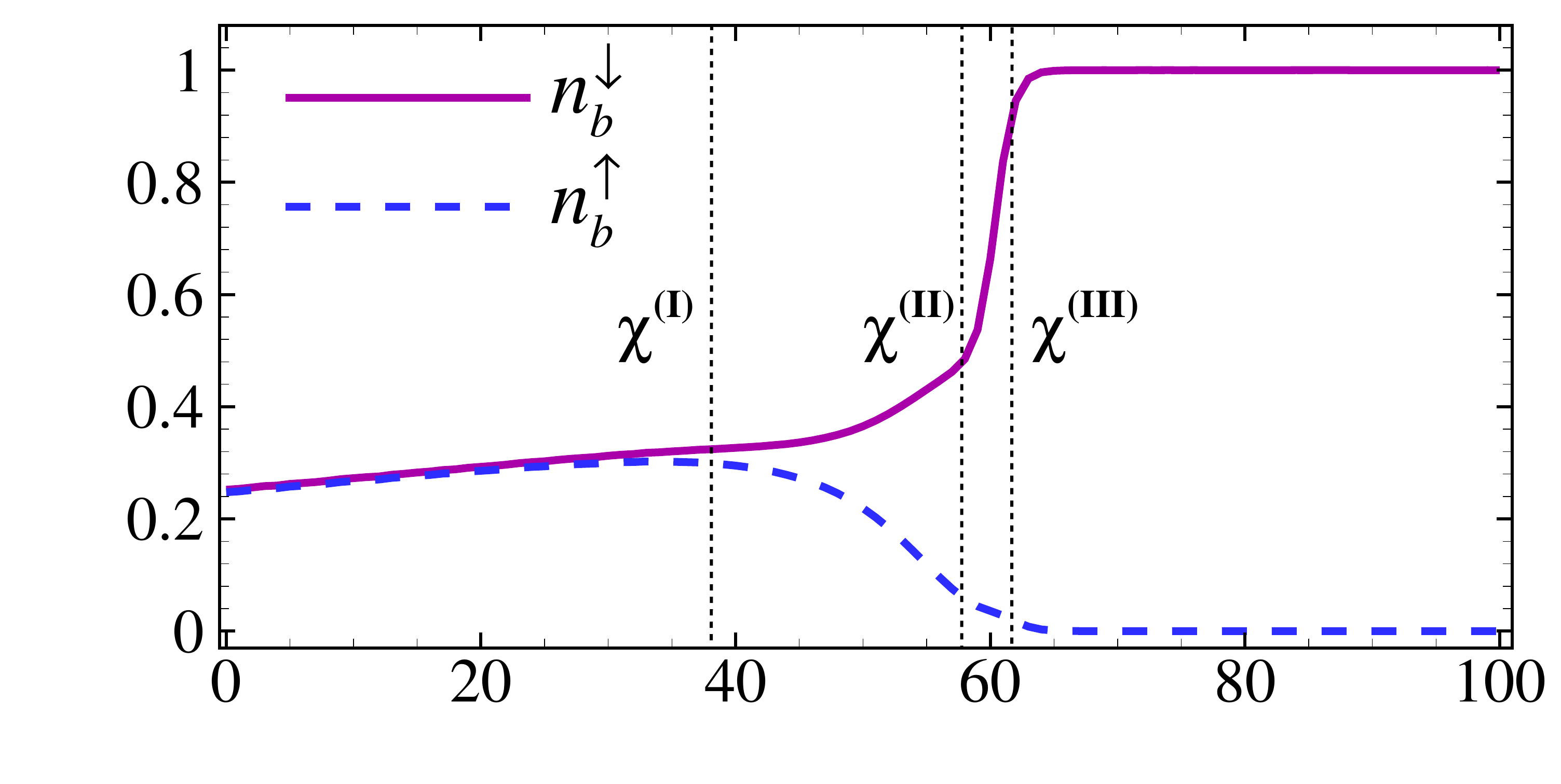}%
}%
\vskip-2mm%
\sidesubfloat[]{%
\hskip-1mm%
\includegraphics[width=0.75\textwidth]{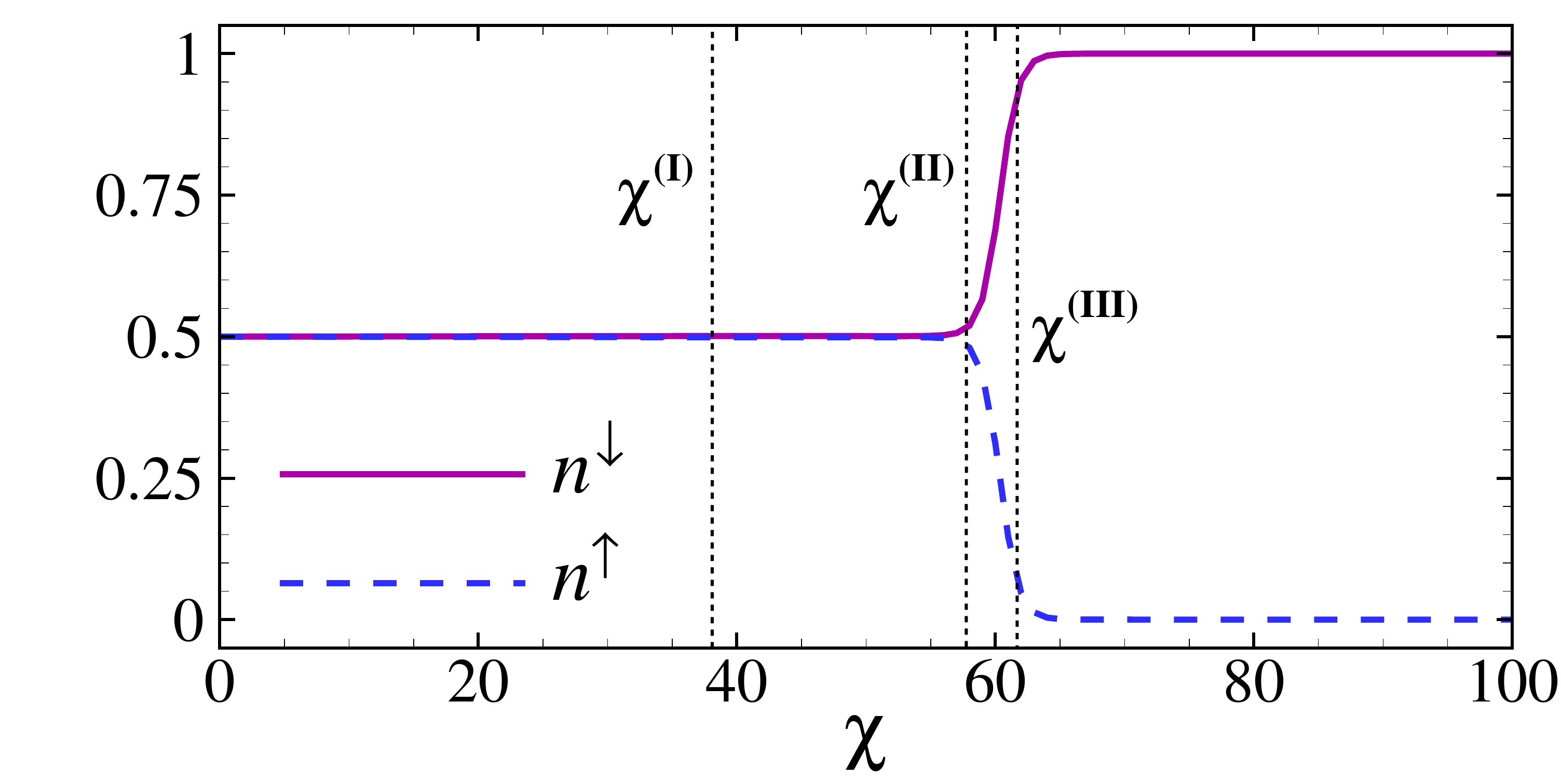}%
}
\vskip-4mm
	\caption{Downward- and upward-pointing fractions of oblate swimmers in (a) the top half, $n_t^{\downarrow\uparrow}$, and (b) bottom half, $n_b^{\downarrow\uparrow}$, of the channel, Eq. \eqref{eq:n_t_b_pm_ob}, and (c) within the whole channel, as functions of the field coupling  strength for fixed $Pe_s=10$,  $Pe_f=50$, $\tilde H=20$ and  aspect ratio $\alpha= 1/3$.  See the text for the definitions of $\chi^{(\textrm{I})}$, $\chi^{(\textrm{II})}$ and $\chi^{(\textrm{III})}$ shown on the graph. 
}
\label{fig_r033_submigration}
\end{center}
\end{figure}

By increasing $\chi$ only slightly further, we find a sharp rise to full saturation in $n_b$, yielding the FM regime. The behavior at/beyond the onset of full migration  is found to be nearly independent of the rescaled channel width, $\tilde H$, and the swim P\'eclet number, $Pe_s$. It is however dependent on the flow P\'eclet number, $Pe_f$, and the particle aspect ratio, $\alpha$, enabling a potential route for separation of oblate swimmers of varying aspect ratio (Section \ref{subsec:separation_ob}). 

The overall picture discussed above is obtained based on the bottom-half swimmer fraction, $n_b$, which is an  orientationally averaged quantity. The nonmonotonic features of the cross-stream migration of oblate swimmers can be elucidated further by examining the orientational subpopulations, resulting from the population splitting effect discussed in the preceding section. Hence, we next consider the fractions of upward/downward-pointing subpopulations in the top (bottom) half of the channel, $n_t^{\uparrow}$ and $n_t^{\downarrow}$ ($n_b^{\uparrow}$ and $n_b^{\downarrow}$), respectively; i.e.,    
\begin{equation}
n_t^{\uparrow\downarrow} = \int_{\tilde H/2}^{\tilde H}\!{\mathrm{d}}\tilde y\,\tilde{\phi}_{\uparrow\downarrow}(\tilde y),\quad  n_b^{\uparrow\downarrow} = \int_{0}^{\tilde H/2}\!{\mathrm{d}}\tilde y\,\tilde{\phi}_{\uparrow\downarrow}(\tilde y), 
\label{eq:n_t_b_pm_ob}
\end{equation}
with the density profiles,  $\tilde{\phi}_{\uparrow}(\tilde y)$ and $\tilde{\phi}_{\downarrow}(\tilde y)=\tilde{\phi}(\tilde y)-\tilde{\phi}_{\uparrow}(\tilde y)$, of the upward/downward-pointing subpopulations being 
\begin{equation}
\tilde{\phi}_{\uparrow}(\tilde y) \!= \!\!\int_{0}^{\pi}\!{\mathrm{d}}\theta\, \tilde \Psi(\tilde y, \theta),\quad \tilde{\phi}_{\downarrow}(\tilde y) \!= \!\!\int_{\pi}^{2\pi}\!{\mathrm{d}}\theta\, \tilde \Psi(\tilde y, \theta). 
\label{eq:phi_t_b_pm_ob}
\end{equation}

\subsubsection{Linear response and standard (downward) migration}
\label{subsec:migration_ob_lin}

For the parameters  in Figs. \ref{fig_r033_submigration}a and b, the fractions of majority ($n_t^{\uparrow}$ and $n_b^{\downarrow}$) and minority subpopulations ($n_t^{\downarrow}$ and $n_b^{\uparrow}$ in the top and bottom halves, respectively) turn out to be nearly equal, $n_t^{\uparrow}\simeq  n_t^{\downarrow}$ and $n_b^{\downarrow}\simeq n_b^{\uparrow}$ for a relatively wide range of $\chi<\chi^{(\textrm{I})}$ (in the figures, $\chi^{(\textrm{I})}\simeq 37$, which is shown by a dotted vertical line, whose locus coincides with that of the local peak of the purple, middle, curve in Fig. \ref{fig_contours_full_r033}e). The figures thus indicate the {\em absence} of reverse bimodality for oblate swimmers, which is due to the lack of the necessary mechanism, such as the one discussed for prolate swimmers in Section \ref{subsubsec:rev_bm} and Fig. \ref{fig_r3_submigration}b. In fact, the behavior of the system conforms to one of standard bimodality ($n_t^{\uparrow}\gtrsim n_t^{\downarrow}$ and $n_b^{\downarrow}\gtrsim n_b^{\uparrow}$)
across the whole $\chi$-interval explored, up until the full migration process sets in. 

Figures \ref{fig_r033_submigration}a and b also identify $\chi<\chi^{(\textrm{I})}$ as the linear-response regime for the field-induced behaviors of {\em all} four subpopulations, contrasting the situation with the subpopulations of prolate swimmers in Fig. \ref{fig_r3_submigration}. The origin of linear response is as in the case of prolate swimmers (Section \ref{subsec:pinning} above and Section III of the supplementary material) and it can be quantified similarly using a response factor (Appendix \ref{sec:weak_field}). Also, for $\chi<\chi^{(\textrm{I})}$, the four fractions $n_t^{\uparrow\downarrow}$ and $n_b^{\uparrow\downarrow}$ display monotonically deceasing and increasing behaviors with $\chi$, respectively, in accordance with the {\em standard migration} paradigm; i.e., effective particle migration takes place in the downward direction of the applied field.  For  $\chi>\chi^{(\textrm{I})}$, all subpopulations begin to exhibit nonlinear trends, and migration anomalies appear, which we discuss next. 

\subsubsection{Reverse migration and midchannel depletion}
\label{subsec:migration_ob_reverse}

In the RM regime, $\chi^{(\textrm{I})}<\chi<\chi^{(\textrm{II})}$, the majority and minority subpopulations in either  half of the channel are enhanced and suppressed, respectively (Figs. \ref{fig_r033_submigration}a and b); here, $\chi^{(\textrm{II})}$ is taken as the  locus of the local dip in $n_b$ or that of the  peak in $n_t^{\uparrow}$ (i.e., $\chi^{(\textrm{II})}\simeq 58$ for the purple, middle, curve in Fig. \ref{fig_contours_full_r033}e and the blue dashed curve in Fig. \ref{fig_r033_submigration}a). The noted behaviors indicate that what we identify as the RM regime involves not one but two distinct and physically discernible subprocesses not differentiated by the net fraction $n_b$ shown in Fig. \ref{fig_contours_full_r033}e: A {\em standard migration subprocess} displayed by the downward-pointing subpopulations, whose fractions $n_t^{\downarrow}$ and $n_b^{\downarrow}$ (shown by the purple solid curves in Figs. \ref{fig_r033_submigration}a and b) evolve with increasing $\chi$ in accordance with the standard paradigm, and a {\em reverse (or anomalous) migration subprocess} displayed by the upward-pointing ones (blue dashed curves), which do the `opposite'. In particular, the subpopulation  $n_t^{\uparrow}$ shows a pronounced increase with $\chi$ up to a global maximum, indicating re-uptake of the upward-pointing subpopulations from across the channel width into its top half. 

It turns out, however, that the anomalous subprocess does {\em not} necessarily involve migration of the upward-pointing subpopulations in the {\em exact opposite direction} to that of the field.  In fact, the noted subprocesses are caused by an effective stabilization, or pinning, of the most-probable swim orientation at two disparate angles in the second and third $\theta$-quadrants;  see Section \ref{sec:pinning_ob}. The anomalous subprocess involves swimmer pinning and migration at a second-quadrant orientation angle, varying from around $\pi/2$ (nearly opposing the field) up to $\pi$ (upstream normal-to-field migration).  
Moreover, since the sums $n^{\uparrow\downarrow}=n_t^{\uparrow\downarrow}+n_b^{\uparrow\downarrow}$ remain constant in  the RM regime (Fig. \ref{fig_r033_submigration}c), the said  subprocesses also represent a  {\em midchannel depletion process}, with  the upward-/downward-pointing subpopulations cross-translating as intact populations across the channel  toward the top/bottom walls, respectively. This explains why $n_b$ in  Fig. \ref{fig_contours_full_r033}e levels off  down to nearly its base value of 1/2 but not any lower than that at the upper bound of the RM regime, $\chi^{(\textrm{II})}$. 

The fact that the fractions $n^{\uparrow\downarrow}$ remain constant also within the linear-response regime, $\chi<\chi^{(\textrm{I})}$ (Fig. \ref{fig_r033_submigration}c), is of a different nature than the one discussed above for the RM regime. It indicates that the overall rotational residence times spent by the swim orientation in upward- and downward-pointing configurations are equal, although they vary from one $\theta$-quadrant to another (see Sections II and III of the supplementary material). Figure \ref{fig_r033_submigration}c further shows that the up-down polarization symmetry, $\theta\rightarrow 2\pi-\theta$,  inherent to the deterministic problem, but not to the full problem being solved numerically,  is preserved in the orientationally integrated quantities, $n^{\uparrow\downarrow}$, in the case of oblate swimmers up until the Um regime, which is discussed next.

\subsubsection{Flip-down and full migration to the bottom half} 
\label{subsubsec:full_migration_ob}

Based on Figs. \ref{fig_r033_submigration}a and b, the top-half upward-pointing fraction, $n_t^{\uparrow}$, eventually begins to fall off for  $\chi>\chi^{(\textrm{II})}$. All upward-pointing swimmers thus {\em flip down} 
and self-propel to the bottom half of the channel, strengthening the bottom-half downward-pointing majority, $n_b^{\downarrow}$.  This transient process can be assumed to saturate using a criterion similar to the one in Section \ref{subsubsec:rev_bm}; i.e., for $n^{\downarrow}_b > 0.95$, identifying the Um regime, which occurs here almost concurrently with the FM regime defined through the criterion $n_b > 0.95$, at $\chi^{(\textrm{III})}\simeq 62$. There is thus a rapid crossover between the RM and FM/Um regimes through a narrow {\em transient (down-flipping) regime} for $\chi^{(\textrm{II})}<\chi<\chi^{(\textrm{III})}$.

\subsubsection{Migration pattern of oblate swimmers: A summary}
\label{subsubsec:summary_oblates}

The cross-stream migration of oblate swimmers can be summarized as
\begin{align}
\left\{\begin{array}{ll}
\triangleright\,\,\textrm{Linear response/Standard migration:}&\\
\,\,\,\,\,n_b(\chi)-n_b(\chi=0)={\mathcal O}(\chi)&\hskip-13mm\chi<\chi^{(\textrm{I})},\\ 
\\
\triangleright\,\,\textrm{Reverse migration (RM):}&\\
\,\,\,\,\,\partial n_b/\partial \chi<0,\,\, \partial n_t^{\uparrow}/\partial \chi>0&\hskip-13mm\chi^{(\textrm{I})}<\chi<\chi^{(\textrm{II})},\\
\\
\triangleright\,\,\textrm{Transient (down-flipping) regime:}&\\
\,\,\,\,\,\partial n_b/\partial \chi>0,\,\, \partial n_t^{\uparrow}/\partial \chi<0&\hskip-13mm\chi^{(\textrm{II})}<\chi<\chi^{(\textrm{III})},\\
\\
\triangleright\,\,\textrm{Full migration/Unimodality (FM/Um):}&\\
\,\,\,\,\,n_b\simeq n^{\downarrow}_b\simeq 1&\hskip-13mm\chi>\chi^{(\textrm{III})}.
\end{array}\right.
\end{align}

\begin{figure}[t!]
\begin{center}
\includegraphics[width=0.8\textwidth]{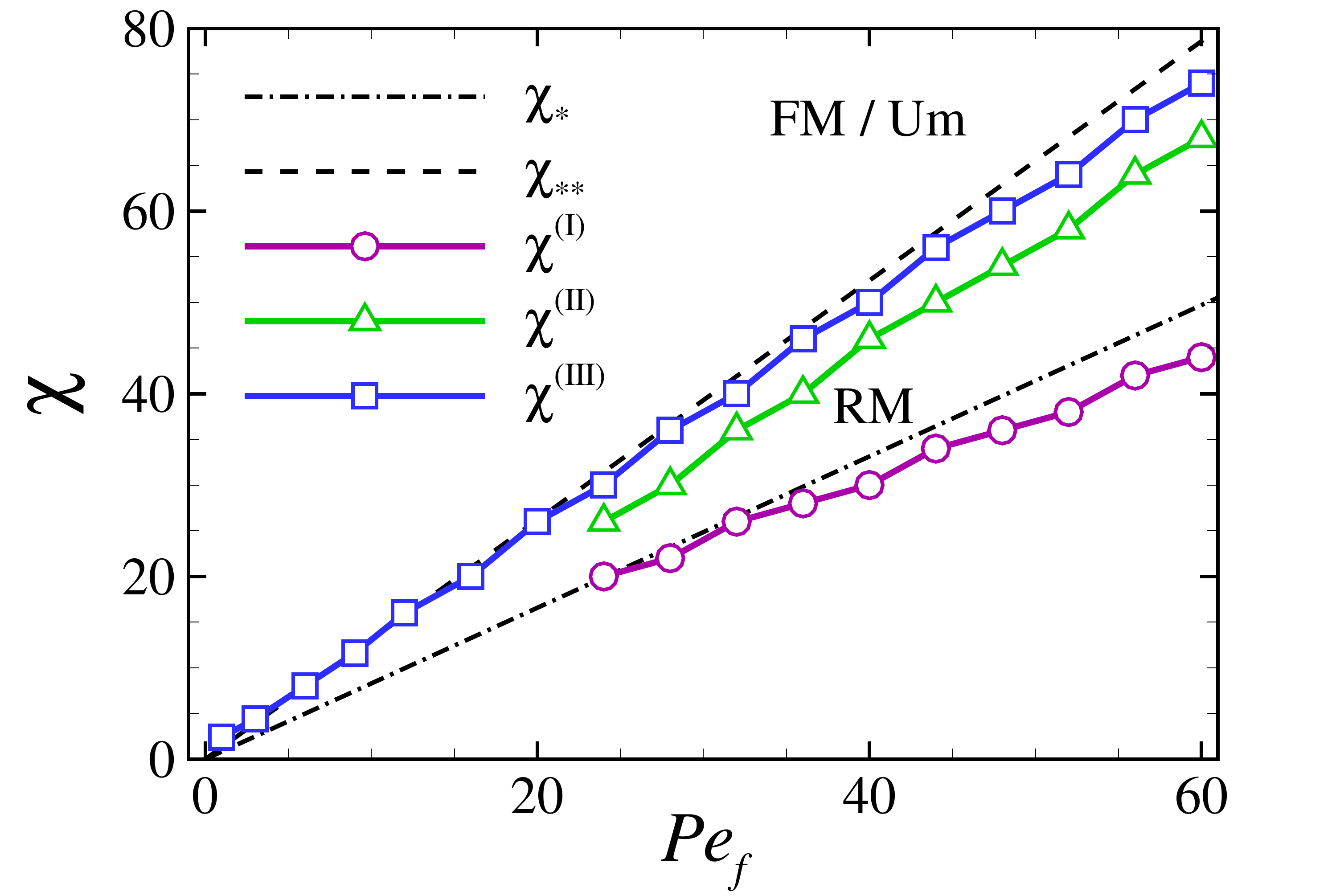}
\vskip-4mm
\caption{Different regimes of behavior for the field-induced migration of oblate swimmers across the channel width in terms of the field coupling strength, $\chi$, and the flow P\'eclet number, $Pe_f$, at fixed $\alpha=1/3$, $Pe_s=10$ and  $\tilde H=20$. The computed crossover couplings $\{\chi^{(\textrm{I})}, \chi^{(\textrm{II})}, \chi^{(\textrm{III})}\}$  are shown by symbols, with solid curves drawn to guide the eye. See the text and the supplementary material for the definitions of $\chi_{\ast}$ and $\chi_{\ast\ast}$ (dot-dashed and dashed lines, respectively).  
}
\label{fig_r033_submigration_PD}
\end{center}
\end{figure}

The crossover values $\{\chi^{(\textrm{I})}, \chi^{(\textrm{II})}, \chi^{(\textrm{III})}\}$ of the field coupling strength can generally vary depending on other system parameters. Figure \ref{fig_r033_submigration_PD} gives a broader view of the cross-stream migration regimes of oblate swimmers across the whole range of $Pe_f$ values explored here, accentuating the interplay between the effects of shear- and field-induced torques  at fixed $\alpha=1/3$, $Pe_s=10$ and $\tilde H=20$. The figure also mirrors the previously noted observation that a  direct crossover to the FM/Um regime (without going through the intervening regimes) is feasible at  small enough flow P\'eclet numbers,  below a  threshold value of $Pe_f^\ast$, where both the boundary lines $\chi^{(\textrm{I})}$ and $\chi^{(\textrm{II})}$ terminate (in the figure, $Pe_f^\ast\simeq 24$). Our numerical inspections show that $Pe_f^\ast$ decreases upon increasing the rescaled channel width, $\tilde H$, but shows no discernible dependence on the swim P\'eclet number, $Pe_s$, within our numerical margin of error. The same trends are found for   $\chi^{(\textrm{I})}$ at a given $Pe_f>Pe_f^\ast$, while $\chi^{(\textrm{III})}$ and $\chi^{(\textrm{II})}$ are found to  show no significant dependence on $\tilde H$ or $Pe_s$.

\subsection{Orientational bistability and double pinning}
\label{sec:pinning_ob}

The deterministic dynamics of the swim orientation can be invoked to shed further light on the regimes of migration found numerically for oblate swimmers. 

As noted before, the swim orientations $\theta =\pi/2$ and $3\pi/2$ are favored by all the three key mechanisms involved here; i.e., the shear flow (favoring nematic alignment of the major body axis with the flow,  minimizing $| \omega_f|$), the magnetic field (favoring downward, normal-to-flow, alignment of the magnetic dipole moment, with vanishing $ \omega_{\mathrm{ext}}$), and the self-propulsion (favoring normal-to-wall swim with prolonged detention times near the walls). This maximally cooperative situation is to be contrasted with the competitive one in the case of prolate swimmers, where $| \omega_{\mathrm{ext}}|$ is maximized at the shear-favored nematic orientations, while $| \omega_f|$ is maximized at the self-propulsion-preferred swim orientations  (Section \ref{subsec:pinning}). 

It turns out that, for a certain regime of field coupling strengths, $\chi_\ast<\chi<\chi_{\ast\ast}$, the net angular velocity, $ \omega(\theta)$, Eq. \eqref{eq:w_tot0},  vanishes at multiple values of $\theta$, as detailed in Section IV of the supplementary material, where explicit definitions of the thresholds $\chi_\ast$ and $\chi_{\ast\ast}$ are given as well. The roots of $ \omega(\theta)=0$  specifically include {\em two} stable fixed points  $\theta_\ast$ and $\theta_{\ast\ast}$ in the third and second $\theta$-quadrants, respectively, indicating a regime of {\em double pinning} or orientational {\em bistability} in the deterministic dynamics of the swim orientation. As $\chi$ is increased within the interval $\chi_\ast<\chi<\chi_{\ast\ast}$, $\theta_\ast$ varies rather weakly within the third $\theta$-quadrant, increasing from an initial value close to $3\pi/2$ at $\chi=\chi_\ast$  and reaching a value only slightly closer to the said angle at $\chi=\chi_{\ast\ast}$. By contrast, $\theta_{\ast\ast}$ varies more rapidly within the second $\theta$-quadrant,  starting at a value closer to $\pi/2$ at $\chi=\chi_\ast$ and reaching $\pi$  at $\chi=\chi_{\ast\ast}$. When $\chi$ is taken beyond the double-pinning regime, $\chi>\chi_{\ast\ast}$, the second-quadrant fixed point, $\theta_{\ast\ast}$, disappears abruptly, leaving the swim orientation pinned only at an angle $\theta_\ast$ in the third $\theta$-quadrant that tends to $3\pi/2$ as $\chi$ is increased further; see Fig. 3b of the supplementary material for an illustration. 

The emergence of a field-induced double-pinning with a second-quadrant fixed point at intermediate field strengths and its subsequent disappearance  in stronger fields  place the proposed mechanisms of reverse and full migration/unimodality on a firmer ground; i.e., the second-quadrant angular pinning is responsible for the anomalous upward migration of a macroscopic fraction of swimmers to the top half, while its elimination leads to full downward migration of all swimmers, coincident with the unimodal regime of behavior caused by the pinning of the swim orientation at a single third-quadrant angle. This is further supported by the quantitative agreements established between the crossover field strengths obtained numerically and those predicted by the deterministic dynamics of the swim orientation,  as depicted in Fig. \ref{fig_r033_submigration_PD}; compare $\chi^{(\textrm{I})}$ and $\chi_\ast$ (circles and the dot-dashed line) and $\chi^{(\textrm{III})}$ and $\chi_{\ast\ast}$ (squares and the dashed line). The anomalous increase in $n_t^{\uparrow}$ (Fig. \ref{fig_r033_submigration}a) can also be understood using an explanation based on the basin of attraction of the fixed point $\theta_{\ast\ast}$, which is given in Section IV of the supplementary material.

In our model, the deterministic dynamics of the swim orientation and, hence, also the predicted thresholds  $\chi_\ast$ and $\chi_{\ast\ast}$, are independent of the swim P\'eclet number, $Pe_s$. The computationally determined boundaries of the RM regime, $\chi^{(\textrm{I})}$ and $\chi^{(\textrm{III})}$, can however depend  on  $Pe_s$. As noted in Section \ref{subsubsec:summary_oblates}, our numerical results indicate that, in the case of oblate swimmers,  $\chi^{(\textrm{I})}$ and $\chi^{(\textrm{III})}$ do not show any discernible dependence on $Pe_s$,  facilitating meaningful comparisons between the analytical and computational results, as discussed above. Recall that the situation is different for prolate swimmers, where the computationally determined crossover values of $\chi$ can depend rather strongly on $Pe_s$ and, e.g., the full migration there is caused by strong field-induced modifications in Jeffery oscillations  (Section \ref{subsec:pinning}), well before the unimodal regime sets in. Just as in the case of prolate swimmers, the predicted pinning angles of oblate swimmers are found to be close to their computed most-probable orientations at the walls within a relative margin of only a few percent (not shown).

\subsection{Shape-based separation of oblate swimmers}
\label{subsec:separation_ob}

\begin{figure}[t!]
\begin{center}
		\includegraphics[width=0.8\textwidth]{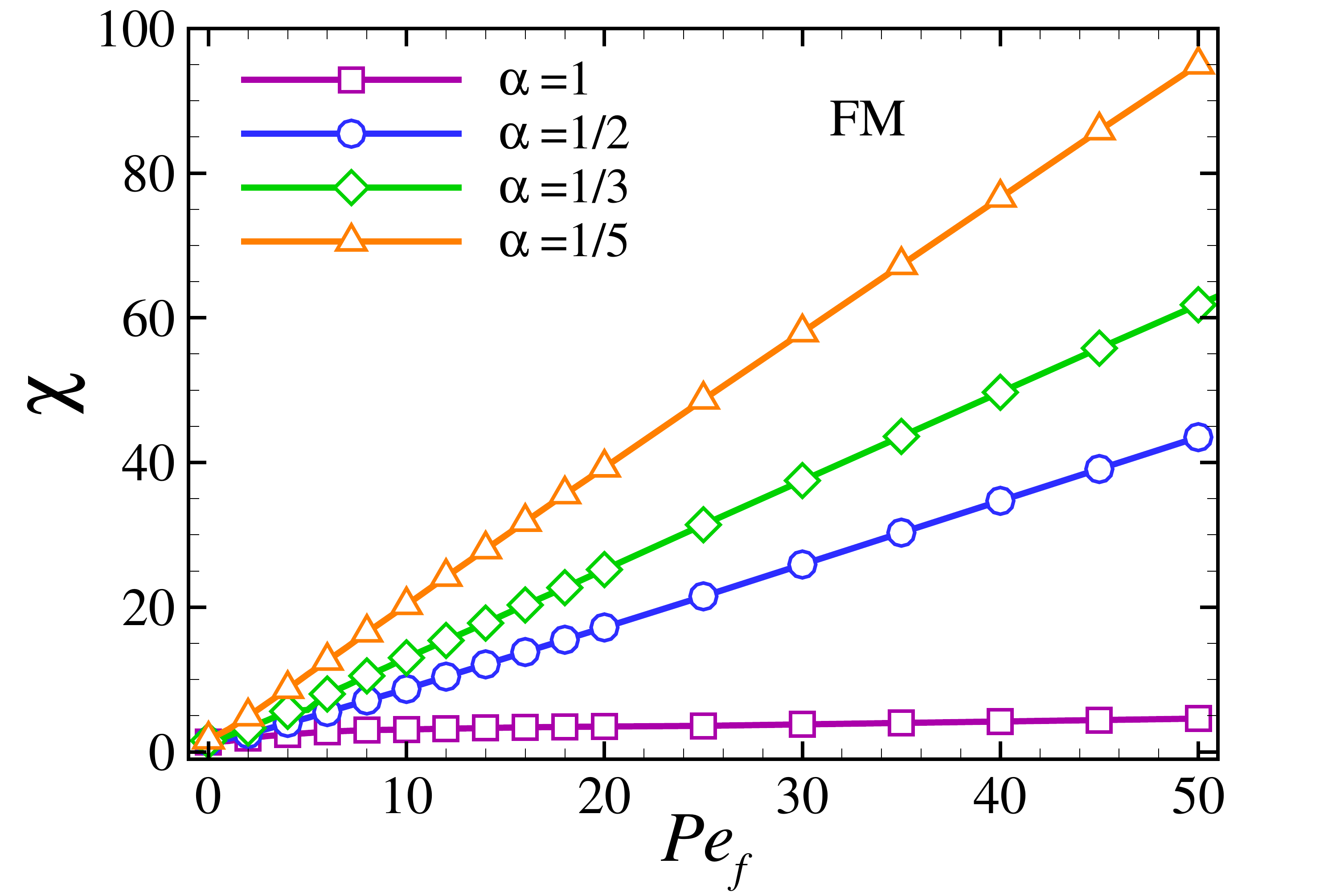} 
\vskip-4mm\caption{Boundaries of the full migration (FM) regime for oblate swimmers of different aspect ratio as indicated on the graph at fixed  $\tilde H=100$ and $Pe_s=50$. Symbols are computed data and curves are drawn to guide the eye.
}
\label{fig:separation_ob} 
\end{center}
\end{figure}

The rapid  crossover to the FM regime in the case of oblate swimmers enables a potential means for particle separation based on the aspect ratio $\alpha\leq 1$. In Fig. \ref{fig:separation_ob}, we show the boundary lines indicating the FM regime  on the left side for each of the curves, obtained here for $\alpha=1$ (spherical), $1/2$, $1/3$ and $1/5$ (disklike swimmers) for $Pe_s=50$ and $\tilde H=100$. As noted before, the FM boundary (i.e., $\chi^{(\textrm{III})}$ in Fig. \ref{fig_r033_submigration_PD}) is largely independent of $Pe_s$ and $\tilde H$ (see also Fig. \ref{fig_contours_full_r033}e), making the following discussions also applicable to other parameter values as explored in this work. Figure \ref{fig:separation_ob}  indicates that by increasing the field coupling strength, $\chi,$ at a fixed $Pe_f$, swimmers of different aspect ratios, from $\alpha=1$ to the more flattened ones with smaller $\alpha<1$, undergo full migration sequentially and at widely separated values of $\chi=\chi^{(\textrm{III})}$; e.g., in the figure, we have $\chi^{(\textrm{III})}\simeq 4.6, 43.5, 62, 95$ 
for  $\alpha=1, 1/2, 1/3, 1/5$  at fixed $Pe_f=50$. Hence, in a sheared mixture of magnetic swimmers with different aspect ratios $\alpha\leq 1$,  each type of swimmers can wholly and separately be guided to the bottom half of the channel for particle-separation purposes by tuning the external field over a relatively wide interval of values. 

There are two further features that make the foregoing mechanism an especially efficient one  as a potential route  to particle separation. First, it turns out that, at or beyond the onset of full migration, not only the swimmers---especially the non-spherical ones with $\alpha<1$---fully migrate to the bottom half of the channel, but they also  become confined to  a thin layer (with a thickness comparable to $R_{\mathrm{eff}}$) next to the bottom wall.  Second, when the full-migration onset for a specific swimmer aspect ratio is reached,  the other swimmers (i.e., those with smaller aspect ratios that have not yet undergone full migration) remain well dispersed  across the channel width; e.g., at fixed $Pe_f=50$, when spheres fully migrate to the bottom half, the swimmers with $\alpha=1/2$, gain only around 10\% increase in their total fraction in the bottom half, and similarly when the latter type of swimmers fully migrate to the bottom half, those with $\alpha=1/3$ gain around 10\% increase in their total fraction in the bottom half. This is a beneficial consequence of the fact that the sequential values of $\chi^{(\textrm{III})}$ are such that as one type of swimmer enters the FM regime, the other types of swimmers with smaller  $\alpha$  still remain well within their linear-response and/or reverse migration regimes.

\section{Net upstream particle current}
\label{sec:flux}

We now consider the regime of intermediate to large field coupling strengths and evaluate the current density and the total current produced in the channel by prolate/oblate swimmers of different aspect ratio.  
The rescaled particle current density relative to the fixed {\em laboratory frame} (or the stationary lower boundary of the channel, where the {\em no-slip} boundary condition is imposed; see Section \ref{sec:Smoluchowski_eq}), is defined as 
\begin{equation}
\tilde j (\tilde y) = \int_{-\pi /2}^{3\pi /2} {\mathrm{d}}\theta\, \tilde \Psi(\tilde y, \theta) \left[ Pe_s  \cos \theta + Pe_f \left(\tilde y+\tilde b(\alpha)\right) \right], 
\label{C122}
\end{equation}
where the first term in the brackets is due to swimmer self-propulsion  (involving  the polarization of swimmers along the $x$-axis) and the second term is due to particle advection by the flow. The net particle current, passing through the channel in $x$-direction relative to the lab frame, is obtained in dimensionless form as 
\begin{equation}
\tilde J = \int_{0}^{\tilde H} {\mathrm{d}} \tilde y\, \tilde j (\tilde y).
 \label{C125}
\end{equation}

\begin{figure}[t!]
\begin{center}
\sidesubfloat[]{%
\hskip0mm%
\includegraphics[width=0.75\textwidth]{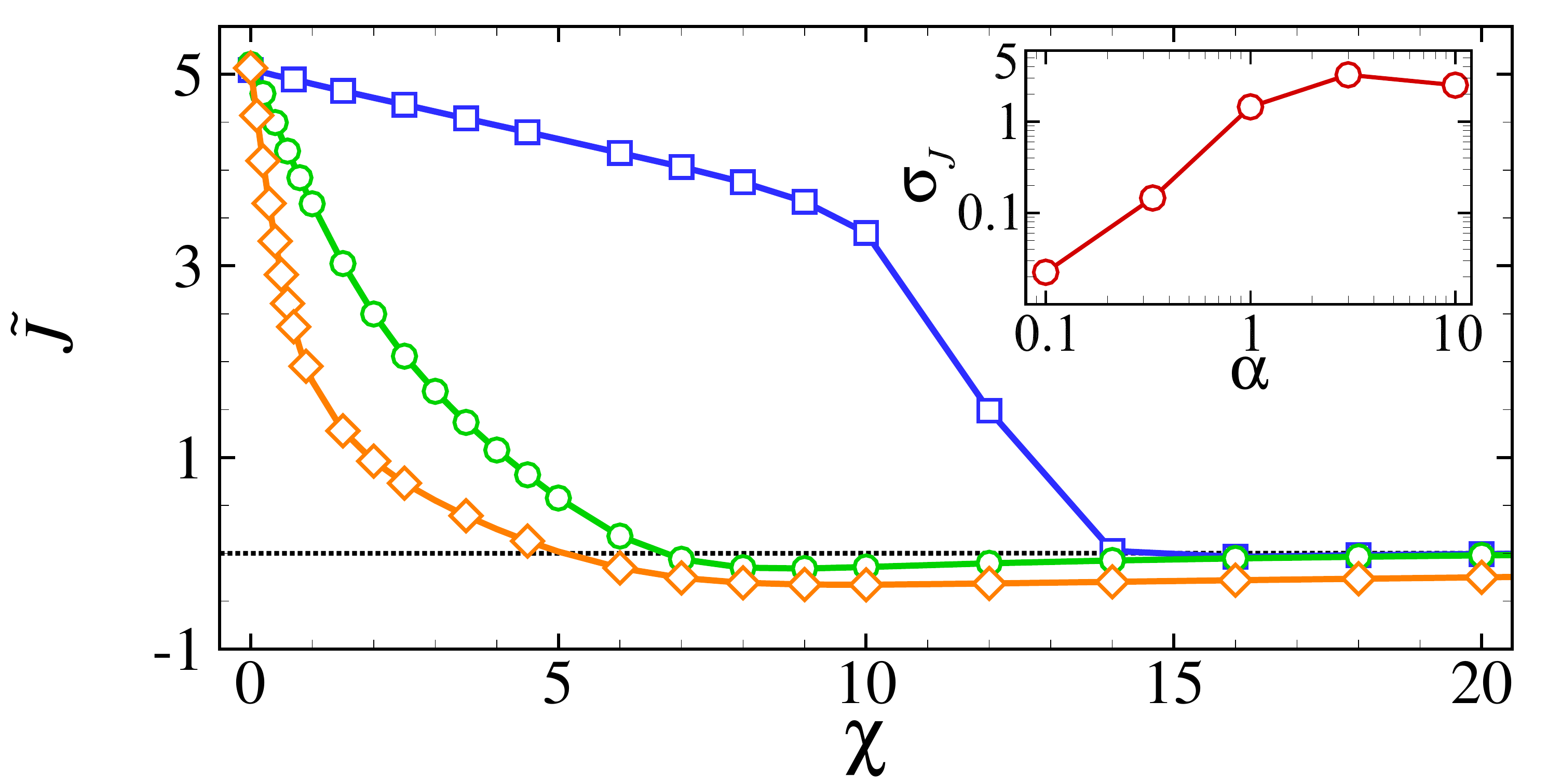}%
}%
\vskip0mm%
\sidesubfloat[]{%
\hskip0mm%
\includegraphics[width=0.75\textwidth]{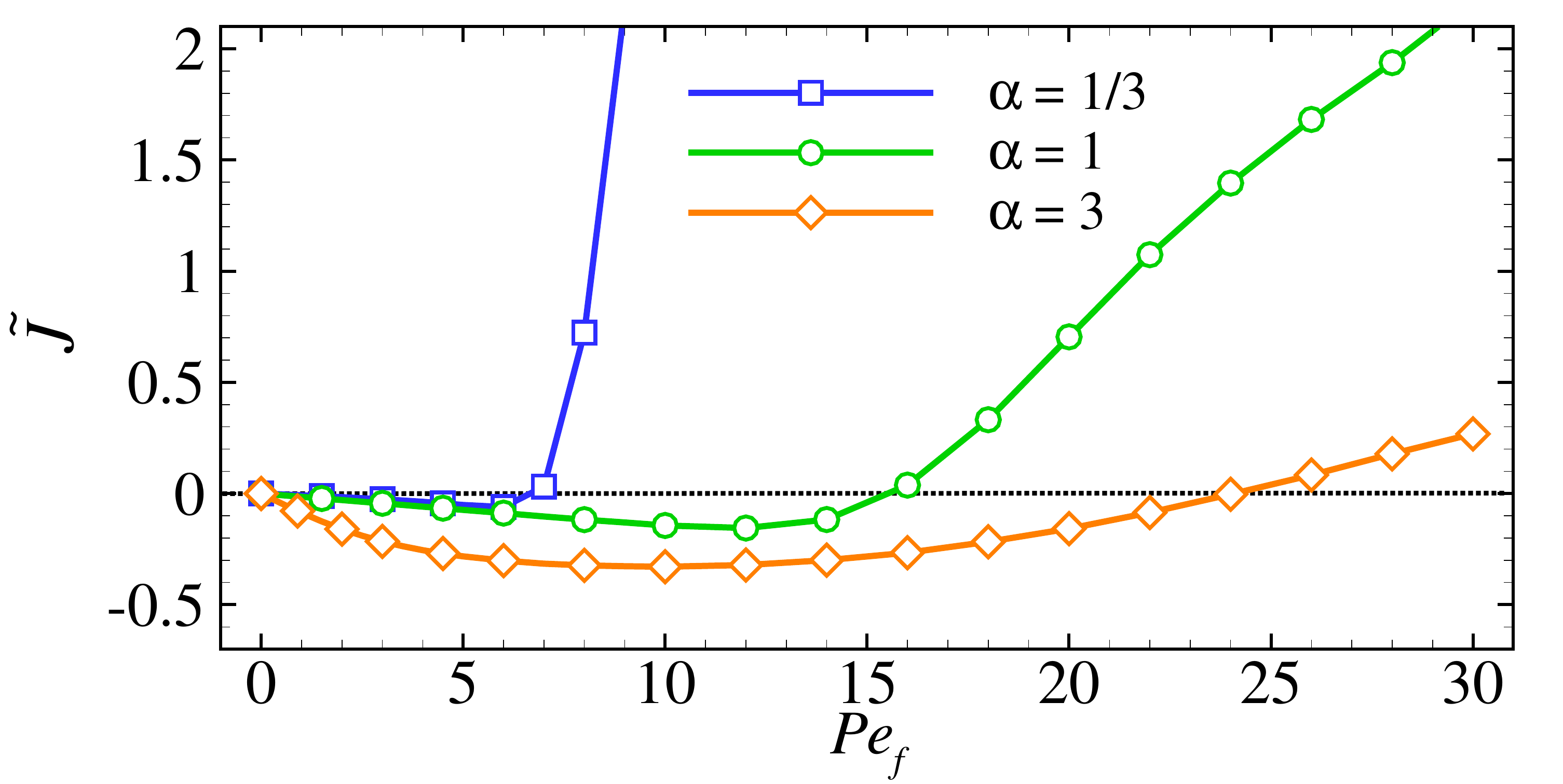}%
}%
\vskip-4mm\caption{(a) The net particle current relative to the laboratory frame (stationary bottom wall of the channel) as a function of $\chi$ for fixed  $Pe_f=10$, and (b) as a function of $Pe_f$ for fixed  $\chi=10$ and different aspect ratios,  as indicated in the legends in (b). In both cases, $Pe_s=50$ and $\tilde H=100$. Inset of panel a shows the linear particle-current response factor (slope of the curves) in the weak-field regime.}
\label{fig:j_profiles} 
\end{center}
\end{figure}

In Fig. \ref{fig:j_profiles}a, the typical behavior of this quantity as a function of the field coupling strength, $\chi$, is compared for prolate ($\alpha=3$), spherical ($\alpha=1$) and oblate ($\alpha=1/3$) swimmers  and at fixed $Pe_s=50$, $Pe_f=10$ and $\tilde H=100$. The net current $\tilde J$ is positive for sufficiently small $\chi$ and, for the parameter values chosen in the figure, it is nearly equal for different aspect ratios for vanishing $\chi$. In this regime, particle advection is dominant and the effective particle polarization along the channel is negligible.

As $\chi$ is increased, $\tilde J$ decreases due to the enhanced swimmer migration to the bottom half, where the advection contribution is minimized. The decrease in $\tilde J$ occurs linearly for small $\chi$; hence, the (negative) slope of the curves can be interpreted as a {\em linear particle-current response factor} $\sigma_J=-\delta \tilde J/\delta \chi$ for sufficiently small $\chi$. This quantity is shown in the inset of Fig. \ref{fig:j_profiles}a for different values of the aspect ratio at fixed $Pe_s=50$, $Pe_f=10$  and $\tilde H=100$. As seen, the current response factor of prolate swimmers can be larger than that of oblate swimmers by more than an order of magnitude. Since the external field is applied transversely (in $- y$ direction) and the net particle current is along the channel length (in $\pm x$ directions), the particle-current response factor is generally of a tensorial nature, whose off-diagonal components (e.g., $\sigma_J$ that gives the $xy$ component) are due to the coupling to the shear flow. 

At elevated field couplings, $\tilde J$ can become negative, indicating that swimmers display a net upstream particle flux even relative to the laboratory frame. In the figure, this occurs for $\chi\gtrsim 5$, 6.5 and 14 for $\alpha=3$, 1 and $1/3$, respectively. Clearly, as $\chi$ tends to infinity, one expects the swimmers accumulate strongly on the bottom wall  and take the downward-pointing orientation $\theta=-\pi/2$ because of strong field alignment, suppressing both  the self-propulsion and the advection contributions to the net current. This explains why the curves in Fig. \ref{fig:j_profiles}a display a weak nonmonotonic behavior, showing a global (negative) minimum for $\tilde J$ at some intermediate value of $\chi$ and approaching to zero for large $\chi$. This behavior is discernible in the figure for prolate and  spherical swimmers  and not for oblate swimmers that have a stronger preferential tendency to downward-pointing polarization. 

In Fig. \ref{fig:j_profiles}b, the typical behavior of $\tilde J$ is compared for the three mentioned aspect ratios  as a function of $Pe_f$ for fixed $Pe_s=50$, $\chi=10$ and $\tilde H=100$. The net current obviously vanishes when the shear strength is zero, $Pe_f=0$. In the opposite limit of large $Pe_f$, the net current increases linearly with  $Pe_f$ as the relative effects of  self-propulsion and external field  diminish. Again, in an intermediate regime of shear strengths, the net current becomes negative, being now discernible also for oblate swimmers. In all cases, $\tilde J$  attains a global minimum, whose depth is larger  for prolate swimmers. 

It is important to note that the regime, where the net current becomes negative does not necessarily coincide with any of the regimes identified based on swimmer migration behavior  in the previous sections. In general, to obtain a negative current, the flow strength ($Pe_f$) must be large enough to produce a sufficiently strong  upstream swimmer polarization (i.e., by giving the swimmers a most-probable orientation angle  sufficiently close to $\pi$), but not so large as to counteract (by its advective effect) the actual upstream particle flux produced by the swimmer self-propulsion. 

\subsection{Potential application to particle separation}
\label{subsec:separation_current}

\begin{figure}[t!]
\begin{center}
\includegraphics[width=0.8\textwidth]{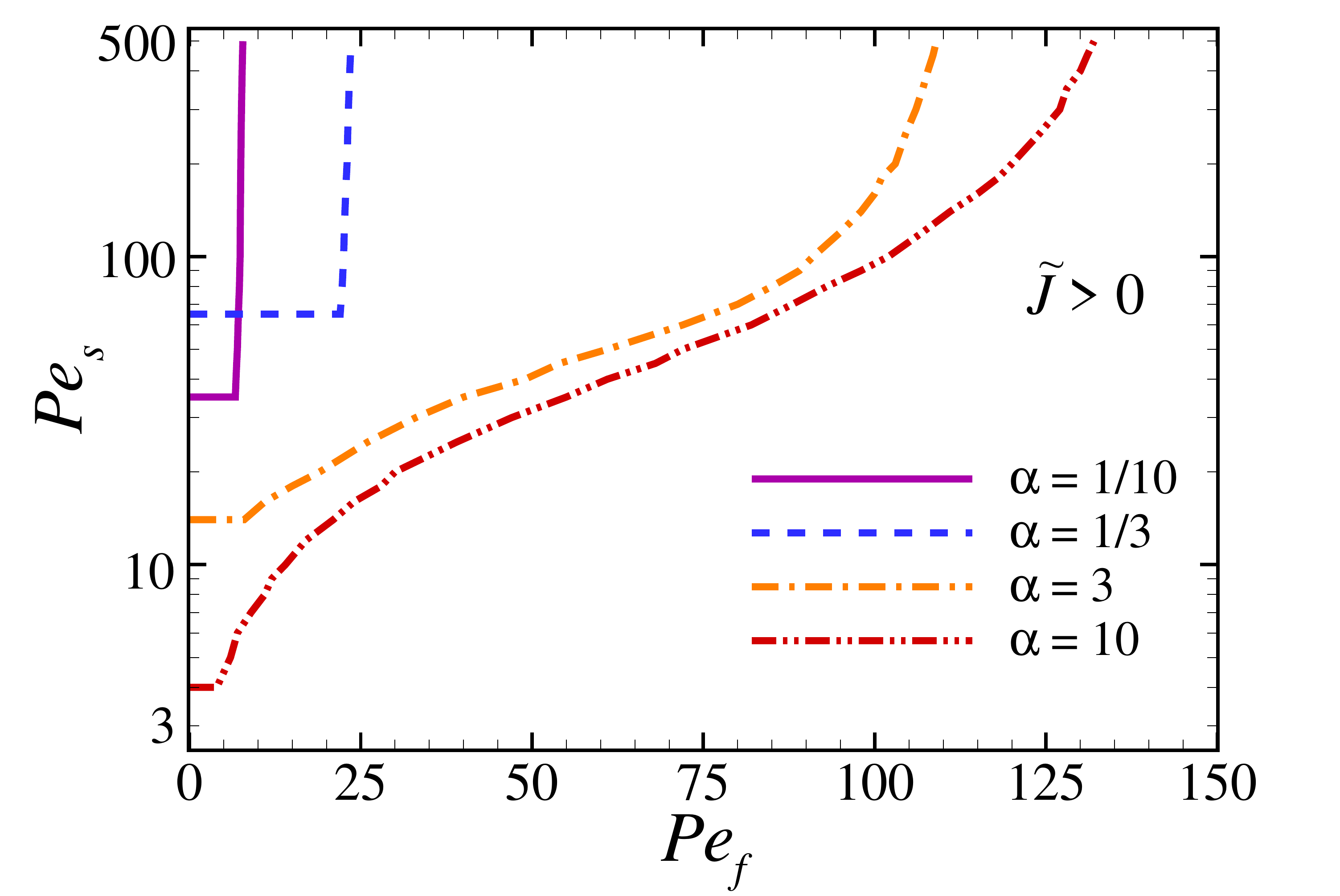}
\vskip-4mm\caption{The zero-current ($\tilde J=0$) boundary curves in the $(Pe_f,Pe_s)-$plane, separating the regimes of positive net particle current ($\tilde J>0$, the regions below and to the right) and negative net particle current   relative to the laboratory frame  ($\tilde J<0$, the regions above and to the left of each curve) for a few different aspect ratios, as indicated on the graph. Here, we have fixed $\tilde H=100$ and $\chi=30$.}
\label{fig:j_phase} 
\end{center}
\end{figure}

Being able to vary the system parameters to obtain positive and negative net currents provides another potential strategy to separate swimmers based on their aspect ratio and/or their self-propulsion strengths. By numerically scanning the parameter space for given particle aspect ratios, we can determine the regimes of positive ($\tilde J>0$) and negative ($\tilde J<0$)  net current relative to the laboratory frame. An example is shown in Fig. \ref{fig:j_phase}, where we have fixed $\chi=30$ and $\tilde H=100$ to enable positive and negative net currents for swimmers of widely varying aspect ratios; here the zero-current separating boundaries are shown for the cases with $\alpha=10$ (rodlike swimmers) and $1/10$ (disklike swimmers) as well as the more realistic values of $\alpha=3$ and $1/3$. In each case, the region above  (below) and to the left (right) side of the boundaries represents the regime of parameters, where we have $\tilde J<0$ ($\tilde J>0$); at this particular set of parameters, spherical swimmers represent an exceptional case for which $\tilde J$ always remains positive. 

For sufficiently large values of $Pe_s$, swimmers of differing aspect ratio show net upstream current, $\tilde J<0$, at small $Pe_f$ (note that for the strict case of $Pe_f=0$, we have  $\tilde J=0$ in all cases). Thus, in a mixture of such swimmers, our results in Fig. \ref{fig:j_phase} propose that, by gradually increasing the shear strength,  the dislike swimmers with $\alpha=1/10$  are the first to assume a net {\em downstream} current, $\tilde J>0$, as $Pe_f$ is increased beyond the threshold indicated by the  purple solid curve, while swimmers of other aspect ratios still hold a net upstream current. This is true until $Pe_f$ is increased beyond the next threshold shown by the blue dashed curve, where oblate swimmers with $\alpha=1/3$ exhibit $\tilde J>0$. It is interesting to note that the threshold curves ($\tilde J=0$) for oblate swimmers are nearly independent of $Pe_s$, a behavior that is similar to the one mentioned for the threshold $Pe_f$ in Section \ref{subsubsec:summary_oblates}. Further increasing $Pe_f$, one can cross the threshold curves corresponding to prolate swimmers. This sequential change of sign in the net current relative to the lab frame can be utilized to separate  swimmers of different aspect ratios in suitably designed nano-/microfluidic devices. The foregoing procedure can be reversed  starting from a large $Pe_f$, where  $\tilde J>0$ for all aspect ratios; then the sign of  $\tilde J$ can  be reversed by decreasing $Pe_f$, in which case rodlike swimmers with $\alpha=10$ and then prolate swimmers with $\alpha=3$ are the first ones to give {\em upstream} currents. Clearly, if $Pe_s$ is smaller than around 30 in Fig. \ref{fig:j_phase}, oblate swimmers always show  $\tilde J>0$, and the noted sign change cannot be used to separate them from the other types of swimmers. 

Figure \ref{fig:j_phase} can also be used to separate swimmers of equal aspect ratio and different self-propulsion strengths, specifically, in binary mixtures of swimmers with two distinct value of $Pe_s$ that lie below and above their corresponding zero-current boundary, giving net positive and negative particle currents, respectively.

\section{Summary}
\label{sec:summary}

We study the steady-state behavior of magnetic prolate/oblate swimmers in a planar channel subjected to an imposed Couette flow and a transversally applied (`downward')  magnetic field. The swimmers are modeled as active Brownian spheroids, self-propelling at constant speed along their symmetry axis (major/minor body axis for prolate/oblate particles), also taken as their axis of permanent magnetic dipole moment, pointing in the swim direction. The system is studied using a numerical continuum approach based on a noninteracting Smoluchowski equation on the joint position-orientation probability distribution function of swimmers in the channel.

We show that the competitive/cooperative nature of the shear-/field-induced torques exerted on swimmers, and the tendency of swimmers to accumulate with normal orientation at the channel walls, lead to a host of interesting effects  in their cross-stream migration, not explored before for this system. We uncover different regimes of behavior by identifying distinct orientational subpopulations of swimmers (created by the imposed shear), whose distinct responses to external field explain the salient aspects of the migration regimes of the overall swimmer population. We also elucidate  major differences that emerge between prolate/oblate swimmers.

For prolate swimmers, the regime of vanishing or weak field portrays two orientationally distinct majority/minority subpopulations in each channel half. This {\em standard bimodality} regime \cite{Nili} involves a majority subpopulation in the top (bottom) channel half swimming in down- (up-) stream direction and  a corresponding minority subpopulation swimming in the opposite direction.  As the field is strengthened, swimmers partially migrate to the bottom  half of the channel through a well-defined sequence of subpopulation changes: The top-half (downstream) majority subpopulation is the first to diminish in response to the applied field, creating a regime of {\em reverse bimodality},  where the downstream population becomes smaller than the upstream one in the top half of the channel, followed by a regime of {\em full migration}, where the remaining top-half upstream subpopulation also migrates to the bottom half. The swimmers still show a partially bimodal distribution in the full migration regime due to a {\em minority persistence} effect, which is eliminated only in stronger fields, where all  swimmers are focused in a single, highly polarized,  upstream population at the bottom wall, giving a final regime of orientational {\em unimodality}.

For oblate swimmers, the bimodality manifests itself in the form of two predominantly up/downward-pointing populations developed across the central regions of the channel. Each of these populations  originates as a continuous extension of a single population of wall-accumulated,  oppositely and normal-to-wall-pointing, swimmers. Thus, as one moves away from the walls into the channel,  the most-probable orientation of oblate swimmers flips in  clockwise and downward (upward) direction from its typical upward (downward) orientation at the top (bottom) wall. The field-induced migration occurs here in a  {\em nonmonotonic} fashion, as  a macroscopic fraction of oblate swimmers migrate back to the top channel half in a certain, intermediate, regime of field strengths. This peculiar {\em reverse migration} behavior is superseded by the full swimmer migration  to the bottom channel half at larger field strengths. The reverse migration arises due to a {\em double-pinning} of the swim orientation at two disparate angles, one  in the second and the other in the third angular quadrant  (see the supplementary material). This leads to two physically discernible subprocesses, with the whole upward- (downward-) pointing swimmer population migrating to the top (bottom) half of the channel, creating strong midchannel depletion. In the anomalous (`upward') subprocess,  swim orientation is  {\em pinned} at  an acute angle relative to the field axis, with the angle varying  from a value close to $\pi/2$ (reverse migration opposing the field) up to $\pi$ (upstream normal-to-field migration) as the field strength is increased within the reverse migration regime. 

For both prolate and oblate swimmers, the partial cross-stream migration in weak fields turns out to be a linear process, explained based on the field-induced modifications in Jeffery oscillations of the swim orientation  (see the supplementary material), and quantified using a {\em linear-response factor}, whose peculiar dependencies  on system parameters are thoroughly explored (Appendix \ref{sec:weak_field}). In strong fields, we determine the regimes of parameters, where  a net upstream particle current occurs in the channel relative to the laboratory frame. Using these regimes, or those found for the full migration of swimmers to the bottom half of the channel, we propose different strategies to separate swimmers of different aspect ratios or motility strengths  in microfluidic setups. 

\section{Discussion}
\label{sec:discussion}

Our analysis of the response of sheared dipolar swimmers to an external field can be relevant to both synthetic magnetic swimmers \cite{Martin2014,Carlsen2014,Magdanz2013,Santon2017,Zhao2012,Schattling2017,Dreyfus2005, Tierno2009, Benkoski2011, Pak2011, Vach2015,Zhang2009_1, Zhang2009_2,Ogrin2008,KlumppReview2019,Kokot_2017} as well as magnetotactic bacteria,  such as  {\em Magnetococcus marinus} (MC-1 strain) and {\em Magnetospirillum  gryphiswaldense} (MSR-1 strain). Of these particular types of bacteria,  the former  shows dipolar magnetotactic behaviors, while the latter shows more complex dipolar/axial behaviors \cite{KlumppReview2019,Lefevre2014}. Their  north-seeking strains can bias their flagellar motion in oxic media to swim in the direction of their magnetic moment, which is passively aligned with the external magnetic field  \cite{Schleifer1991, Frankel1997, FaivreReview2008, Bennet2014, Lefevre2014,Klummp2014BJ, Klumpp2016, KlumppReview2019}; a property that makes these strains more relevant to our model (see Appendix \ref{app:parameters} for a discussion of applicable parameter values). 

Even though our study  specifically focuses  on passive orientational response of magnetic swimmers based on a field-induced angular velocity, Eq. \eqref{eq:w_ext0}, that comes from a standard magnetic torque, cross-product  angular velocities of the type \eqref{eq:w_ext0} have frequently been used in similar contexts.   For instance, gyrotactic swimmers, such as the bottom-heavy {\em Chlamydomonas nivalis}, having an offset between their centers of gravity and buoyancy, align themselves to swim against the direction of  the gravitational field; an effect that can directly be modeled via Eq. \eqref{eq:w_ext0} by invoking a negative gravity-dependent field coupling $\chi<0$; see Refs.  \cite{Kessler1984, Kessler1985, Kessler1986b,Pedley1988,Pedley1990,PedleyReview1992,Pedley2015}  and Appendix \ref{app:parameters}. In this example, fixed swim orientation due to a torque balance in a shear flow (resembling the orientational pinning discussed in the present work) has been noted in Ref. \cite{PedleyReview1992}. Also, while our focus is on the passive type of orientational response, phenomenological cross-product reorientation rates as in Eq. \eqref{eq:w_ext0} have been used to model active response to external stimuli. These include both klinotactic responses (e.g., chemotaxis of sperm    \cite{JulicherSpermChemo,JulicherPRL,Lushi} and phototaxis of {\em C. reinhardtii} \cite{Bennett_Chlamy}), that are more relevant to the present context, and  klinokinetic reponses (e.g.,  chemotaxis of {\em E. coli}) \cite{Taktikos,Liebchen_2016}. Our analysis may thus be relevant to orientational responses to other types of external fields/stimuli. In comparing with the foregoing examples, however, caution must be exercised by noting that the characteristic helical motion of klinotactic swimmers can be captured only in a three-dimensional implementation of the model, and that the proposed forms of the phenomenological cross-product angular velocities are available  mainly for the case of still fluids. 

The continuum approach used here applies only to dilute suspensions  of active particles \cite{Ezhilan, Nili,Saintillan2013_ComptesRendus, Saintillan2015}, where the constituent particles can be modeled effectively as pointlike objects, whose shape and size enter only through the aspect-ratio dependence of their translational/rotational diffusivities and the corresponding shear-/field-induced torques. It is nevertheless  important to examine the effects due to steric and hydrodynamic  (see, e.g., Refs. \cite{Saintillan2013_ComptesRendus, Hernandez-Ortiz1, Meng2018,Potomkin_2017,Lauga:RPP009,goldstein_review,Lauga:ANNREVF2016,Zottl2014,DrescherGoldstein,SaintillanPRL2008,SaintillanPF2008,LushiPRE2012,dunkel2013, lushi2014fluid,Reinken2018,Reinken2019}) as well as magnetic   interactions (see, e.g., Refs. \cite{Klumpp2016,Meng2018,Kokot_2017}) between swimmers in semidilute or dense suspensions.  These interactions can produce additional components in the angular velocity experienced by individual swimmers and can  thus modify  the rotational dynamics of the swim orientation and  the migration patterns of swimmers, when incorporated in a setting similar to the one considered here. Also, of particular relevance to hydrodynamically active particles (e.g., pushers and pullers) are  interparticle interactions induced by active stresses and internal flows that can be a source of instability  and  mesoscale turbulence in concentrated swimmer suspensions (see Refs. \cite{Lowen:EPJST2016,SaintillanPRL2008,SaintillanPF2008,LushiPRE2012,lushi2014fluid,Lushi,Garg2018,Alonso-Matilla2019,Saintillan2013_ComptesRendus, Saintillan2015,SaintillanPRL2008,SaintillanPF2008,dunkel2013,Reinken2018,Reinken2019} and references therein). These phenomena can be studied using an extended continuum approach  \cite{Saintillan2013_ComptesRendus}, but how such instabilities may develop in the presence of externally applied torques, as applicable to the present problem, is a question that remains to be addressed.

In dilute suspensions (of, e.g., volume fractions below 0.1; see, e.g., Refs. \cite{LushiPRE2012,Lowen:EPJST2016}), the interactions mentioned above will be relatively short-ranged, given that steric interactions play a role typically when swimmers come into close contact, and hydrodynamic and magnetic interactions decay rapidly with the interparticle distance, e.g., with its inverse third power, when swimmers are modeled as pointlike hydrodynamic force- and magnetic dipoles. Furthermore, in the shear-/field-dominant regimes of behavior explored here, the ensuing interparticle torques are expected to be masked largely by the externally applied torques. The important exceptions are the loci within the parameter space, where the imposed shear and magnetic field  counteract one another and their contributing torques on active particles sum up to zero, brining any  subleading contributions back into play. The pinning thresholds and the swim-orientation fixed points, and primarily the unstable ones, may thus be prone to possible alteration by interparticle interactions. 

Other interesting factors that can be explored within the present context include  swimmer chirality \cite{Teeffelen:PRE2008, Fily2012, Reichhardt:2013, Xue:EPJST2014, Xue:EPL2015, Mijalkov:2015, Wykes_2016, Lowen:EPJST2016,Liebchen_2016,QuinckeRotor2019} and steric/hydrodynamic interactions between swimmers and  the channel walls \cite{Hill_Chirality_Upstream,Chilukuri,wallattraction, wallattraction2,wallattraction3, ardekani, Mathijssen:2016a, Mathijssen:2016b, Mathijssen:2016c,Schaar2015_PRL, ZottlPRL2012,Das_Nature_2015,Uspal_SM_2015,Uspal,KaturieaaoSciAdv,ZottlPRL2012, ZottlPoiseuille2013}. A  more detailed modeling of the near-wall behavior would be desirable  especially for realistically modeled swimmers. The qualitative features of our results may nevertheless remain applicable in such cases, given that our analysis mainly builds on globally defined criteria and quantities (e.g., integrated number of swimmers in an entire channel half), minimizing possible artifacts due to our specific modeling of the near-wall regions. 

\section{supplementary material}
See the supplementary material for an analysis of the deterministic dynamics of the swim orientation in our model, including the origin of linear response in the cross-stream migration of swimmers in weak fields and their orientational pinning in strong fields. 

\section{Author contributions}
M.R.Sh. performed the theoretical derivations, developed the numerical codes and generated the output data and their plots. A.N. conceived the study, supervised the research and wrote the manuscript. Both authors analyzed the results, contributed to the discussions, and reviewed the manuscript. 

\section{Acknowledgements}
A.N. acknowledges partial support from the Associateship Scheme of The Abdus Salam International Centre for Theoretical Physics (Trieste, Italy). We acknowledge useful discussions  with J. Abazari and M. Kheyri. 

\section{Conflicts of interest}
There are no conflicts of interest to declare. 

\appendix

\section{Expressions for $\Delta_\parallel(\alpha)$, $\Delta_\perp(\alpha)$ and $\Delta_R(\alpha)$}
\label{app:diffusivities}

The shape functions $\Delta_\parallel(\alpha)$, $\Delta_\perp(\alpha)$ and $\Delta_R(\alpha)$ can be expressed  as \cite{Perrin,koenig,kim_microhydrodynamics}
\begin{equation}
\Delta_{\parallel,\perp}(\alpha) =\alpha^{1/3}G_{\parallel,\perp}^{-1}(\alpha),\quad \Delta_{R}(\alpha) =G_R^{-1}(\alpha), 
\end{equation}
with $G_{\parallel,\perp}(\alpha)$ and $G_R(\alpha)$ being as follows.  For prolate spheroids $(\alpha>1)$, 
\begin{eqnarray}
&&G_\parallel
= 
\frac{8}{3} 
\Bigg[
\frac{-2\alpha}{\alpha^2-1}
+\frac{2\alpha^2-1}{\left(\alpha^2-1\right)^{3/2}} \ln{ \left( \frac{\alpha+\sqrt{\alpha^2-1}}{\alpha-\sqrt{\alpha^2-1}} \right) }
\Bigg]^{-1},
 \nonumber\\
&&G_\perp
=
\frac{8}{3} 
\Bigg[
\frac{\alpha}{\alpha^2-1}
+\frac{2\alpha^2-3}{\left(\alpha^2-1\right)^{3/2}} \ln{ \left( {\alpha+\sqrt{\alpha^2-1}}\right) }
\Bigg]^{-1},
 \nonumber\\
&&G_R =
\frac{2}{3} \frac{\alpha^4-1}{\alpha}
\Bigg[
\frac{2\alpha^2-1}{\sqrt{\alpha^2-1}} \ln\left({\alpha+\sqrt{\alpha^2-1}}\right)-\alpha
\Bigg]^{-1}\!\!\!\!\!\!,
\end{eqnarray}
while, for oblate spheroids $(\alpha<1)$, 
\begin{eqnarray}
&&\!\!\!G_\parallel
= 
\frac{8}{3} 
\Bigg[
\frac{-2\alpha}{\alpha^2-1}
-\frac{2\left(2\alpha^2-1\right)}{\left(1-\alpha^2\right)^{3/2}} \tan^{-1}{ \left(\frac{\sqrt{1-\alpha^2}}{\alpha} \right) }
\Bigg]^{-1},
 \nonumber\\
&&\!\!\!G_\perp
=
\frac{8}{3} 
\Bigg[
\frac{\alpha}{\alpha^2-1}
-\frac{2\alpha^2-3}{\left(1-\alpha^2\right)^{3/2}} \sin^{-1}{\sqrt{1-\alpha^2}   }
\Bigg]^{-1},
 \nonumber\\
&&\!\!\!G_R =
\frac{2}{3} \frac{\alpha^4-1}{\alpha}
\Bigg[
\frac{2\alpha^2-1}{\sqrt{1-\alpha^2}} \tan^{-1}{ \left(\frac{\sqrt{1-\alpha^2}}{\alpha} \right) }-\alpha
\Bigg]^{-1}\!\!\!\!\!\!\!.
\end{eqnarray}

\section{Choice of parameter values}
\label{app:parameters}

The typical choices of dimensionless parameter values in our study ($Pe_s=0-500$, $Pe_f=0-150$, $\chi=0-100$, $\tilde H=20-100$) fall within experimentally realizable ranges of parameters for synthetic/biological swimmers. Before giving a few such examples, we emphasize that our model is aimed at generic aspects of swimmer motion and overlooks swimmer-specific (shape/motility) features, which need to be included or assessed before quantitative comparison with any specific real system is attempted. 

For {\em E. coli} with the swim speed $V_s \simeq 22.3\, \mu{\mathrm{m}}\cdot{\mathrm{s}}^{-1}$ (see Ref. \cite{KAYA2012BJ}), the mean aspect ratio $\alpha\simeq  2.5$ and the major axis length $\simeq 2.1\,\mu{\mathrm{m}}$ (see Ref. \cite{Kaya2009PRL}), we find a reference sphere radius of $R_{\mathrm{eff}} \simeq  0.57 \, \mu{\mathrm{m}}$ and Stokes rotational diffusivity of $D_{0R} \simeq 0.88\,{\mathrm{s}}^{-1}$ and, hence, $Pe_s \simeq 44.5$. In this case, our choices of values for $Pe_f$ and $\tilde H$ are mapped to shear rates of $\dot\gamma\simeq0- 132\,{\mathrm{s}}^{-1}$, and actual channel widths of  $H\simeq 11.4- 57\, \mu$m, respectively. See Ref. \cite{Rusconi2014} for other experimental examples of motile bacteria in shear flow. 

For magnetotactic bacteria such as {\em M. gryphiswaldense} MSR-1, one has  $V_s \simeq  23\, \mu{\mathrm{m}}\cdot{\mathrm{s}}^{-1}$ and a mean aspect ratio of around $\alpha\simeq 4$  (see Refs. \cite{Klumpp2016,Klummp2014BJ}), whence we find $R_{\mathrm{eff}} \simeq  0.68\, \mu$m, $D_{0R} \simeq 0.57\,{\mathrm{s}}^{-1}$ and, thus, $Pe_s \simeq  59.3$. This type of bacteria contains around 20 magnetosomes of iron-oxide magnetite and a net magnetic dipole moment of $m \simeq 6.2 \times 10^{-16}$A\,m$^2$  (see Refs. \cite{Klumpp2016,Klummp2014BJ}). For the magnetic field of the earth ($B \simeq  50\, \mu$T), we have a field coupling strength of $\chi\simeq  7.6 $ at room temperature ($T=300$~K). In this case, the corresponding shear rates and actual channel widths follow as $\dot\gamma\simeq 0- 85\,{\mathrm{s}}^{-1}$ and $H\simeq 13.6- 68\, \mu$m, respectively. 

For gyrotactic microorganisms such as {\em C. nivalis}, one has a typical cell volume of ${\cal V}_0 \simeq 500 \, \mu{\mathrm{m}}^3$ (hence, $R_{\mathrm{eff}} \simeq 4.9 \, \mu{\mathrm{m}}$),  aspect ratio of $\alpha \simeq 1.38$ ($\beta \simeq 0.31$), cell mass density of $\rho_{\mathrm{cell}} \simeq 1.05 \rho_{\mathrm{water}}$, and an offset $h \simeq 0.1 \, \mu{\mathrm{m}}$ between centers of gravity and buoyancy \cite{Kessler1984, Kessler1986b, Pedley1988}. In this case, the field coupling parameter can be derived as $\chi = -{4 \pi  \rho_{\mathrm{cell}}\, g hR_{\mathrm{eff}}^3}/{(3 k_{\mathrm{B}} T)} \simeq - 12.8$.
These particles have swim speeds of about $V_s \simeq 100 \, \mu{\mathrm{m}}\cdot{\mathrm{s}}^{-1}$ in still water  and a typical sedimentation speed of $V_g \simeq 3 \, \mu{\mathrm{m}}\cdot{\mathrm{s}}^{-1}$  (see Refs. \cite{Kessler1984, Kessler1986b}); since $V_g\ll V_s$, the total speed of these swimmers can be assumed to be nearly equal to $V_s$. Their typical rotational diffusion is rather small, $D_{0R} \simeq  1.39\times 10^{-3}\,{\mathrm{s}}^{-1}$, giving an unusually large swim P\'eclet number $Pe_s \simeq 1.47\times 10^5$. The shear rates and channel widths corresponding to our choices of dimensionless parameters follow as $\dot\gamma\simeq 0- 0.21\,{\mathrm{s}}^{-1}$ and $H\simeq 98-490\, \mu$m, respectively. Evidently, gyrotactic swimmers, due to their large size and self-propulsion strength, provide limiting examples, with $Pe_f$ and $\chi$ values deep into the orientational pinning regime. 

For human sperm cells, one has $ V_s \simeq  40.2\, \mu{\mathrm{m}}\cdot{\mathrm{s}}^{-1}$ and head dimensions of $4.5 \times 2.8 \times 1.1 \, \mu$m$^3$  (see Ref. \cite{smith2009}), giving   $R_{\mathrm{eff}} \simeq  1.2 \, \mu m$, $D_{0R} \simeq  0.094\,{\mathrm{s}}^{-1}$ and $Pe_s \simeq  356$. The corresponding shear rates and  channel widths follow as $\dot\gamma\simeq 0- 14.1\,{\mathrm{s}}^{-1}$ and $H\simeq 24- 120\, \mu$m, respectively. While being a motivating example, one should bear in mind that our model may not be applied to sperm cells without accounting for their extended tail that can cause additional hydrodynamic effects. Sperm cells actively align themselves with the direction of Ca$^{2+}$ gradient from the source (egg cell) by changing the local curvature of their tail \cite{JulicherSpermChemo,Alvarez,JikeliSpermHelical}. This creates a klinotactic response based on temporal sensing, modeled with the  angular velocity $|{\omega}_{\mathrm{ext}}| =  (150 \,\mu {\mathrm{m}}\cdot{\mathrm{s}}^{-1}) |{\nabla C}|/{C}$ for small deviations from the gradient direction \cite{JulicherSpermChemo,JulicherPRL}.  
Since we keep $\chi$ below 100, we find  $C/|{\nabla C}| \gtrsim 16\,\mu$m. This indicates that the estimated decay length of the concentration field is within the experimentally relevant range. For sea urchin spermatozoa, one finds $C \sim 1 \,$pM $- 10 \, \mu$M and the maximum lengthscale over which sperm cells can sense the chemical gradient is around 5~mm  (see Ref. \cite{Kashikar2012}). 

\section{Quantifying the linear response}
\label{sec:weak_field}

As noted in the text, cross-stream migration of swimmers to the bottom half of the channel in response to the downward external field shows an extended regime of linear response in sufficiently weak fields, where the bottom-half swimmer fraction, $n_b$, varies linearly with the field coupling strength, $\chi$. To quantify this behavior,  we use the difference, $\delta n_b(\chi)=n_b(\chi)-n_b(0)$, of the bottom-half swimmer fraction from its base value,  $n_b(0)=1/2$, to define the {\em linear migration response factor}  
\begin{equation}
  \sigma_\chi = \frac{\delta n_b(\chi)}{\chi},  
\end{equation}
which is generally a function of  $Pe_s, Pe_f, \alpha$ and $\tilde H$, but {\em not} of $\chi$, as  $\chi$ can be chosen as any (computationally suitable) value within the  linear-response regimes identified in the text ($ \sigma_\chi$ gives the slope of the linear segment of the curves shown in, e.g., Figs. \ref{fig_contours_full_r3}e and \ref{fig_contours_full_r033}e, in weak fields). Unless otherwise mentioned, we take $\chi=0.5$ and 5 for prolate and oblate swimmers, respectively.  

\begin{figure}[t!]
\includegraphics[width=0.405\linewidth]{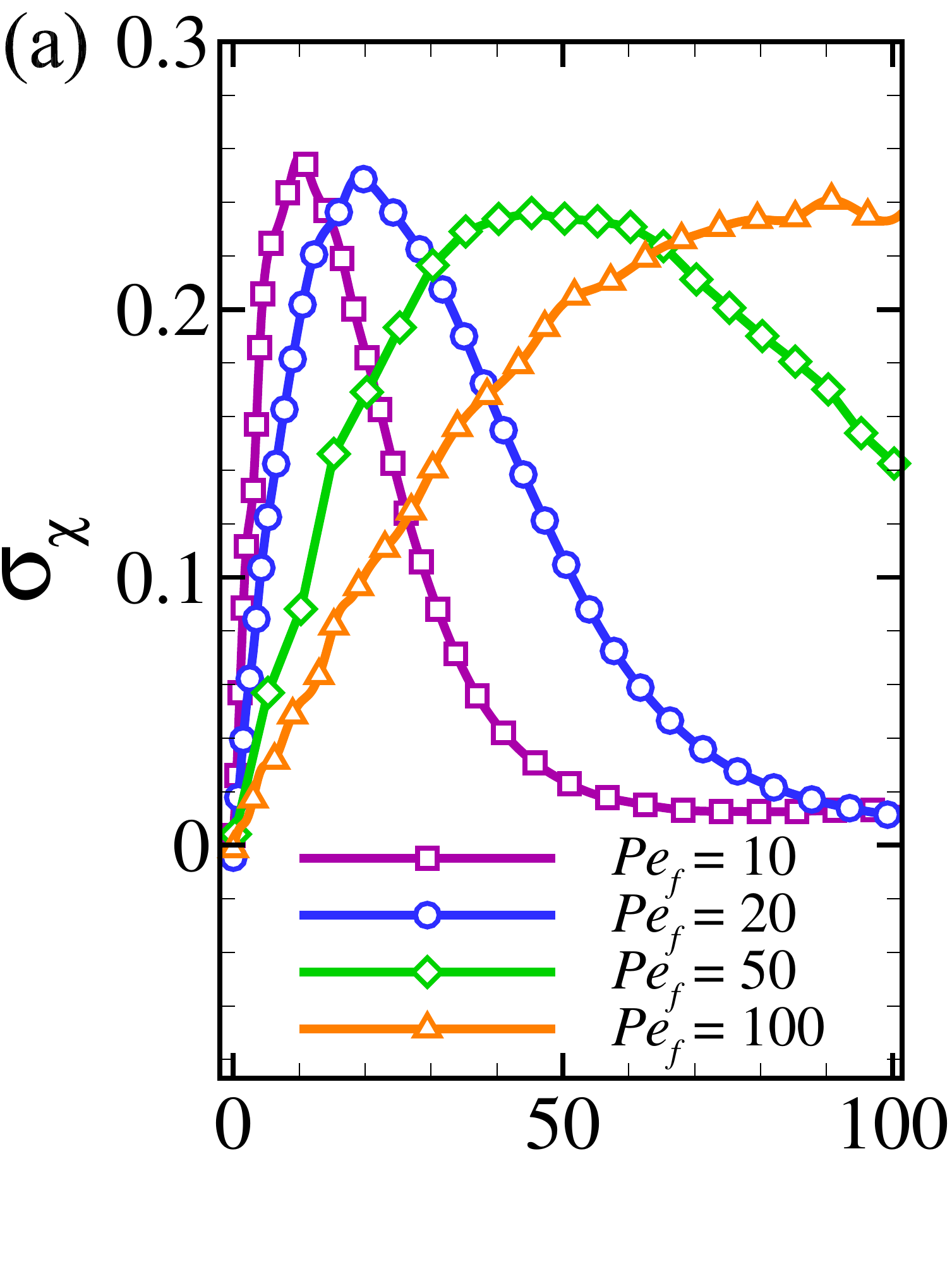}%
\includegraphics[width=0.405\linewidth]{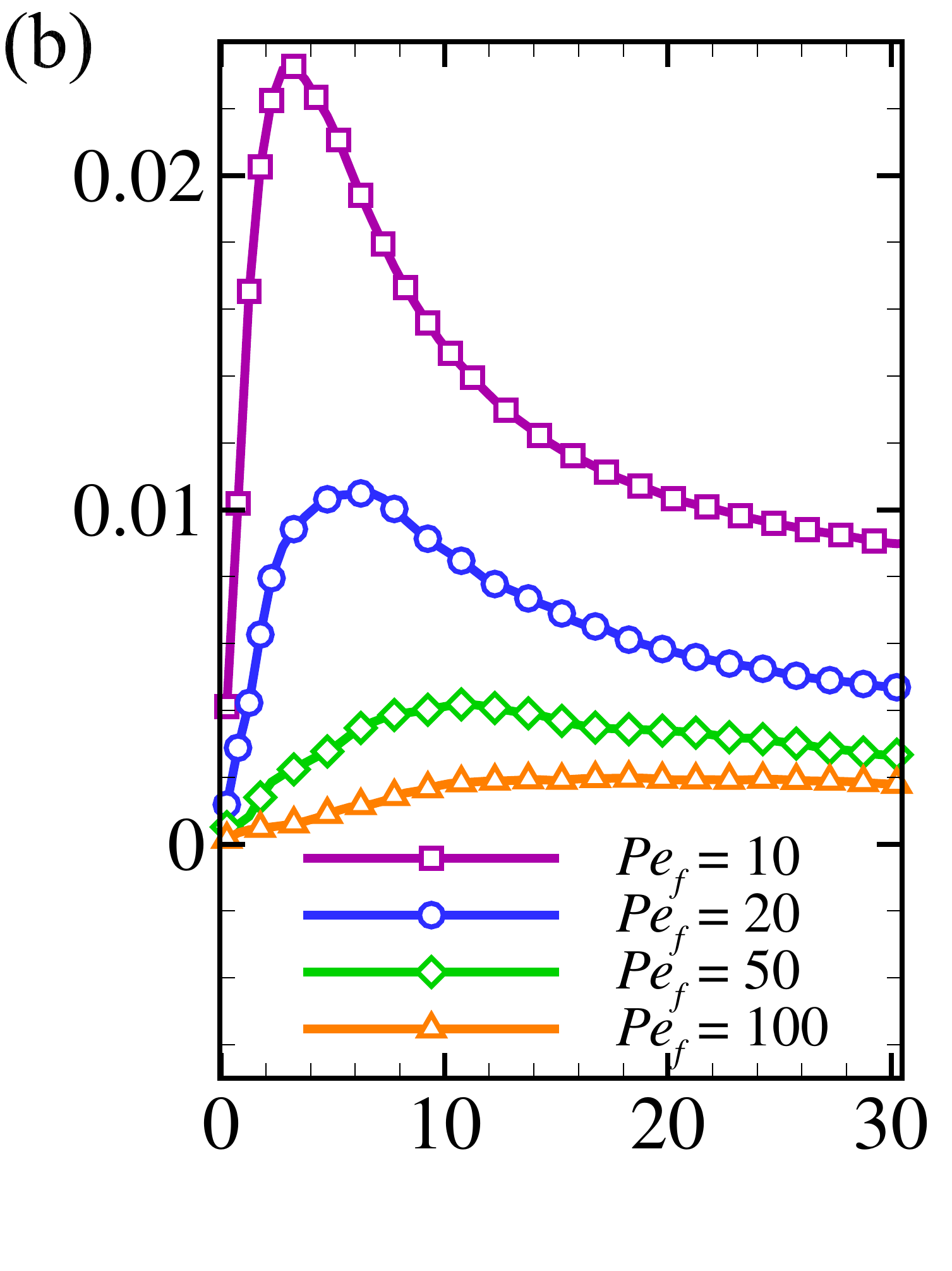}\vskip-4mm
\includegraphics[width=0.405\linewidth]{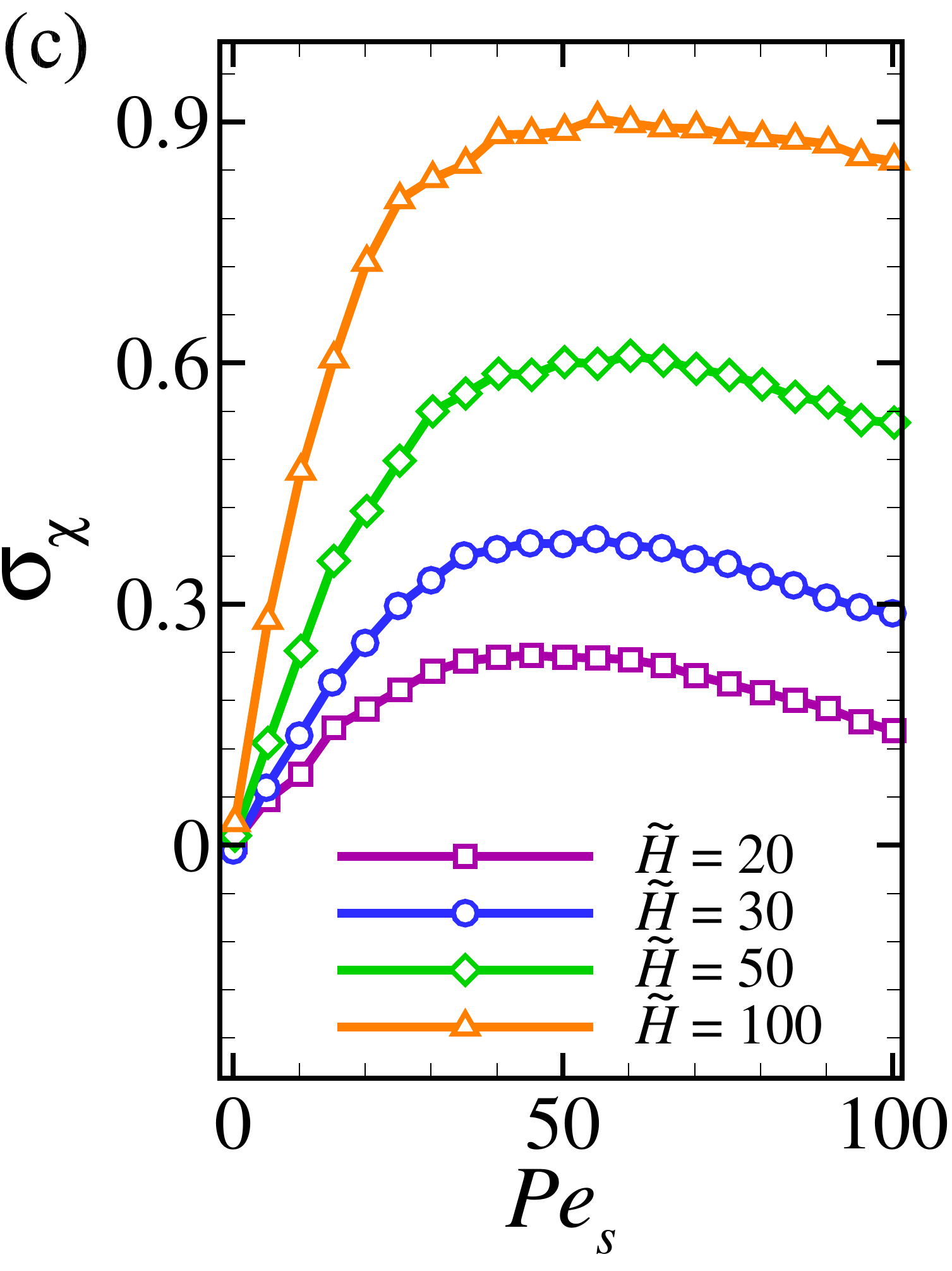}%
\includegraphics[width=0.405\linewidth]{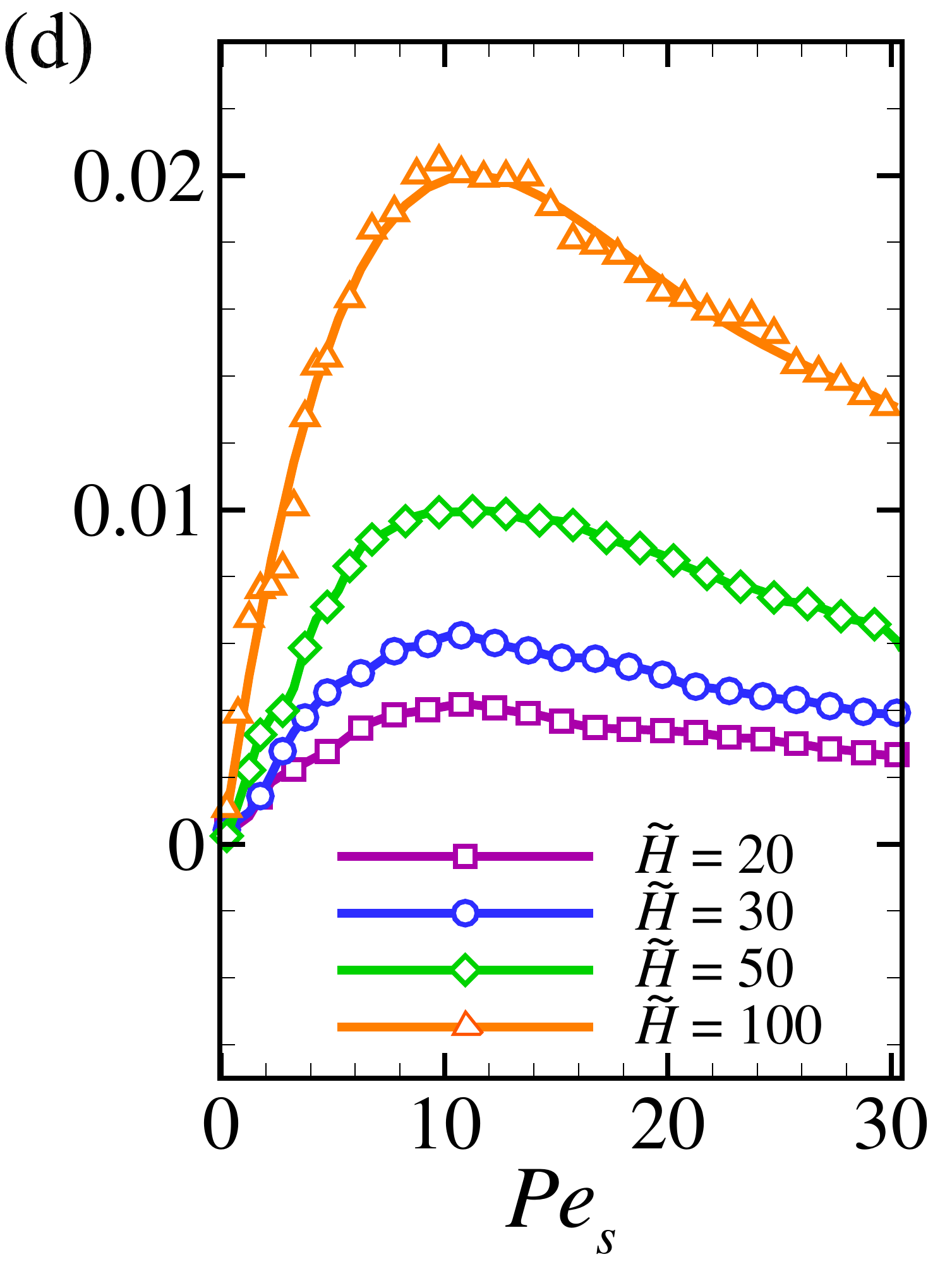}%
\vskip-2mm
\caption{Panels a and b show the linear migration response factors, $\sigma_\chi$, of prolate and oblate swimmers of aspect ratios $\alpha=3$ and $1/3$, respectively, as functions of $Pe_s$ for different values of $Pe_f$ at fixed $\tilde H=20$. Panels c and d show the same quantity for different values of $\tilde H$ at fixed $Pe_f=50$ for   $\alpha=3$ and $1/3$, respectively. In all cases, symbols/curves represent the computed data, except for the orange curve in d, which is a fit drawn to guide the eye for the data set with $\tilde H=100$ that displays more apparent numerical errors. 
}
\label{fig:response}
\end{figure}

Figure \ref{fig:response} shows the migration response factor for different values of the system parameters  at fixed $\alpha=3$ (panels a and c) and $\alpha=1/3$ (panels b and d). The response factor vanishes for vanishing swim P\'eclet number, $Pe_s$, in accordance with the fact that non-active particles do not migrate across the channel width in response to the external field. It also diminishes at sufficiently large $Pe_s$, when $Pe_f$ and $\chi$ are fixed, indicative of the fact that swimmers of sufficiently high motility strength are predominantly accumulated on the walls, showing negligible influence to imposed shear and/or applied external field.

\begin{figure}[t!]
\includegraphics[width=0.405\linewidth]{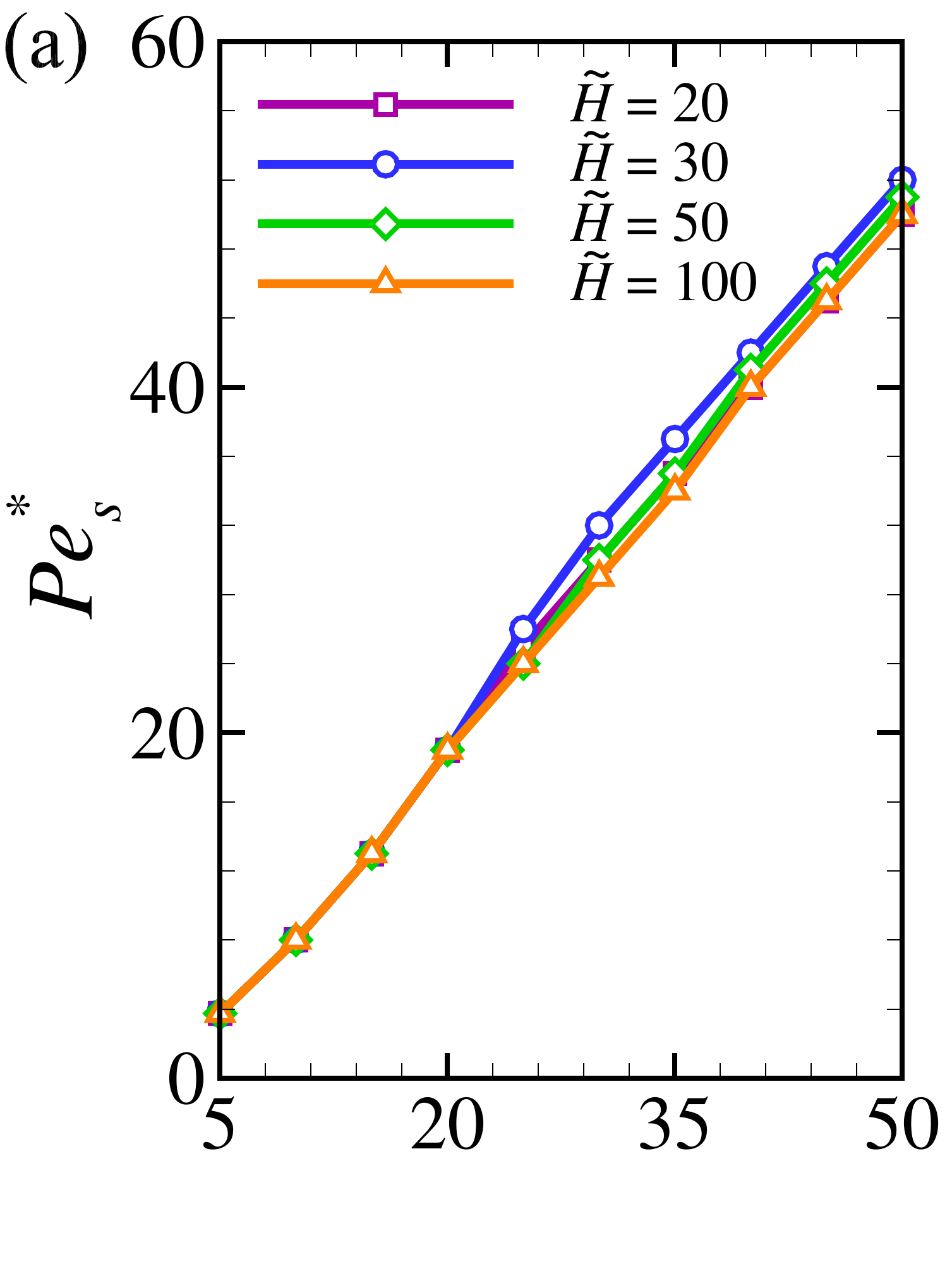}%
\includegraphics[width=0.405\linewidth]{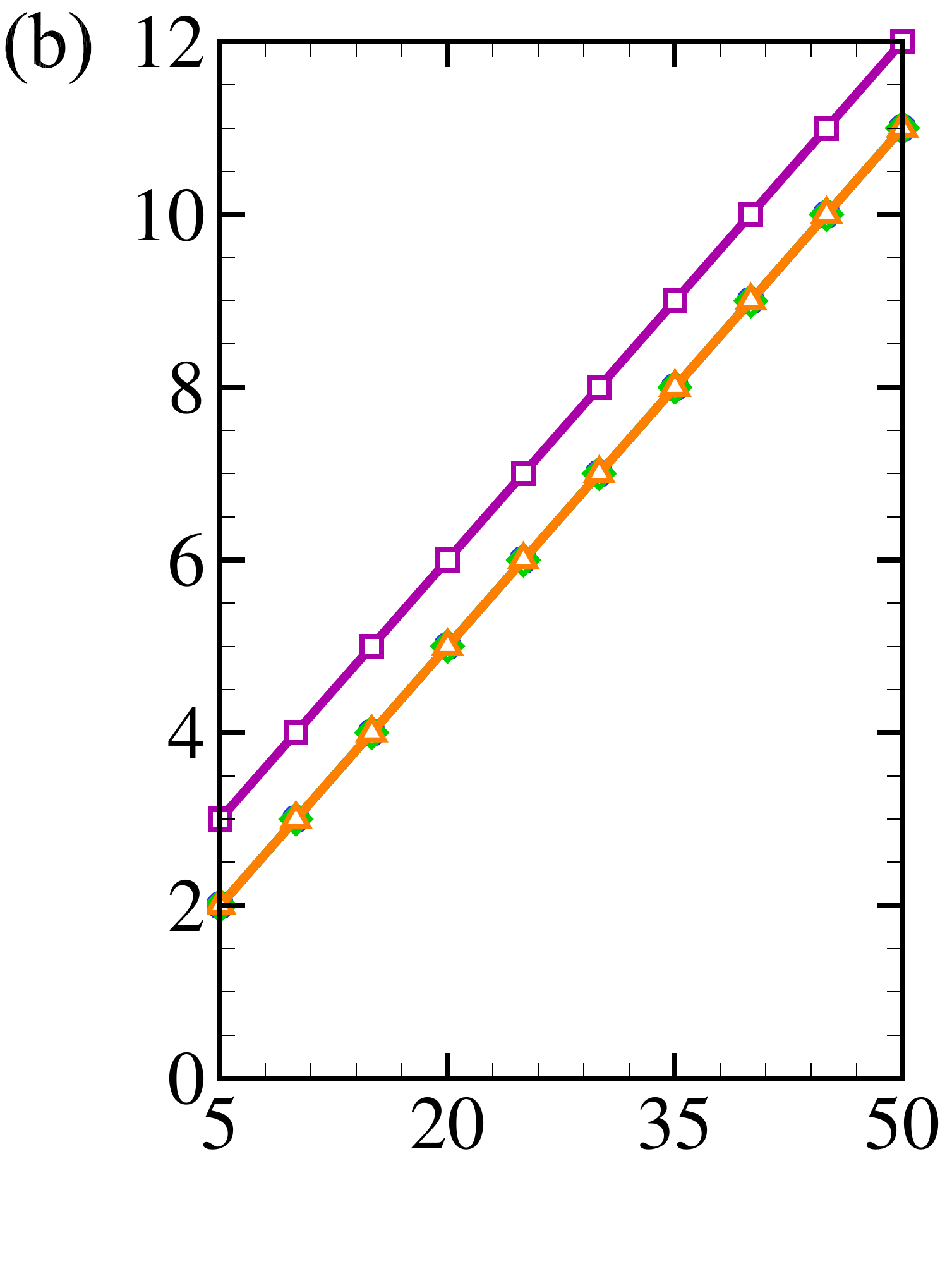}\vskip-4mm
\includegraphics[width=0.405\linewidth]{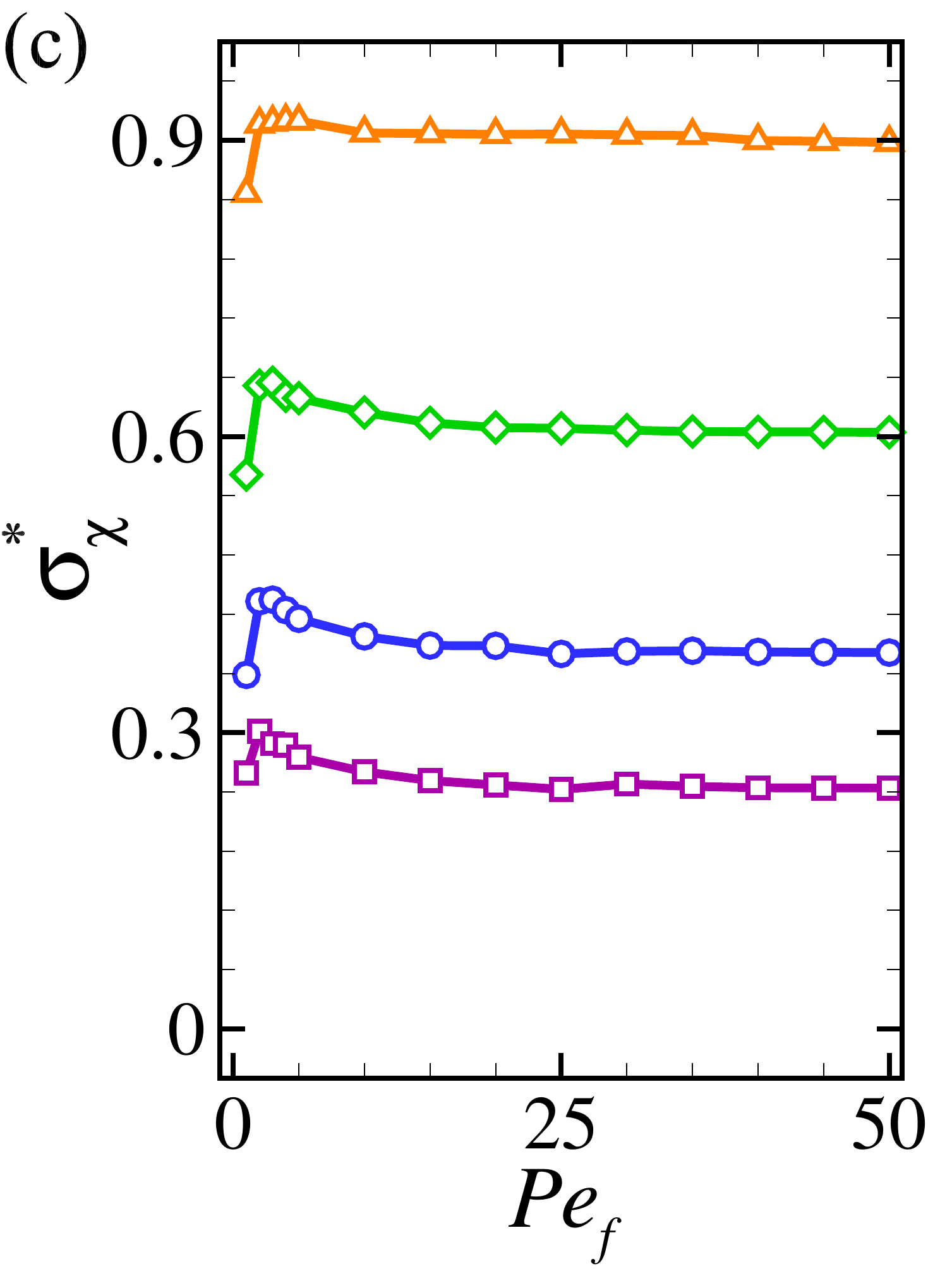}%
\includegraphics[width=0.405\linewidth]{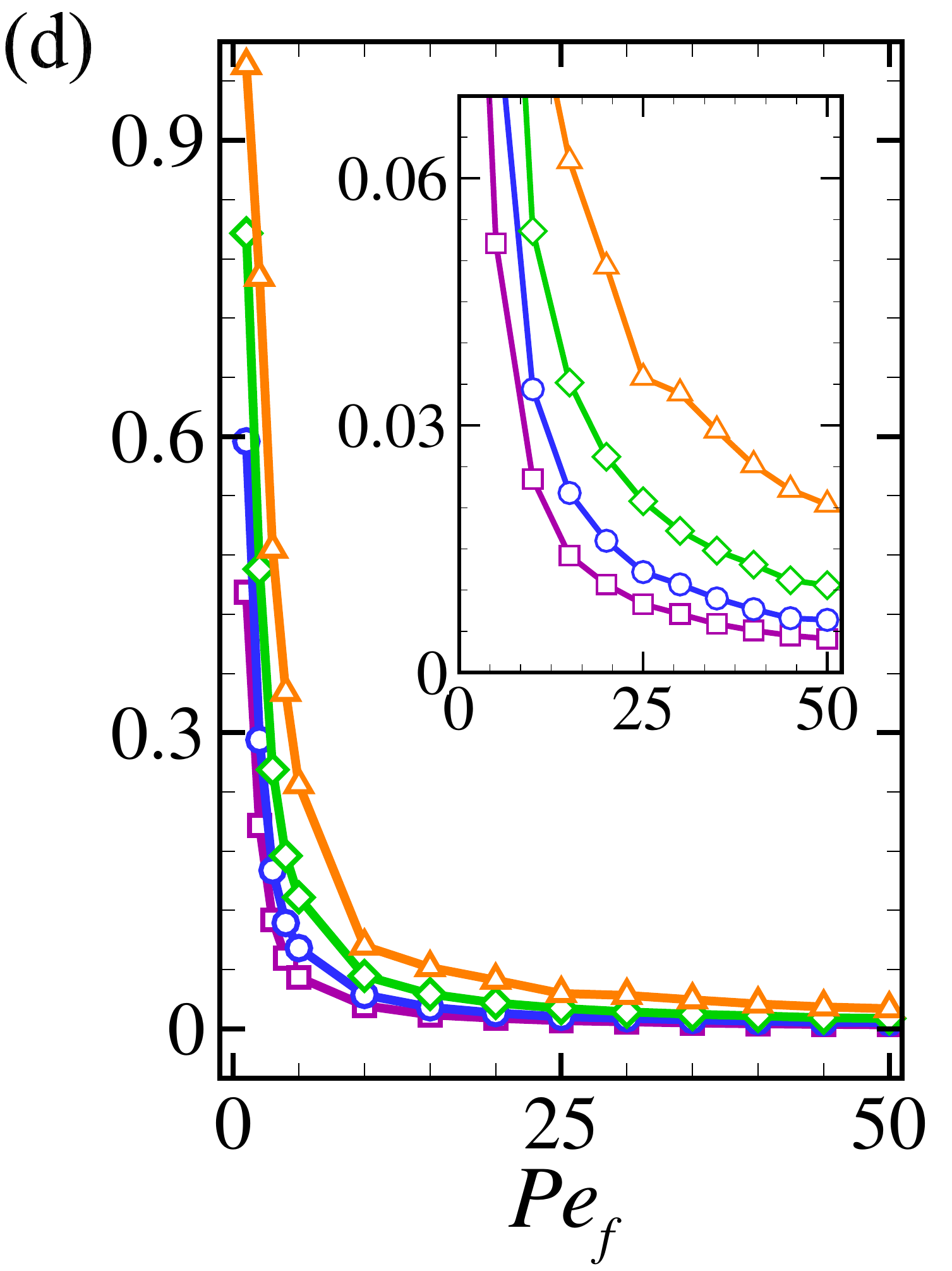}%
\vskip-2mm
\caption{Panels a and b show the optimal motility strengths, $Pe_s^\ast$, where the migration response factors  of prolate and oblate swimmers (of aspect ratios $\alpha=3$ and $1/3$)  are maximized within the linear-response regime.   Panels c and d show the maximal response factor, $\sigma_\chi^\ast$, corresponding to panels a and b, respectively. The mentioned quantities are plotted as functions of $Pe_f$ for different values of $\tilde H$, with the legends relevant to all panels shown only in panel a. Symbols are computed data and curves are drawn to guide the eye. The differences between different data sets in panels a and b fall within the numerical errors.
}
\label{fig:response_max}
\end{figure}

In all cases, the response factor is found to have a {\em nonmonotonic} behavior with $Pe_s$, showing  a pronounced global maximum at an intermediate value of $Pe_s$, representing an {\em optimal motility strength}, $Pe_s^\ast$, for swimmers to undergo efficient cross-stream migration in the weak-field regime. The existence of $Pe_s^\ast$ mirrors  the balance between the downward wall-to-wall migration of swimmers and their tendency to remain accumulated at the top walls, both being enhanced by the self-propulsion strength. The locus of the maximum thus identifies two different subregimes of $Pe_s<Pe_s^\ast$ and  $Pe_s>Pe_s^\ast$, where  the said wall-to-wall migration and wall-accumulation factors dominate, respectively. $Pe_s^\ast$ shows a clear dependence on the flow P\'eclet number, $Pe_f$, but not on the rescaled channel width, $\tilde H$. These behaviors are depicted in Figs.   \ref{fig:response_max}a (prolate) and b (oblate), which show that the optimal motility strength, $Pe_s^\ast$, increases {\em linearly} with the shear strength, $Pe_f$, while the maximal response, $\sigma_\chi^\ast$, remains nearly intact, as the rescaled channel width is increased from intermediate ($\tilde H=20$) to large values ($\tilde H=100$); see Figs.   \ref{fig:response_max}c (prolate) and d (oblate). 

Despite the above similarities, there are qualitative and quantitative differences between the response factors of prolate and oblate swimmers. First, prolate swimmers show substantially larger response factors as compared with the oblate ones. This could be expected given the competitive nature of the interplay between different factors (shear, field and self-propulsion) contributing to swimmer reorientation in the case of prolate swimmers as opposed to the oblate ones, as discussed in the text. It can also be understood based on the larger field-induced changes produced in the rotational residence times of prolate swimmers than the oblate ones in different angular quadrants (see the supplementary material). 

Second, the optimal motility, $Pe_s^\ast$, increases at a much faster rate with $Pe_f$ in the case of prolate swimmers than in the case of oblate ones (compare Figs. \ref{fig:response}a and b, and Figs. \ref{fig:response_max}a and b, where the slope of the curves  in panel a is about five times larger than that of those in panel b). This is indicative of the shear-trapping of prolate swimmers, i.e., a larger $Pe_f$ necessitates a larger $Pe_s$ to achieve the maximum level of downward migration. 

Third, the peak value of the  response factor, $\sigma_\chi^\ast$, which is obtained at $Pe_s^\ast$, remains nearly unchanged as $Pe_f$ is varied by an order of magnitude in the case of prolate swimmers (Fig. \ref{fig:response}a). This contrasts  the case of oblate swimmers (Fig. \ref{fig:response}b), where the peak value rapidly drops to zero. This behavior is shown in Figs. \ref{fig:response_max}c (prolate) and d (oblate), where the dependence on $\tilde H$ is also discernible. In both cases, $\sigma_\chi^\ast$ increases with $\tilde H$, but the increase is much faster in the case of prolate swimmers.

\bibliography{Magnetic_Couette_rv}

\clearpage 

\section*{Supplementary Material}

\subsection*{Deterministic orientational dynamics}
 \label{supp:rot_dyn}

The deterministic, or noise-free, orientational dynamics of the spheroidal swimmers in our model is given by the overdamped equation $\dot \theta(t)= \omega\left(\theta(t)\right)$,  where $\dot \theta(t)={\mathrm d}\theta(t)/{\mathrm d}t$ is the time derivative of the swim orientation angle, $\theta(t)$, measured from the $x$-axis at time $t$, and $ \omega(\theta)$ is the net angular velocity, obtained using  Eqs. (4), (6) and (10) of the main paper, as
\begin{equation}
 \omega(\theta) = \frac{\dot\gamma }{2} \bigg(\beta(\alpha) \cos2\theta-1 \bigg)-\chi D_R(\alpha) \cos\theta. 
\label{eq:thetaDot0}
\end{equation}

Using the nondimensionalization of the system parameters discussed in the main paper, which includes rescaling all units of time with the inverse rotational diffusivity of the reference sphere (as, e.g., $\tilde t=t D_{0R}$), we arrive at the rescaled equation $\tilde {\dot \theta}(\tilde t)=\tilde \omega\left(\theta(\tilde t)\right)$, where 
\begin{equation}
 \tilde \omega(\theta) = \frac{Pe_f}{2}\bigg(\beta(\alpha) \cos 2 \theta -1\bigg)- \chi \Delta_R(\alpha) \cos{\theta}
\label{eq:thetaDot}
\end{equation}
is the rescaled net angular velocity of the spheroidal particles due to the shear flow (first term, involving the flow P\'eclet number, $Pe_f$) and the external field applied in the  `downward' direction normal to the flow (second term, involving the field coupling strength, $\chi$). 

\subsection*{Field-modified Jeffery oscillations}
\label{supp:mod_Jeff}

\begin{figure}[t!]
\begin{center}
\sidesubfloat[]{%
\hskip-1mm%
\includegraphics[width=0.75\textwidth]{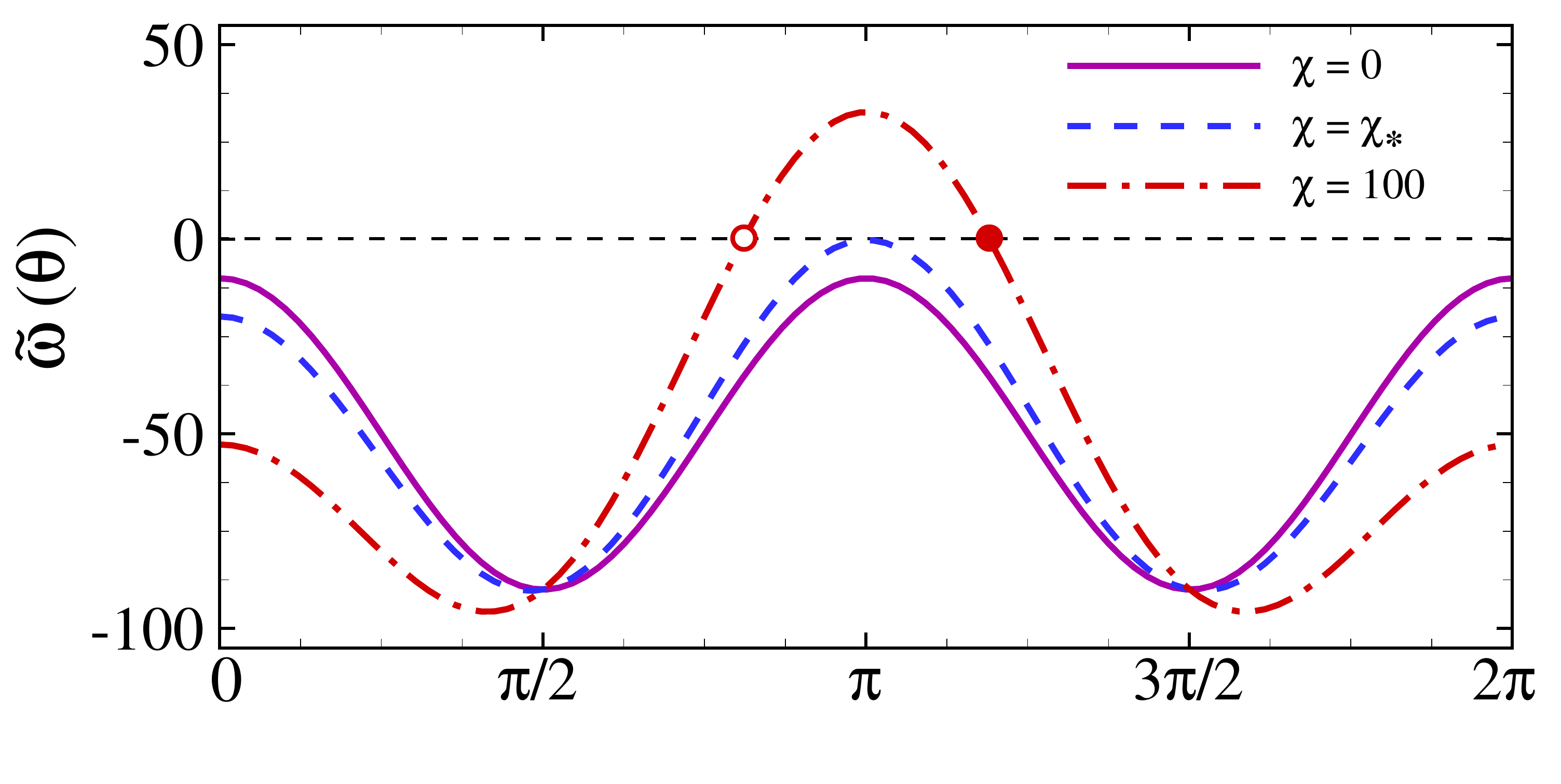}%
}%
\vskip-3mm%
\sidesubfloat[]{%
\hskip-1mm%
\includegraphics[width=0.75\textwidth]{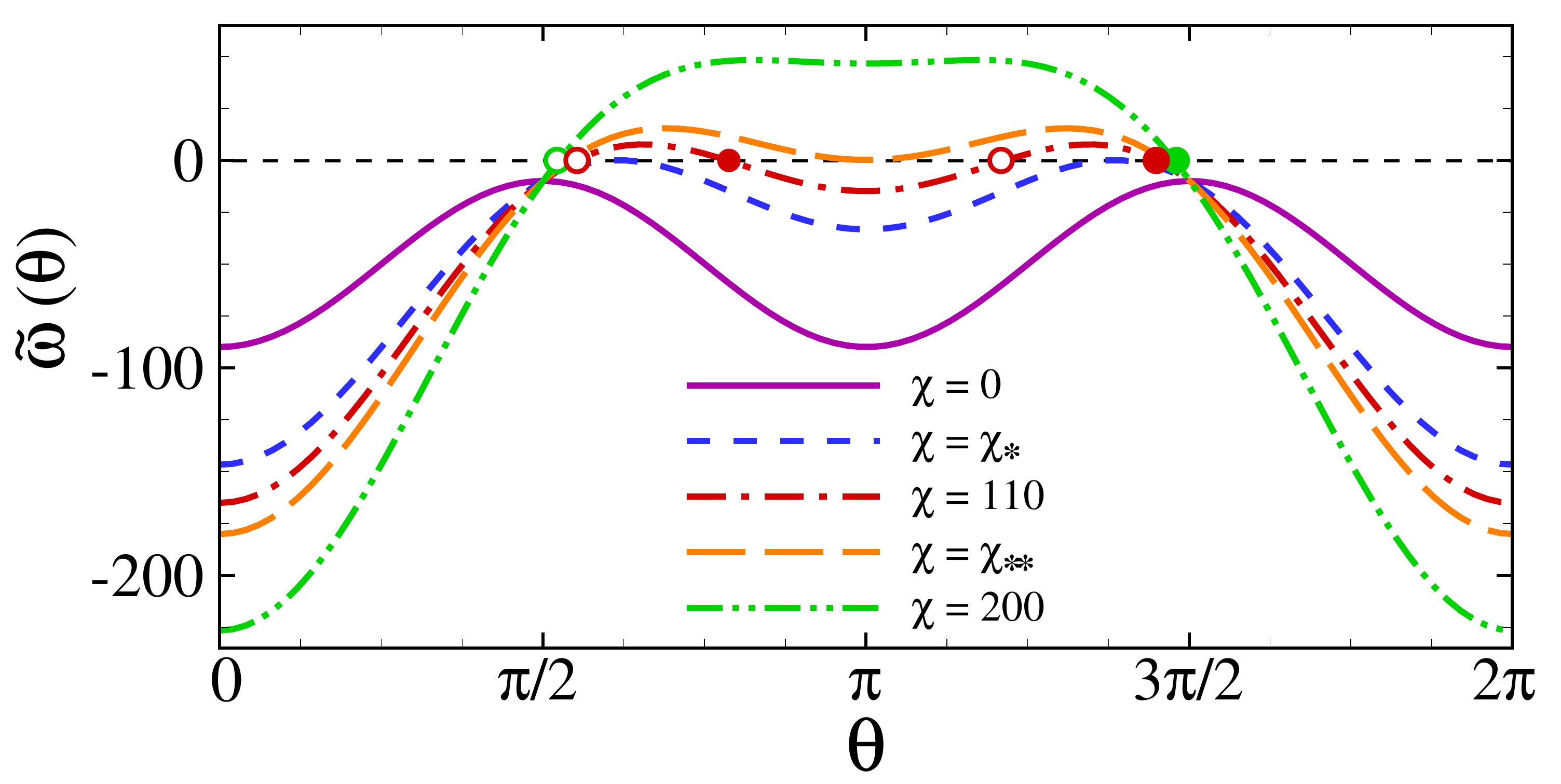}%
}%
\vskip-4mm
\caption{Rescaled net angular velocity of spheroidal particles, $\tilde \omega(\theta)$, for (a) $\alpha=3$ and (b) $\alpha=1/3$, respectively, as a function of the orientation angle, $\theta$, for $Pe_f=100$ and different field coupling strengths, $\chi$, as shown on the graphs. The filled (open) circles show the stable (unstable) fixed points.
}
\label{fig:thetaDot} 
\end{center}
\end{figure}

In the absence of an external field ($\chi=0$), $\tilde \omega(\theta)$ is always negative and takes a simple harmonic form as shown by purple solid curves for aspect ratios $\alpha= 3$ and $ 1/3$  in Figs. \ref{fig:thetaDot}a and b, respectively. It thus generates the standard (field-free) oscillations of the orientation angle of sheared spheroids  in clockwise direction that we refer to as  Jeffery oscillations \cite{Jeffery}. The swim orientation spends {\em equal} rotational `residence' times, $\tau_i$,  in each of the four $\theta$-quadrants (labeled by  $i=1,\ldots,4$). The residence time in, e.g., the first $\theta$-quadrant is $\tau_{1}\!=\!\!\int_0^{\pi/2}\! {\mathrm{d}} \theta\, |\tilde \omega(\theta)|^{-1}$. The residence times sum up to a full Jeffery  period $\tau_0\!=\!\sum_i\!\tau_i\!=\!2\pi(\alpha+\alpha^{-1})/Pe_f$; see Refs.    \cite{Jeffery,Bretherton1962,kim_microhydrodynamics}.

In the presence of an external field ($\chi\!>\!0$), $\tilde \omega(\theta)$ varies differently in different $\theta$-quadrants, but remains invariant at the upward-/downward-pointing orientations, $\theta\! =\! \pi/2$ and $3\pi/2$. In a sufficiently weak  field (below the pinning threshold to be determined later), we still have $\tilde \omega(\theta)\!<\!0$ (clockwise oscillations) but with $|\tilde \omega(\theta)|$ increased (decreased) monotonically with $\chi$ in the first/fourth (second/third) $\theta$-quadrants (Figs. \ref{fig:thetaDot}a and b). Hence, the residence times in the first/fourth (second/third) $\theta$-quadrants become {\em shorter} ({\em longer}) than their corresponding field-free values, $\tau_1\!<\!\tau_0/4\!<\!\tau_2$ and $\tau_4\!<\!\tau_0/4\!<\!\tau_3$. The up-/down-polarization symmetry of the deterministic dynamics ($\theta\rightarrow 2\pi-\theta$)  still ensures equal residence times in the first and fourth (second and third) $\theta$-quadrants,  $\tau_1\!=\!\tau_4$ ($\tau_2\!=\!\tau_3$), with or without an applied field.

The main contribution to the combined residence times $\tau_{14}$ and $\tau_{32}$ comes from the orientation angles that minimize $|\tilde \omega(\theta)|$. For prolate particles (Fig. \ref{fig:thetaDot}a), $\tau_{32}$ is increased with $\chi$ because the local maximum of $\tilde \omega(\theta)$ (minimum of $|\tilde \omega(\theta)|$) at $\theta=\pi$ progressively develops into a global maximum, with a $|\tilde \omega(\theta)|$ value approaching zero. The same holds for oblate particles (Fig. \ref{fig:thetaDot}b), in which case the local maxima of $\tilde \omega(\theta)$ (minima of $|\tilde \omega(\theta)|$) at  $\theta = \pi/2$ and $3\pi/2$ turn into global maxima, with $|\tilde \omega(\theta)|$ values approaching zero. In the latter case, the  loci  of the said extrema  also symmetrically shift into the second and third $\theta$-quadrants. 

The foregoing features portray what we refer to in the main paper as {\em field-modified Jeffery oscillations} of the swim orientation. These oscillations have a period of $\tau=\tau_{14}+\tau_{32}$, which turns out to be {\em bigger} than its corresponding field-free value, $\tau_0$.

\subsection*{Origin of the linear-response behavior}
\label{supp:linear_res}

Our considerations in the preceding section can be used to shed further light on the cross-stream migration of swimmers toward the bottom half of the channel  in weak fields and also on the corresponding linear-response regime discussed thoroughly in the main paper. 

The quantities $\tau_{14}$ and $\tau_{32}$ can be interpreted as the inverse of the {\em flip-down} and  {\em flip-up}  rates of the swim orientation, while rotating in clockwise direction over the angular intervals $\theta\in (\pi/2, -\pi/2]$ and $(3\pi/2, \pi/2]$, respectively. Thus, the aforementioned field-induced decrease (increase) in $\tau_{14}$ ($\tau_{32}$) translates into an increased likelihood of finding swimmers in the bottom half of the channel in the steady state. There is an additional mechanism due to the confinement that concurrently comes into play to enhance the bottom-half fraction of swimmers  (see also Figs. 2e and 5e of the main paper): For swimmers accumulated at the bottom wall, the time required by the shear-induced torque to reorient them (in clockwise direction away from their typical downward-pointing, normal-to-wall, orientation $\theta=3\pi/2$) and to make them swim away from the wall (at an angular orientation  $\theta\geq \pi$) is also increased in an applied field. This is because the mentioned time has a lower bound of $\tau_2=\tau_3=\tau_{32}/2$ (realized in the bulk with no wall effects being present), which increases with $\chi$. As evident,  the shear effects are still relevant in this regime of weak fields and the migration to the bottom half of the channel can only be partial and of an effective nature different from the full migration triggered by the orientational pinning to be considered in the next section. 
 
\begin{figure}[t!]
\begin{center}
\includegraphics[width=0.75\textwidth]{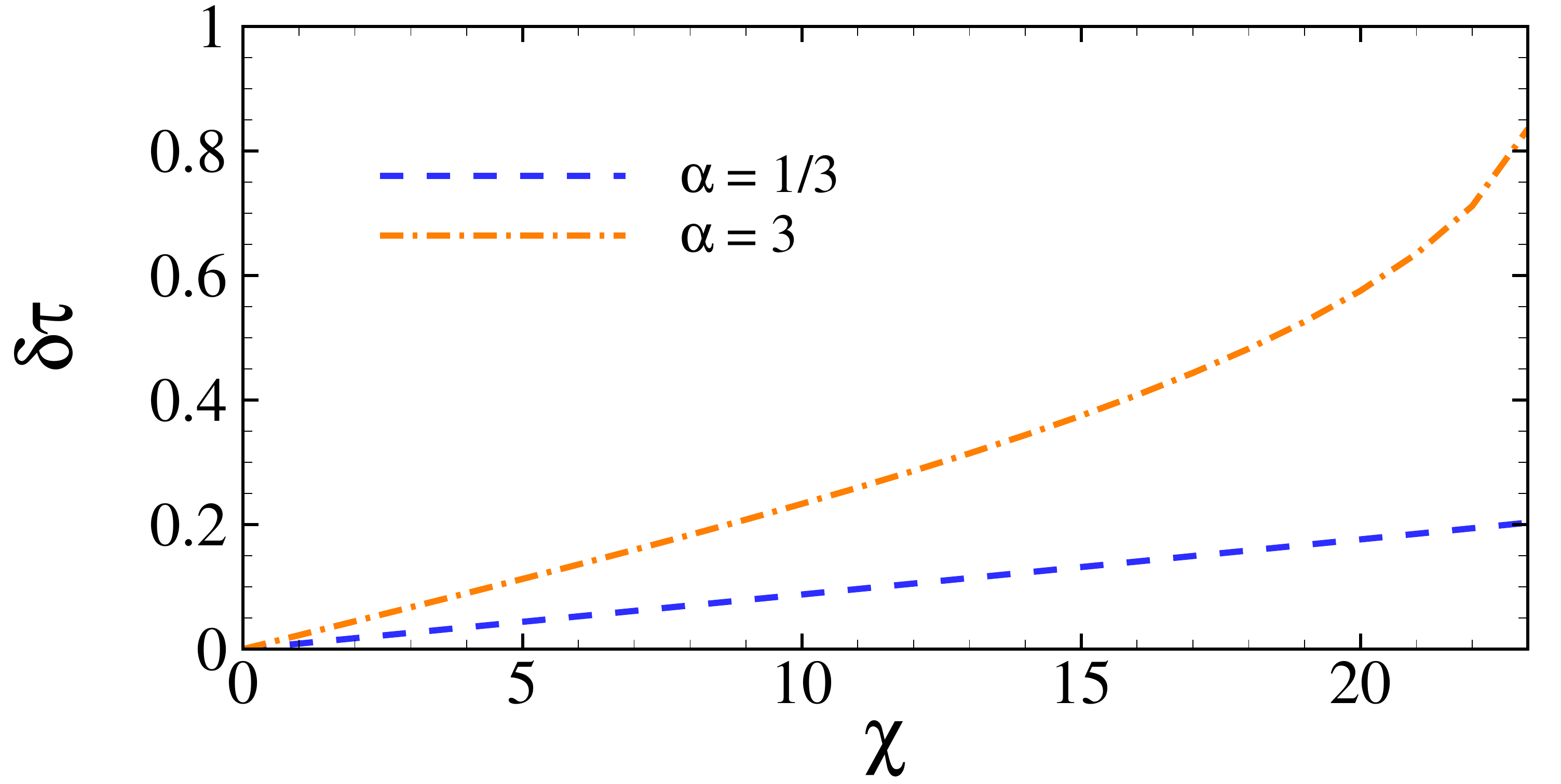}
\vskip-4mm
\caption{Relative difference, $\delta \tau=(\tau_{32} - \tau_{14})/\tau$, in the rotational residence times spent in the second/third and first/fourth $\theta$-quadrants, $\tau_{32}$ and $\tau_{14}$, respectively,  in the weak-field regime as a function of $\chi$ for prolate (oblate) swimmers of aspect ratio $\alpha=3$ ($1/3$)  at fixed $Pe_f=100$. }
\label{fig:timeIntegrals} 
\end{center}
\end{figure}

The difference between the flip-up/-down rates turns out to vary {\em linearly} with $\chi$  over a relatively wide range of values as $\chi$ is increased from zero. This is shown in Fig. \ref{fig:timeIntegrals}, where we plot   $\delta \tau =(\tau_{32} - \tau_{14})/\tau$ as a function of $\chi$ for $\alpha= 3$ and $1/3$  at fixed $Pe_f=100$. This intuitive observation justifies the appearance of a linear-response regime at sufficiently small $\chi$, as established by our numerical solutions. 

The field-induced increase in the period $\tau$ (not shown)  grows more rapidly with $\chi$ in the case of prolate particles of aspect ratio $\alpha$ than in the case of oblate particles of aspect ratio $\alpha^{-1}$ (recall that the field-free period $\tau_0$ is the same in both cases).  The same is true for the magnitude of the field-induced decrease and increase in $\tau_{14}$ and $\tau_{32}$, respectively (not shown), and also for the relative difference, $\delta \tau $, which is shown in Fig. \ref{fig:timeIntegrals} for the exemplary cases with aspect ratios $\alpha=3$ and $1/3$. These findings corroborate those in Appendix C of the main paper that the linear-response factor of prolate swimmers is generally larger than that of the oblate ones.

\subsection*{Orientational (bi-)stability and (double-)pinning} 
\label{supp:pinning}

For large field couplings, wholly different behaviors arise as compared with those discussed in the preceding sections, as $\tilde \omega(\theta)$ vanishes at multiple angles, creating stable/unstable fixed points (or orientational pinnings) in the deterministic  dynamics of the swim orientation.

\begin{figure}[t!]
\begin{center}
\sidesubfloat[]{%
\hskip-1mm%
\includegraphics[width=0.75\textwidth]{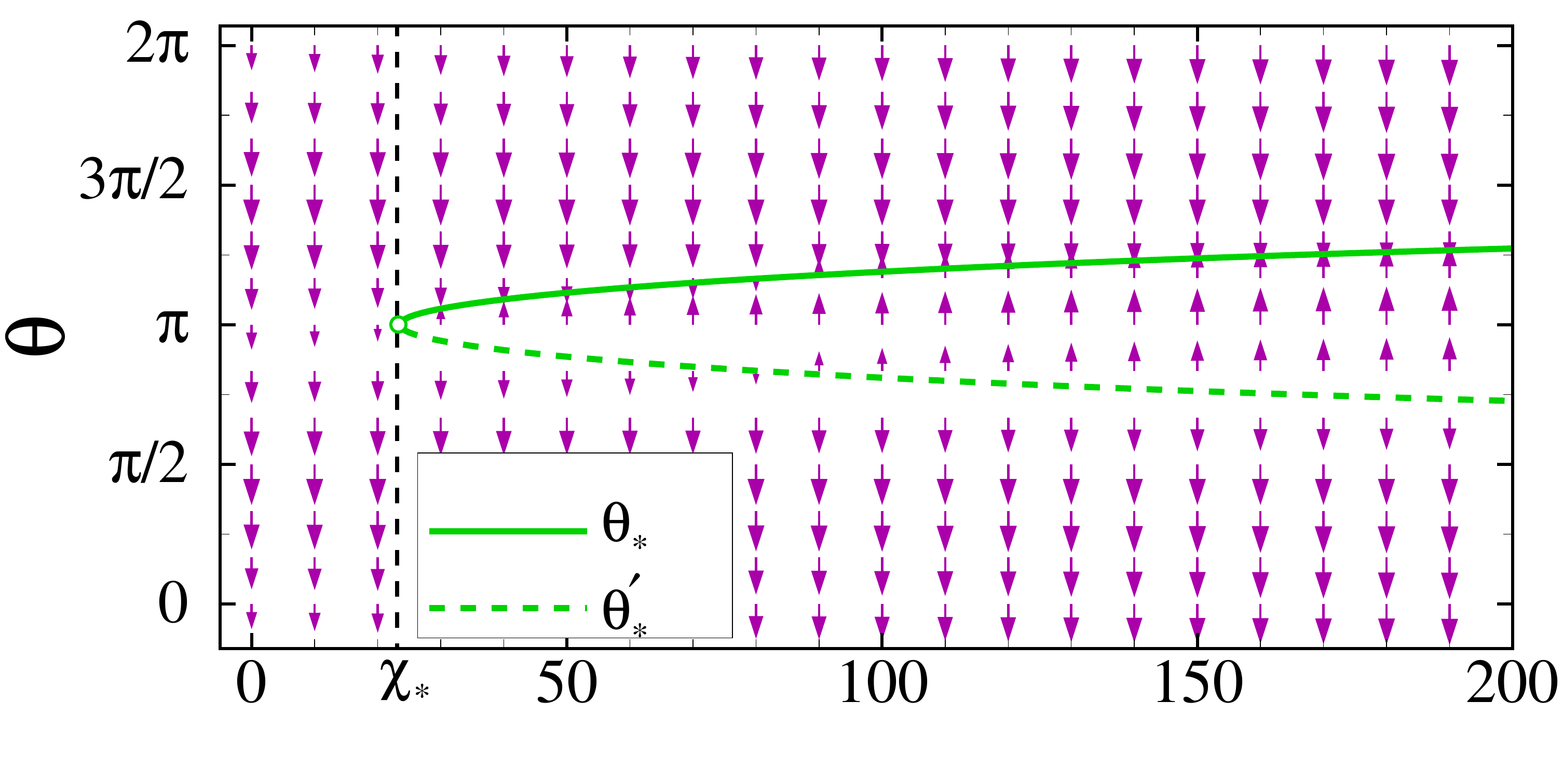}%
}%
\vskip-3mm%
\sidesubfloat[]{%
\hskip-1mm%
\includegraphics[width=0.75\textwidth]{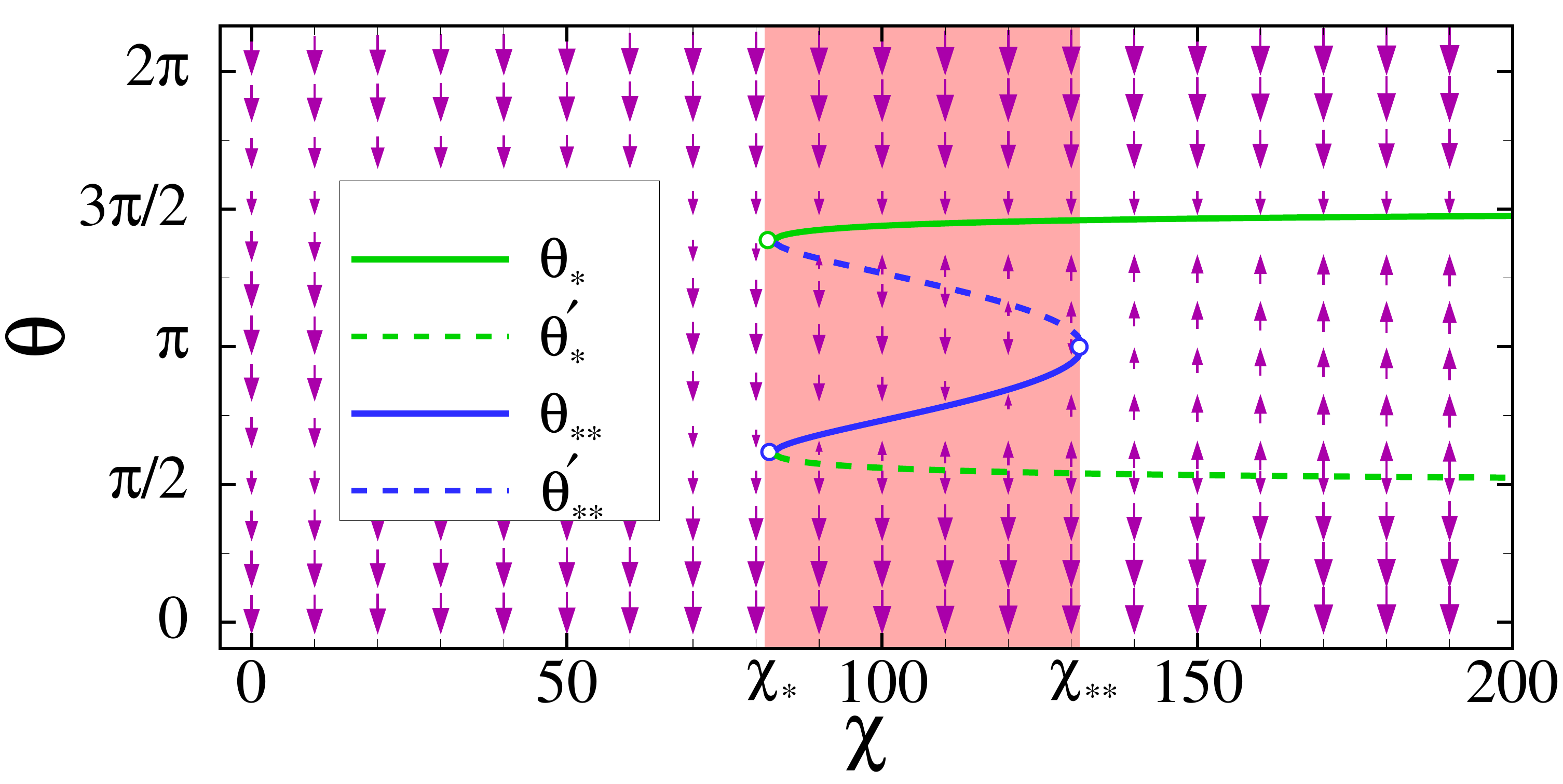}%
}%
\vskip-4mm
\caption{Stable and unstable orientational fixed points (solid and dashed curves, respectively) shown for (a) prolate and (b) oblate swimmers of aspect ratios $\alpha=3$ and $1/3$ as functions of the field coupling strength, $\chi$, for $Pe_f=100$.  The dashed vertical line in (a) shows the threshold $\chi_\ast/Pe_f\simeq 0.234$, and the colored region in (b) shows the interval, $\chi_\ast<\chi<\chi_{\ast\ast}$, with $\chi_\ast/Pe_f\simeq 0.828$ and $\chi_{\ast\ast}/Pe_f\simeq 1.318$.  Purple vertical arrows show the dynamical parameter-space flow $\dot \theta=\tilde \omega(\theta)$. For illustration purposes, the arrow sizes are divided  $|\tilde \omega(\theta)|^{3/4}$ with a suitably chosen numerical factor to enhance visual clarity.}
\label{fig:fixedPoints} 
\end{center}
\end{figure}

For prolate particles ($\alpha>1$, $0<\beta(\alpha)<1$), requiring $\tilde \omega(\theta)=0$, one finds a stable ($\partial \tilde \omega(\theta)/\partial\theta<0$) and an unstable ($\partial \tilde \omega(\theta)/\partial\theta>0$) solution, $\theta=\theta_\ast$ and $\theta'_\ast$, in the third and second $\theta$-quadrants,  respectively, when
\begin{equation}
\chi>\chi_\ast = \frac{1-\beta(\alpha)}{2\Delta_R(\alpha)}Pe_f. 
\label{eq:chi_ast}
\end{equation} 
Figure \ref{fig:thetaDot}a shows a double root $\theta_\ast=\pi$ at exactly $\chi=\chi_\ast$ (blue dashed curve), which upon increasing $\chi$ (red dot-dashed curve) splits into the stable/unstable fixed points (red solid/open circles), representing a saddle-node bifurcation \cite{Strogatz2000} (such an effect has also been reported in the case of sheared, nonactive, magnetic ellipsoids in an external field  \cite{Sobecki2018}). The loci of the fixed points $\theta_\ast$ and $\theta'_\ast$ are plotted as functions of  $\chi$ in Fig. \ref{fig:fixedPoints}a. The dynamical parameter-space flow, $\dot \theta=\tilde \omega(\theta)$, which intuitively indicates how swiftly the orientational states  of the system evolve toward or away from the said fixed points, is depicted by purple vertical arrows. As seen, the basin of attraction of the stable fixed point is the whole angular domain. The orientations of swimmers are thus predicted to be stabilized at $\theta_\ast$ for $\chi>\chi_\ast$, with $\theta_\ast$ increasing from $\pi$ and approaching to $3\pi/2$ as $\chi$ is increased. This gives rise to the unimodal regime of behavior, with swimmers accumulating at the bottom channel wall at a third-quadrant angle (i.e., while they are  inclined in the upstream direction), consistently with the steady-state numerical results discussed in the main paper.

For oblate particles ($\alpha<1$, $-1<\beta(\alpha)<0$), fixed points are found, when 
\begin{equation}
\chi>\chi_\ast = \sqrt{\frac{2|\beta(\alpha)|\left(\beta(\alpha)+1\right)}{[\Delta_R(\alpha)]^2}}Pe_f.
\label{eq:chi_ast_ob}
\end{equation}  
Figure \ref{fig:thetaDot}b shows that, at exactly $\chi=\chi_\ast$ (blue dashed curve), there are two distinct double roots in the second and third  $\theta$-quadrants, which, upon increasing $\chi$ slightly further (red dot-dashed  curve), split into two pairs of stable/unstable fixed points (red solid/open circles). The stable fixed points, denoted by $\theta_\ast$ and $\theta_{\ast\ast}$, are found in the third and second  $\theta$-quadrants, respectively, and the unstable ones, denoted by $\theta'_\ast$ and $\theta'_{\ast\ast}$, are found in the second and third  $\theta$-quadrants, respectively. There is a secondary threshold $\chi=\chi_{\ast\ast}$ (orange dashed curve in Fig.  \ref{fig:thetaDot}b) beyond which  the fixed points $\theta_{\ast\ast}$ and $\theta'_{\ast\ast}$ disappear (green dot-dashed curve); that is, when 
\begin{equation}
\chi>\chi_{\ast\ast} = \frac{1-\beta(\alpha)}{2\Delta_R(\alpha)}Pe_f. 
\label{eq:chi_astast}
\end{equation} 
The fixed-point loci are plotted as functions of $\chi$ in Fig. \ref{fig:fixedPoints}b with the dynamical parameter-space flow again shown by  purple vertical arrows. For $\chi_\ast<\chi<\chi_{\ast\ast}$, the basin of attraction of the stable fixed point $\theta_\ast$ is limited to the angular domain  $\theta'_{\ast\ast}<\theta<2\pi-\theta'_\ast$ and  that of the stable fixed point $\theta_{\ast\ast}$ is limited to the angular domain $\theta'_\ast<\theta<\theta'_{\ast\ast}$. Since $\theta_\ast$ has a wider basin of attraction involving larger parameter-space flow magnitudes as compared with $\theta_{\ast\ast}$, one expects a larger fraction of swimmers to be attracted by $\theta_\ast$. While this produces the driving mechanism for the migration of a larger fraction of orientationally pinned swimmers to the bottom half of the channel, the presence of $\theta_{\ast\ast}$ also produces a substantially large fraction of orientationally pinned swimmers migrating to the top half, generating the {\em reverse migration} behavior discussed in the main paper. For $\chi>\chi_{\ast\ast}$, $\theta_{\ast\ast}$ disappears and one enters the unimodal regime of behavior, as discussed in the main paper. 

The anomalous increase in $n_t^{\uparrow}$ (see Fig. 6a of the main paper), which takes place within the interval $\chi_\ast<\chi<\chi_{\ast\ast}$, can also be understood based on the fact that the basin of attraction of the stable fixed point $\theta_{\ast\ast}$ (i.e., the gap between the blue solid curve and the green dashed curve in Fig. \ref{fig:fixedPoints}b) expands  from below in such a way that it encompasses the orientation angles sufficiently close to $\pi/2$. This means that, in top half of the channel, a larger fraction of swimmers (that would naturally accumulate near the top wall) will be attracted by the second-quadrant fixed point $\theta_{\ast\ast}$, giving a larger $n_t^{\uparrow}$  as $\chi$ approaches $\chi_{\ast\ast}$. 

Finally, we note that, while the unstable branches in Figs. \ref{fig:fixedPoints}a and b can be affected by the thermal noise, the results related to the stable branches can only marginally be affected by the noise, which is omitted here but included in our probabilistic analysis in the main paper. 

\end{document}